
%
%
\input amstex
\input amsppt.sty
\CenteredTagsOnSplits
\NoBlackBoxes
\def\today{\ifcase\month\or

January\or February\or March\or April\or May\or June\or
 July\or August\or September\or October\or November\or December\fi
 \space\number\day, \number\year}
\define\({\left(}
\define\){\right)}

\define\Aut{\operatorname{Aut}}
\define\CC{{\Bbb C}}

\define\Diff{\operatorname{Diff}}

\define\HH{{\Bbb H}}
\define\Hom{\operatorname{Hom}}
\define\Map{\operatorname{Map}}

\define\RR{{\Bbb R}}

\define\Tr{\operatorname{Tr}}
\define\ZZ{{\Bbb Z}}
\define\[{\left[}
\define\]{\right]}

\define\chiup{\raise.5ex\hbox{$\chi$}}
\define\cir{S^1}
\define\coker{\operatorname{coker}}

\define\exertag #1#2{\removelastskip\bigskip\medskip\eightpoint\noindent%
\hbox{\rm\ignorespaces#2\unskip} #1.\ }

\define\inv{^{-1}}
\define\mstrut{^{\vphantom{1*\prime y}}}
\define\protag#1 #2{#2\ #1}

\define\res#1{\negmedspace\bigm|_{#1}}
\define\temsquare{\raise3.5pt\hbox{\boxed{ }}}

\define\theprotag#1 #2{#2~#1}

\define\zmod#1{\ZZ/#1\ZZ}

\define\Adgi{\Ad_{g\inv}}
\define\Ad{\operatorname{Ad}}
\define\Ess#1#2#3{\specseq E{#1}{#2}{#3}}

\define\Fss#1#2#3{\specseq F{#1}{#2}{#3}}

\define\HfBG{H^4(BG)}
\define\Lie{\operatorname{Lie}}

\define\Mor{\operatorname{Mor}}
\define\Obj{\operatorname{Obj}}
\define\Ob{\boldsymbol\Omega}
\define\PGp{P_{G'}}
\define\Qb{\bold{Q}}
\define\Qfix{{Q;h_1,\dots,h_k}}
\define\RZ{\RR/\ZZ}
\define\Rb{{\bold R}}
\define\TT{{\Bbb T}}
\define\Tb{\boldsymbol\Theta}

\define\Xo{\Xi _1}
\define\Xt{\Xi _2}
\define\Yos{\hbox{$\ssize Y$\kern-.5em\raise1.1ex\hbox{$\ssize\circ$}}}
\define\Yo{\hbox{$Y$\kern-.6em\raise1.55ex\hbox{$\ssize\circ$}}}
\define\ac#1#2{S_X(#1,#2)}
\define\acT{S_X(\Theta)}
\define\ad{\operatorname{ad}}
\define\altiab#1#2{e\(\int_{#1}\br\alpha\mstrut _{\vphantom{/}#2}\)}

\define\bQ{\partial Q}
\define\bX{\partial X}
\define\bY{\partial Y}
\define\bconn#1{\overline{\Cal A}_{#1}}
\define\bd{\dot{b}}
\define\bfld#1{{\Cal C}^\prime_{#1}}
\define\bfldb#1{\overline{\bfld{#1}}}
\define\br#1{<\!\!#1\!\!>}

\define\ce#1{\widehat{\Map(#1,G)}}
\define\cfrm#1{\Omega^#1_{\Yos}(\gQ)_c}

\define\conn#1{{\Cal A}_{#1}}

\define\cut{^{\text{cut}}}
\define\delc#1{\HH^{#1}}
\define\dtz{\frac{d}{dt}\res{t=0}}
\define\eac#1#2{e^{2\pi iS_{#1}({#2})}}
\define\eacT{e^{2\pi i \acT}}
\define\eb{\boldsymbol\eta}
\define\eint#1#2{\exp\( 2\pi i\int_{#1}{#2}\)}

\define\etdr{\raise.5ex\hbox{$\dot{\eta}$}}
\define\etd{\dot{\eta}}
\define\et{\eta _t}
\define\ev{\operatorname{ev}}
\define\fconn#1{\conn{#1}^{\operatorname{flat}}}
\define\ficonn#1{\conn{#1}^{*,\operatorname{flat}}}
\define\fld#1{{\Cal C}_{#1}}
\define\fldb#1{\overline{\fld{#1}}}
\define\fldg#1#2{\Cal{C}_{#1}^{#2}}
\define\form{\langle \cdot \rangle}
\define\frm#1{\Omega^#1_{Y}(\gQ)}
\define\gQ{\frak{g}_Q}
\define\gauge#1{{\Cal G}_{#1}}

\define\hb{\bar{h}}
\define\hol{\operatorname{hol}}
\define\ia#1#2{\exp\(\tpi\int_{#1}\alpha_{#2}\)}
\define\iab#1#2{\exp\(\tpi\int_{#1}\br\alpha _{#2}\)}
\define\id{\operatorname{id}}
\define\iline#1{I\mstrut _{#1,<\alpha>}}
\define\image{\operatorname{image}}
\define\im{\operatorname{im}}
\define\intline#1#2{I\mstrut _{#1,#2}}
\define\isline#1{I\mstrut _{#1,i^*<\alpha>}}
\define\lb{\bar{\ell}}

\define\mflat#1{\Cal{M}_{#1}}
\define\pM{\path(M)}
\define\pN{\path(N)}
\define\partrans{\operatorname{PT}}
\define\path{\Cal{P}}
\define\phibar{\bar{\varphi}}
\define\phitil{\tilde{\varphi}}
\define\qb{\bold{q}}
\define\ray{[0,\infty )}
\define\sect#1{\Cal{S}_{#1}}
\define\specseq#1#2#3#4{{#1}^{#2}_{#3#4}}
\define\tpi{2\pi i}

\define\unit#1{N_{#1}}

\define\ver{\operatorname{vert}}
\define\xo{\xi _1}
\define\xt{\xi _2}
\define\zi{[0,\infty)}
\define\zo{[0,1]}
\NoRunningHeads
\loadbold
\refstyle{A}
\widestnumber\key{DasW}
\topmatter
\title\nofrills Classical Chern-Simons Theory, Part 1 \endtitle
\author Daniel S. Freed  \endauthor
 \thanks The author is supported by NSF grant DMS-8805684, an Alfred P.
Sloan Research Fellowship, a Presidential Young Investigators award,
and by the O'Donnell Foundation.\endthanks
 \affil Department of Mathematics \\ University of Texas at Austin\endaffil
 \address Department of Mathematics, University of Texas, Austin, TX
78712\endaddress
 \email dafr\@math.utexas.edu \endemail
 \date June 4, 1992\enddate
\endtopmatter

\document

The formulations of Classical Mechanics by Lagrange and Hamilton are the
modern foundation of classical physics~\cite{Ar}.  Not only do these theories
describe the motion of systems of {\it particles}, but Maxwell's theory of
electromagnetism, as well as other {\it field} theories, can also be
formulated in Lagrangian and Hamiltonian terms.  A Lagrangian field theory is
defined by a local functional of the fields, called the {\it lagrangian\/},
and its integral over spacetime,\footnote{Here we refer to relativistic field
theories with the Euclidean signature.  A theory has a fixed dimension~$d+1$
for the ``spacetimes'', which consist of $d$~space dimensions and 1~time
dimension.  Since we are in the Euclidean framework, the term `spacetime' is
a misnomer, but we use it anyway.} called the {\it action\/}.  The classical
solutions of the field theory are the critical points of the action.  In
particular, the minima satisfy the ``least action principle'' of
Maupertius.\footnote{See ~\cite{Fey,\S19} for a delightful exposition.
Feynman is adamant that the principle of least action is the fundamental
physical principle of classical field theories; his path integral is the
corresponding approach to quantum field theories.} The Hamiltonian theory is
defined by a function, the {\it hamiltonian\/}, on phase space, or more
generally on a symplectic manifold.  The classical motion of the system is
then described by Hamilton's equations, whose solutions are integral curves
of the symplectic gradient vector field of the hamiltonian.  For many
mechanical systems of particles, which should be regarded as
$0+1$~dimensional field theories, there is both a Lagrangian and Hamiltonian
formulation.  Then the relationship between them is expressed by the Legendre
transform, if the lagrangian is nondegenerate.  A typical example is the
motion of a particle on a Riemannian manifold~$M$.  The action of the
Lagrangian theory is the energy of a path in~$M$. The hamiltonian is the norm
square function on the tangent bundle~$TM$, which obtains a symplectic
structure from the metric.  For higher dimensional field theories the
Hamiltonian formulation only makes sense on spacetimes which are globally a
product of space and time, that is, on $(d+1)$-dimensional manifolds which
are the product of a $d$-manifold with a 1-manifold.  A field on this product
manifold is a path of fields in space.  Symmetry plays an important role in
classical physics; many mechanical systems admit groups of symmetries.
Noether discovered that symmetries are linked to conservation principles.
This leads to the study of Noether currents in the Lagrangian theory and
moment maps in the Hamiltonian theory.

The Chern-Simons form of a connection, introduced in~\cite{CS}, can be viewed
as the lagrangian of a classical field theory.  In this paper we study the
$3=2+1$~dimensional case.  We are principally motivated by recent work of
Witten~\cite{W}, who demonstrates that the quantization of this classical
theory leads to new topological invariants.  Our study of the classical
theory clarifies certain features which are reflected in the quantum theory.
In the course of our study we found that the classical Chern-Simons theory
has many subtle aspects which provoke some refinements of the usual
mathematical formulations of field theories.  However, one should be aware
that the Chern-Simons theory differs from most field theories in two
important respects.  First, the theory is topological (generally covariant)
in that it is defined on oriented manifolds without any choice of metric.
Hence it admits diffeomorphism groups as symmetries, whereas the usual
examples admit only isometries.  Second, the lagrangian~\thetag{1.26}
involves only first powers of first derivatives of the fields, which leads to
a first order Euler-Lagrange equation~\thetag{3.4}; the usual examples
involve second derivatives or first derivatives squared, and so lead to
second order equations of motion.

{\it Everything is derived from the action.\/} This principle, articulated by
many theoretical physicists, places the Lagrangian formulation ahead of the
Hamiltonian formulation.  In fact, the latter is derived from the former.  In
many field theories one writes down the action and verifies its fundamental
properties immediately; by contrast in~\S{2} we take many pages to define the
Chern-Simons action.  If $X$~is a closed, oriented 3-manifold, then the
action is a complex number with unit norm.  This is the exponential of
$\tpi$~times the usual action, which is a real number determined modulo the
integers.  The fact that the action not a real number, but is only defined up
to integers, leads to the interesting geometry of the theory.  The reality of
the action, which implies that the exponential has unit norm, reflects the
{\it unitarity\/} of the theory.  If $X$~has a nonempty boundary, then the
action is an element of unit norm in an abstract metrized complex line which
depends on the restriction of the field to~$\bX$.  We call this line the {\it
Chern-Simons line\/}.\footnote{Instead of complex lines with inner products
we can restrict to the elements of unit norm, which form a principal
homogeneous space for the circle group.  Many discussions of classical and
quantum Chern-Simons theory replace the Chern-Simons line bundle (over the
moduli space of flat connections; see below) with a determinant line bundle.
Although these two bundles may be isomorphic if appropriately chosen, the
Chern-Simons action leads directly to the Chern-Simons line bundle, not the
determinant line bundle.  From this perspective the determinant line bundle
is extraneous to the theory.  Observe that the Chern-Simons line bundle is
defined by a cohomology class in~$\HfBG$, whereas the determinant line bundle
is defined by a representation of~$G$.  However, the determinant line bundle
most likely enters into another theory, whose action is the $\eta $~invariant
of Atiyah-Patodi-Singer.} Thus the Lagrangian theory assigns a metrized line
to every field on a space (closed, oriented 2-manifold)~$Y$, and a unit
element in the line of the boundary field for every field on a spacetime
(compact, oriented 3-manifold)~$X$.  In most field theories these lines are
canonically trivial; the nontriviality of these lines in the Chern-Simons
theory is why this theory illuminates the mathematical structure of classical
field theory so vividly.  We remark that it is crucial to construct these
lines precisely, and not just up to isomorphism. The characteristic
properties of the Lagrangian theory are spelled out in \theprotag{2.19}
{Theorem}, which is an axiomatization of topological classical field theory.
The Chern-Simons action is invariant under a large symmetry group (of gauge
transformations), which we discuss below.

The most important property of a Lagrangian field theory is {\it locality\/}.
Fields are local; they can be cut and pasted.  (A formal statement looks very
similar to the definition of a sheaf.) The action is local in that it adds
when we glue fields together.  We call this a {\it gluing law\/}; it is
stated precisely in \theprotag{2.19(d)} {Theorem}.  All of our constructions
in this paper are local, and so lead to gluing laws: \theprotag{2.19(d)}
{Theorem}, \theprotag{4.4(d)} {Proposition}, \theprotag{4.11} {Proposition},
\theprotag{4.26(d)} {Theorem}, and \theprotag{5.29} {Proposition}.

So far we have discussed general properties of field theories.  Chern-Simons
theory is a gauge theory, so in~\S{1} we review the geometry of connections
on principal bundles with compact structure group~$G$.  Our exposition is
slanted toward field theory.  For example, we emphasize cutting and pasting
of connections.  At the end of this section we introduce the Chern-Simons
3-form of a connection, which is our lagrangian.  Chern-Simons forms are
parametrized by invariant bilinear forms on the Lie algebra of~$G$.  To get
an action which is well-defined up to integers, we must restrict to a certain
lattice of {\it integral\/} forms.  For simple, simply connected groups
like~$SU(2)$ these forms (or cohomology classes) are parametrized by the
integers~\thetag{6.2}.  This integer is usually called the ``level'' of the
theory.  This is sufficient to define the Chern-Simons action if $G$~is
connected and simply connected, as we assume throughout this paper.  But for
a general compact Lie group~$G$ we need to choose a refinement of this
bilinear form to define the Chern-Simons action.  This is (up to isomorphism)
a cohomology class in~$H^4(BG)=H^4(BG;\ZZ)$, where $BG$~is the classifying
space of~$G$; the integral bilinear forms are in natural 1:1~correspondence
with the image of~$\HfBG$ in~$H^4(BG;\RR)$.  The theories for arbitrary
compact Lie groups are the subject of Part 2 of this paper~\cite{F2}.  The
importance of the general case in the study of rational conformal field
theories has been emphasized by Moore and Seiberg~\cite{MS}.  The definition
of the action for the general case was sketched by Dijkgraaf and
Witten~\cite{DW}.  However, they use singular cochains, and it is not
possible to discuss smoothness (of the Chern-Simons line bundle, for example)
in their framework.

The classical solutions in Chern-Simons theory are flat connections.  We
derive the Euler-Lagrange equations at the beginning of~\S{3}, where we also
discuss a little of the geometry of the space of flat connections.  Mostly
in~\S{3} we study the Chern-Simons action on product manifolds\footnote{We
write `$\zo$' first in this product so that the orientations work out
properly.  This should be done quite generally for homotopies and
bordisms.}~$X=\zo\times Y$; that is, we study the Hamiltonian theory.  Then
the action can be interpreted as the parallel transport of a unitary
connection on the Chern-Simons line bundle formed over the space of fields on
the boundary (\theprotag{3.17} {Proposition}).  The abstract argument of
Appendix~B shows that the existence of this connection depends only on the
basic properties of the action, in particular the gluing law, not on
particular features of this theory, and so should exist in any Lagrangian
field theory.  However, our proof in~\S{3} proceeds by direct calculation.
The curvature of this connection (times ~$i/2\pi $) is the symplectic
structure~\thetag{3.18} of the Hamiltonian theory.\footnote{This geometric
point of view on the relationship between the Lagrangian and Hamiltonian
theories appears to be well-known (see ~\cite{Ax}, for example), though it is
not often expressed in the literature.} Notice that the symplectic form
automatically satisfies an integrality condition, the integrality condition
in ``geometric quantization'' theory, as it is the curvature of a line
bundle.  Since the Chern-Simons theory is topological, the hamiltonian
function is identically zero---there is no local motion.  The symmetries in
the theory---gauge transformations---lift to the Chern-Simons line bundle
over the space of fields; this lift is constructed directly from the
Chern-Simons action.  From this lift one computes a moment map~\thetag{3.19}
for the symmetry on the space of fields.  Not only does the Lagrangian
formulation guarantee the existence of a moment map, but it also constructs
one (via this action on the line bundle) in situations where the moment map
is not unique.\footnote{This occurs in Chern-Simons theory if the center of
the gauge group~$G$ has positive dimension, which does not happen in the
groups we consider here, but certainly there are compact Lie groups
(e.g.~$G=\TT$, the circle group) where this occurs.  They will be considered
in Part~2.} The space of classical solutions on the infinite
cylinder~$\zi\times Y$ is also a symplectic manifold, in this case the
symplectic quotient of the space of fields.  The space of classical solutions
on an arbitrary 3-manifold with boundary, which is not necessarily a product,
maps onto a Lagrangian submanifold of the symplectic manifold attached to the
boundary (\theprotag{3.27} {Proposition}).  Not only the symplectic form, but
also the Chern-Simons line bundle with connection are trivial on this
submanifold.  The Hamiltonian theory is then the assignment of a symplectic
manifold to every space~$Y$ and a Lagrangian submanifold to every
spacetime~$X$.

One feature of the Chern-Simons theory which is not usually considered in
standard field theories is the geometry of the theory on spaces (as opposed
to spacetimes) with boundary, here 2-manifolds with boundary.  In the quantum
theory~\cite{W} these are especially important and lead to an ``exact
solution'' of the path integral.  In~\S{4} and~\S{5} we construct the
classical theory on surfaces with boundary.  Our constructions depend on a
closely related $(1+1)$~dimensional field theory, the Wess-Zumino-Witten
theory, which is discussed in Appendix~A.  In these constructions we
generalize the theory on closed surfaces and so construct complex lines
associated to connections on surfaces with boundary.  To accomplish this we
must introduce certain trivializing data on the boundary of the surface.
Hence in~\S{4} we make a digression into the geometry of connections on the
circle.  In particular, we discuss the universal bundle and properties of the
holonomy map (\theprotag{4.18} {Lemma}).  We then construct the Chern-Simons
lines, but only if we suitably restrict the values of the boundary holonomies
(\theprotag{4.26} {Theorem}).  These lines fit together to form a
(Chern-Simons) line bundle over the moduli space of flat connections on the
surface with boundary, now with fixed boundary holonomies.  We discuss this
moduli space at the end of~\S{4} and derive a formula for its
dimension~\thetag{4.44}.  In~\S{5} we construct a connection on the
Chern-Simons line bundle and compute its curvature (\theprotag{5.9}
{Proposition}).  This line bundle and connection depends on the trivializing
boundary data.  This dependence will be more clear in Part~2, where we also
construct geometric objects similar to line bundles with connection which
exist without restriction on the boundary holonomies.

In many ways the classical Chern-Simons theory unifies the gauge-theoretic
geometrical objects in low dimensions that are associated with a compact Lie
group~$G$ and an integral bilinear form on its Lie algebra:\footnote{More
precisely, a class in~$\HfBG$.} a characteristic class of $G$~bundles over a
closed oriented 4-manifold, the Chern-Simons invariant of a connection over a
closed oriented 3-manifold, and the line bundle with connection over the
moduli space of flat connections on a closed oriented 2-manifold.  There are
corresponding geometric objects over manifolds with boundary, as we discussed
earlier.  There is even something one can say in 1~dimension, but this is
more complicated.  These constructions are all fit in nicely with {\it smooth
Deligne cohomology\/}.\footnote{Gaw\c edzki~\cite{Ga} treated the action in
Wess-Zumino-Witten theory using smooth Deligne cohomology.  Related ideas
also appear in work of Brylinski.} In fact, they can be constructed by
integration of Deligne cocycles over these various manifolds.  We develop
this integration theory in~\cite{F1}, and discuss the Chern-Simons theory
from this point of view in Part~2 of this paper~\cite{F2}.  Finally, then,
the Chern-Simons theory looks like standard field theories---the action is an
integral over spacetime.  A different integration theory (for singular
cocycles) can be used when the gauge group is finite~\cite{FQ}.  There is
probably an analogous integration theory for $K$-theory and superconnections.
The outstanding question is to construct integration theories which give the
corresponding objects in {\it quantum\/} Chern-Simons theory.  (When the
gauge group is finite, the standard path integral is a finite sum, and the
theory can be constructed directly~\cite{FQ}.)

We conclude with a final comment about the symmetries in Chern-Simons theory,
or more generally in {\it any\/} classical gauge theory.  Namely, the
symmetries form a {\it groupoid\/} rather than a group.  A groupoid is a
``group with states''.  In gauge theory the states are the various principal
bundles over a given manifold, and the group elements are isomorphisms
between principal bundles (which cover the identity map on the base).  For
each principal bundle the automorphisms form a group---the group of gauge
transformations.  However, the larger symmetry of the groupoid enters since
there is no canonical identification of two isomorphic bundles with
connections; in other words, connections have automorphisms.  This is
particularly important when we glue together bundles and connections.  This
extra symmetry has nontrivial consequences, some of which are manifest in the
quantum theory, and we find that they justify our use of categorical
language.

Several closely related manuscripts appeared while this work was in progress.
Among them is Axelrod's thesis~\cite{Ax}, which also treats the classical
Chern-Simons theory (and other topics).  He considers more examples of
classical theories, and he goes further towards formulating axioms for
classical topological field theories in general.  Daskalopoulos and
Wentworth~\cite{DasW} also construct line bundles with connection over the
moduli spaces of flat connections on a surface with boundary.  Their
connection differs from ours; they found a relationship between boundary
holonomies and representations that our connection does not exhibit
(cf.~\S{5}).  Chang~\cite{Ch} constructs a determinant line bundle with
connection over these moduli spaces.

Over the two years this work has been in progress, I talked to many people
about the contents of this paper.  I thank all of them, and especially single
out Scott Axelrod, Joseph Bernstein, Sheldon Chang, George Daskalopoulos,
Frank Quinn, Karen Uhlenbeck, Alan Weinstein, and Richard Wentworth for
illuminating conversations.  This paper grew out of lectures given at the
University of Texas in the spring of~1990.  I thank the audience for their
feedback.  I'd also thank the Aspen Center for Physics and the Regional
Geometry Institute in Park City for their hospitality while some of this work
was carried out.  Finally, my thinking about field theories has been
stimulated greatly by Graeme Segal's work in this area.

\newpage
\heading
\S{1} Connections on principal bundles
\endheading
\comment
lasteqno 1@ 31
\endcomment

In this section we review some basics about Lie groups and connections on
principal bundles.  We consider arbitrary compact Lie groups.  Although
subsequent sections of this paper only treat connected and simply connected
groups, in Part~2 ~\cite{F2} we take up the general case.  Beyond textbook
material, which we review to establish notation, we discuss carefully how to
cut and paste connections (\theprotag{1.24} {Proposition}).  This establishes
the fact that connections can serve as local fields in a field theory.  Then
we introduce the lagrangian, the Chern-Simons form~\thetag{1.26}, and state
its basic properties (\theprotag{1.27} {Proposition}).  This form is only
defined on the total space of a bundle, not on the base, and this leads to
some of the rich geometry of the theory.  Since connections have
automorphisms, we use the language of categories to describe the space of all
connections.

Let $G$~be a compact Lie group.  The identity component~$G_0$ is a normal
subgroup with quotient the finite group~$\Gamma =\pi _0G$ of components:
  $$ 1 @>>> G_0 @>>> G @>>> \Gamma @>>> 1. \tag{1.1}$$
Then there is a finite covering
  $$ 1 @>>> A @>>> \widetilde{G_0}@>>> G_0 @>>> 1 \tag{1.2}$$ 
for some finite abelian group~$A$ such that
  $$ \widetilde{G_0}\cong T \times K_1 \times \cdots \times K_s, \tag{1.3}$$
where $T$~is a torus and $K_i$~are connected, simply connected simple groups.
The basic examples of compact Lie groups are finite groups, the circle
group~$\TT$, and the connected, simply connected simple groups (of type
A,B,C,D,E,F,G); equations~\thetag{1.1} and~\thetag{1.2} show that all others
are products of these up to finite covers and finite extensions.

The {\it Lie algebra\/} $\frak{g}$ of~$G$ is the vector space of {\it left\/}
invariant vector fields on~$G$ endowed with the Lie bracket.  The Lie
algebras of~$G$, $G_0$, and~$\widetilde{G_0}$ coincide.  There is a natural
$\frak{g}$-valued 1-form~$\theta $ on~$G$ which assigns to each vector its
left invariant extension.\footnote{It is often denoted $g\inv dg$ for matrix
groups.} This {\it Maurer-Cartan form\/} satisfies
  $$ \aligned
     L_g^*\theta &= \theta ,\\
     R_g^*\theta  &= \Adgi\theta ,\endaligned \tag{1.4}$$
where $L_g\:G\to G$ is left multiplication by~$g\in G$, the map $R_g\:G\to
G$~is right multiplication, and $\Ad_g = L_g\circ R_{g\inv }$ is the adjoint
map.  Its differential satisfies the Maurer-Cartan equation
  $$ d\theta  + \frac{1}{2}[\theta \wedge \theta ] =0; $$
it is a simple consequence of the relationship of~$d$ to the Lie bracket.

A {\it principal $G$ bundle\/} ~$P\to X$ is a manifold~$P$ with a free {\it
right\/} $G$~action; the quotient~$X$ is then also a manifold.  Notice that
if $G$~is finite, then $P$~is a regular covering space
of~$X$.\footnote{However, the categories of principal bundles and regular
covering spaces differ:  Maps of principal bundles commute with the group
action, whereas maps of covering spaces need not.}  In general, the
fiber~$P_x$ over~$x$ is a space on which $G$~acts simply transitively by
right multiplication.  Hence for each $p\in P_x$ there is an isomorphism of
right $G$-spaces
  $$ \aligned
     \tau _p\:G&\longrightarrow P_x\\
     g&\longmapsto p\cdot g.\endaligned $$
The ``coordinate change'' $\tau _{p'}\inv \tau _p=L_g$, where $g\in G$ is the
unique element with $p'=p\cdot g$.  Hence left invariant tensors on~$G$
induce tensors on~$P_x$.  Thus we identify $T_pP_x\cong \frak{g}$ for
all~$p\in P_x$.  Dually, the Maurer-Cartan form~$\theta $ induces a
form~$\theta _x$ on~$P_x$ which satisfies
  $$ \gather
     R_g^*\theta _x = \Adgi\theta _x, \\
     d\theta _x+\frac{1}{2}[\theta _x\wedge \theta _x] = 0.
     \tag{1.5}\endgather $$
Here $R_g\:P_x \to P_x$ is the right action of~$g\in G$.  If there is a
global section $s\:X\to P$ then $P$~is trivial;  the map
  $$ \aligned
     \tau _s\:X\times G&\longrightarrow P\\
     \langle x,g\rangle &\longmapsto s(x)\cdot g\endaligned $$
is a trivialization.  Any principal $G$	~bundle is locally trivial.

Suppose $h\:G\hookrightarrow G'$ is an inclusion of compact Lie groups and
$P\to X$ a principal $G$~bundle.  Then the induced $G'$~bundle $\PGp\to X$
is the quotient
  $$ \PGp = P\times _{G} G' = P\times G' \Bigm/ \langle p,g'\rangle \sim
     \langle p\cdot g,h(g\inv )g'\rangle ,\qquad p\in P,\quad g'\in G',\quad
     g\in G.  \tag{1.6}$$
There is a natural inclusion
  $$ \aligned
     h_P\:P&\longrightarrow \PGp\\
     p&\longmapsto \langle p,e\rangle \endaligned \tag{1.7}$$
which covers the identity map on~$X$.

A map of principal $G$~bundles $\varphi \:P'\to P$ is a smooth map of
manifolds which commutes with the $G$~action.  Hence it induces a map
$\phibar\:X'\to X$ on the quotient spaces.  When $X'=X$ and $\phibar$~is the
identity we term~$\varphi $ a {\it morphism\/} of principal bundles over~$X$.
If, in addition, $P'=P$ then $\varphi $~is termed an automorphism of~$P$, or
a {\it gauge transformation\/}.  In this case there is an associated map
  $$ g_{\varphi }\:P\longrightarrow G $$
defined by the equation
  $$ \varphi (p) = p\cdot g_{\varphi }(p). \tag{1.8}$$
Clearly maps of principal $G$~bundles compose, and the set of gauge
transformations of~$P$ forms a group~$\gauge{P}$.  If $h\:G\hookrightarrow
G_1$ is an inclusion, and $\varphi \:P'\to P$ a map, then there is an induced
map $P'_{G_1}\to P_{G_1}$ on the $G_1$~extensions.  Also, if $P\to X$ is a
principal $G$~bundle and $\phibar\:X'\to X$ a smooth map, then there is an
induced bundle $\phibar^*P\to X'$ and a bundle map $\varphi \:\phibar^*P\to
P$ covering~$\phibar$.

A {\it connection\/} on~$P\to X$ is a $\frak{g}$-valued 1-form~$\Theta $ such
that
  $$ \gather
     i_x^*\Theta =\theta _x, \tag{1.9}\\
     R_g^*\Theta =\Adgi\Theta , \tag{1.10}\endgather $$
where $i_x\:P_x\hookrightarrow P$ is the inclusion of the fiber over~$x\in X$
and $R_g\:P\to P$ is the right translation by~$g\in G$.  Connections can be
constructed locally using local trivializations.  Since \thetag{1.9}
and~\thetag{1.10} are convex conditions, local connections patch together via
a partition of unity into global connections.  Hence connections exist.  In
fact, \thetag{1.9} and~\thetag{1.10} are affine conditions, so the
set~$\conn P$ of all connections on~$P$ is an affine subspace of~$\Omega
^1_P(\frak{g})$, the vector space of $\frak{g}$-valued 1-forms on~$P$.  A
tangent vector to the space of connections is a $\frak{g}$-valued
1-form~$\Theta $ on~$P$ such that
  $$ \gather
     i_x^*\dot{\Theta} =0,  \tag{1.11}\\
     R_g^*\dot{\Theta} =\Adgi\dot{\Theta}.  \tag{1.12}\endgather $$
The vector space of forms satisfying~\thetag{1.11} and~\thetag{1.12} is
denoted~$\Omega ^1_X(\frak{g}_P)$, since $\dot{\Theta }$~is the lift to~$P$
of a 1-form on~$X$ transforming in the {\it adjoint bundle\/}
$\frak{g}_P=P\times _G\frak{g}$.  The {\it curvature\/}~$\Omega $ of a
connection~$\Theta $ is the $\frak{g}$-valued 2-form
  $$ \Omega  = d\Theta  + \frac{1}{2}[\Theta \wedge \Theta ]; \tag{1.13}$$
it satisfies
  $$ \gather
     R_g^*\Omega =\Adgi\Omega , \tag{1.14}\\
     i_x^*\Omega =0. \tag{1.15}\endgather $$
(This last equation follows from \thetag{1.5}.)  Hence the curvature is an
element of~$\Omega ^2_X(\frak{g}_P)$.  Differentiating~\thetag{1.13} we
obtain the {\it Bianchi identity\/}
  $$ d\Omega + [\Theta \wedge \Omega ] = 0. \tag{1.16}$$
Introduce the covariant derivative
  $$ d_{\Theta } = d + \ad(\Theta ). $$
Then the Bianchi identity can be rewritten as
  $$ d_{\Theta }\Omega =0. $$

Suppose $h\:G\to G'$ is an inclusion and $\Theta $~a $G$~connection on $P\to
X$.  Then there is an induced $G'$~connection~$\Theta _{G'}$ on $\PGp\to
X$.  Namely, define the $\frak{g}'$-valued 1-form
  $$ (\Theta _{G'})_{\langle p,g'\rangle } = \dot{h}\(\Ad_{(g')\inv
     }\Theta _p\) + (\theta ')_{g'} \tag{1.17}$$
on $P\times G'$, where $\theta '$~is the Maurer-Cartan form on~$G'$, and
$\dot{h}\:\frak{g}\hookrightarrow \frak{g}'$ the induced inclusion of Lie
algebras.  From~\thetag{1.4} and~\thetag{1.10} we see that
\thetag{1.17}~vanishes along the orbits of the right $G$~action on~$P\times
G'$ and is invariant under that action, so passes to a $\frak{g}'$-valued
1-form on $\PGp=P\times G'\bigm/G$ (cf\.~\thetag{1.6}).
Properties~\thetag{1.9} and~\thetag{1.10} are easily verified, so $\Theta
_{G'}$~is a connection.  The pullback of~$\Theta _{G'}$ under the natural
map $h_P\:P\to \PGp$ ~\thetag{1.7} can be identified with the original
connection~$\Theta $.

Let $\varphi \:P'\to P$ be a map of principal bundles and $\Theta $~a
connection on~$P$; then $\varphi ^*(\Theta )$ ~is a connection on~$P'$.  In
particular, the group~$\gauge P$ of gauge transformations acts (on the right)
on the space of connections~$\conn P$.  Suppose $\varphi \:P\to P$ is a gauge
transformation with associated map $g_{\varphi }\:P\to G$~\thetag{1.8}.  Let
$\phi _\varphi =g_{\varphi }^*(\theta )$ be the pullback of the Maurer-Cartan
form.  Then a basic equation in the theory of connections asserts
  $$ \varphi ^*(\Theta ) = \Ad_{g_\varphi \inv }\Theta +\phi _{\varphi }.
     \tag{1.18}$$
By contrast the curvature transforms as a tensor:
  $$ \varphi ^*(\Omega ) = \Ad_{g_\varphi \inv }\Omega . \tag{1.19}$$
An automorphism of a connection~$\Theta $ is a gauge transformation which
satisfies $\varphi ^*(\Theta )=\Theta $.  Equation~\thetag{1.18} is then a
first order differential equation for~$\varphi $, which asserts that
$g_{\varphi }$~is parallel.  Since parallel sections over connected manifolds
are determined by their value at a single point, we have proved the
following.

     \proclaim{\protag{1.20} {Proposition}}
 If $\varphi $~is an automorphism of~$\Theta $, then $g_{\varphi }$~is
parallel.  In particular, if $X$~is connected and $\varphi $~is the identity
at some point of~$X$, then $\varphi $~is the identity on~$X$.
     \endproclaim

We will often encounter connections over cylinders.  The following lemma
gives a standard form for such connections.

     \proclaim{\protag{1.21} {Lemma}}
 Suppose $Q\to Y$ is a $G$~bundle over a manifold~$Y$, and $[0,\infty )\times
Q\to \ray\times Y$ the pullback to the cylinder over~$Y$.  Let $\Theta $~be a
connection on $\ray\times Q$, which we write as
  $$ \Theta =\eta _t + \xi _t\,dt,\qquad t\in \ray,\quad \eta _t\in \Omega
     ^1_{Q}(\frak{g}),\quad \xi _t\in \Omega _Q^0(\frak{g}). $$
Then there exists a unique gauge transformation~$\varphi $ of $\ray\times Q$
which satisfies
  $$ \gather
     \varphi \res{\{0\}\times Q} = \id, \tag{1.22}\\
     \varphi ^*(\Theta ) = \tilde{\eta}_t. \tag{1.23}\endgather $$
     \endproclaim

\flushpar
 In other words, the transformed connection has no $dt$~component.\footnote{A
section~$s$ of $\ray\times Q$ with the property that $s^*\Theta $~has no
$dt$~component is called a {\it temporal gauge\/}; it exists if $Q$~is
trivializable.}

     \demo{Proof}
 Let $g_t\:Q\to G$ be the map associated to $\varphi \res{\{t\}\times Q}$,
and $\phi _t=g_t^*(\theta )$.  Then \thetag{1.22}~asserts $g_0=\id$, and
by~\thetag{1.18} we see that \thetag{1.23}~is equivalent to
  $$ \Ad_{g_t\inv }\xi _t + \phi _t(\frac{\partial }{\partial t}) = 0.
     $$
In more usual notation this reads
  $$ g\inv \xi g + g\inv \frac{\partial g}{\partial t} = 0, $$
which has a unique solution with initial condition $g_0=\id$ by the standard
theory of first order ordinary differential equations.
     \enddemo

We apply this lemma to glue connections over manifolds with boundary along
boundaries where the connections agree.

     \proclaim{\protag{1.24} {Proposition}}
 Suppose $P\to X$ is a $G$~bundle over an oriented manifold~$X$, and
$Y\hookrightarrow X$ is an oriented codimension one submanifold.  Let
$X\cut$~be the manifold obtained by cutting~$X$ along~$Y$.  There is a gluing
map $\bar{g}\:X\cut\to X$ which is a diffeomorphism off of~$Y$ and maps two
distinct submanifolds ~$Y_1,Y_2$ of~$\partial X\cut$ diffeomorphically
onto~$Y$.  Let $P\cut = \bar{g}^*P$ be the cut bundle and $g\:P\cut\to P$ the
gluing map.  Now suppose $\Theta \cut$~is a connection on~$P\cut$ such that
there exists a connection~$\eta $ on~$P\res Y$ with $g^*(\eta )= \Theta
\res{Y_1\sqcup Y_2}$.  Then $\eta $~extends to a connection~$\Theta $ on~$X$
such that $g^*(\Theta )$~is gauge equivalent to~$\Theta \cut$.  The gauge
equivalence class of~$\Theta $ is uniquely determined.
     \endproclaim

\flushpar
 The hypothesis asserts that $\Theta \cut\res{Y_1}$ and~$\Theta \cut\res{Y_2}$
agree under the given identification of the bundles.  Just as a smooth
function on~$X\cut$ whose restrictions to~$Y_1$ and~$Y_2$ agree does not
necessarily glue into a smooth function on~$X$, since the resulting function
may not be smooth in the transverse direction, so too the connection~$\Theta
\cut$ does not glue directly.  \theprotag{1.24} {Proposition} asserts that we
can glue smoothly if we make a gauge transformation.

     \demo{Proof}
 Choose tubular neighborhoods $N_i\cong \ray\times Y_i$ of~$Y_i$ in~$X\cut$
and bundle isomorphisms $P\res{N_i}\cong \ray\times P\res{Y_i} $.  Let
$\varphi _i$~be the gauge transformations over~$N_i$ guaranteed by
\theprotag{1.21} {Lemma}.  Let $\rho _1\:[0,1]\to[0,1]$ be a monotone
increasing smooth function with $\rho _1(t)=t$ for~$0\le t \le 0.5$ and~$\rho
_1([0.9,1])=1$.  Let $\rho _2\:[0,1]\to[0,1]$ be a monotone increasing smooth
function with $\rho _2([0,0.1])=0$ and $\rho _2([0.9,1])=1$.  Define
$\phitil_i$ by
  $$ \phitil_i(t,y) = \cases \varphi _i\bigl(\rho_1(t),y\bigr) ,&0\le t\le
     1;\\
     \varphi _i(2 - \rho_2\bigl(t),y\bigr),&1\le t\le 2;\\
     \id,&2\le t,\endcases $$
and extend $\phitil_i$ to a smooth gauge transformation ~$\phitil $
on~$X\cut$ which is the identity outside of~$N_i$.  Then $\phitil ^*(\Theta
\cut)$~glues to form a smooth connection~$\Theta $ on~$X$, since near~$Y_i$
the transverse component of~$\phitil ^*(\Theta \cut)$ vanishes.
     \enddemo

Let $\form \:\frak{g}\otimes \frak{g}\to\RR$ be an Ad-invariant symmetric
bilinear form on the Lie algebra~$\frak{g}$.  Since
  $$ \frak{g}\cong \frak{t}\oplus \frak{k}_1\oplus \dots \oplus \frak{k}_s
     \tag{1.25}$$
is a direct sum of an abelian Lie algebra~$\frak{t}$ with simple Lie
algebras~$\frak{k}_i$, from \thetag{1.1}--\thetag{1.3}, the invariant form
also decomposes as a direct sum.  Any invariant form on~$\frak{k}_i$ is a
multiple of the Killing form~$b_i$, and any symmetric form on~$\frak{t}$ is
invariant.  Thus the space of forms on~$\frak{g}$ is $S^2\frak{t}^*\oplus \RR
b_1\oplus \dots \oplus \RR b_s$.

Suppose $\Theta $~is a $G$ connection on $P\to X$ with curvature~$\Omega $.
Let $\langle \Omega \wedge \Omega \rangle $ be the {\it Chern-Weil\/} 4-form
associated with the bilinear form~$\form$.  It follows from~\thetag{1.14}
and~\thetag{1.15} that this form on~$P$ is the lift of a 4-form on~$X$, which
we also denote $\langle \Omega \wedge \Omega \rangle$.  Furthermore, a simple
computation using~\thetag{1.16} shows that this form is closed.  The
fundamental result of Chern-Weil theory is that the de Rham cohomology class
of this form is a certain characteristic class of~$P$.  The form~$\langle
\Omega \wedge \Omega \rangle $~is gauge invariant by~\thetag{1.19} and the
fact that $\form$~is Ad-invariant.

The {\it Chern-Simons form\/}~\cite{CS} is an antiderivative of~$\langle
\Omega \wedge \Omega \rangle$ on~$P$.  Set
  $$ \alpha  = \alpha (\Theta ) = \langle \Theta \wedge \Omega \rangle -
     \frac{1}{6}\langle \Theta \wedge [ \Theta \wedge \Theta ]\rangle .
     \tag{1.26}$$

     \proclaim{\protag{1.27} {Proposition}}
 The 3-form~$\alpha $ satisfies:\newline
 \rom(a\rom)\  $i_x^*\alpha = -\frac{1}{6}\langle \theta _x\wedge [\theta
_x\wedge
\theta _x]\rangle $;\newline
 \rom(b\rom)\ $d\alpha  = \langle \Omega \wedge \Omega \rangle $;\newline
 \rom(c\rom)\ $R_g^*\alpha =\alpha $;\newline
 \rom(d\rom)\ If $\varphi \:P'\to P$ is a bundle map and $\Theta $~a connection
on~$P$, then $\alpha (\varphi ^*\Theta ) = \varphi ^*\alpha (\Theta
)$.\newline
 \rom(e\rom)\ If $\varphi \:P\to P$ is a gauge transformation with associated
map
$g=g_{\varphi }\:P\to G$, and $\phi =\phi _{\varphi }=\varphi ^*(\theta )$,
then
  $$ \varphi ^*\alpha  = \alpha  + d\langle \Adgi\Theta \wedge \phi \rangle
     - \frac{1}{6}\langle \phi \wedge [\phi \wedge \phi ]\rangle .
     \tag{1.28}$$
     \endproclaim

     \demo{Proof}
 (a) and~(c) follow easily from \thetag{1.9}, \thetag{1.10}, \thetag{1.14},
\thetag{1.15}, and the Ad-invariance of~$\langle \cdot ,\cdot \rangle $.
(d)~is also trivial.  (b) ~and (e)~require some calculation; we
illustrate~(b) and leave~(e) to the reader:
  $$ \aligned
     d\alpha  &= \langle d\Theta \wedge \Omega \rangle  - \langle \Theta
     \wedge d\Omega \rangle  - \frac{1}{2}\langle d\Theta \wedge [\Theta
     \wedge \Theta ]\rangle  \\
     &= \langle (\Omega  - \frac{1}{2}[\Theta \wedge \Theta ])\wedge \Omega
     \rangle  + \langle \Theta \wedge [\Theta \wedge \Omega ]\rangle  -
     \frac{1}{2}\langle \Omega \wedge [ \Theta \wedge \Theta ]\rangle\\
     &= \langle \Omega \wedge \Omega \rangle .\endaligned $$
Here we use
  $$ \langle [\Theta \wedge \Theta ]\wedge [\Theta \wedge \Theta ]\rangle =
     \langle [[\Theta \wedge \Theta ] \wedge \Theta ] \wedge \Theta \rangle =
     0 \tag{1.29}$$
which follows from Ad-invariance and the Jacobi identity.
     \enddemo

It is useful for us to think about all connections at once.  In a technical
sense we shouldn't think of the {\it set\/} of connections since the set of
``all'' of anything leads to contradictions.  Also, we want to account for
the automorphisms of connections, since that is our basic symmetry.  Thus we
consider the {\it category\/} of all connections and morphisms of
connections.  As the word `category' is anathema for many mathematicians and
most physicists, but is a useful concept for us, we insert a few general
remarks.  The simplest algebraic structure on a set~$S$ is a binary
associative composition law $S\times S\to S$.  Such a structure is termed a
{\it monoid\/}.\footnote{If there is an identity element as well, $S$~is
termed a {\it semigroup\/}.  If, in addition, there are inverses, then $S$~is
a group, which is perhaps the most familiar algebraic type.  Categories are
usually defined as generalizations of semigroups, but we will not insist on
identity elements.} The simplest example is the set of natural numbers under
addition.  A {\it category\/}~$\Cal{C}$ is nothing more than a ``monoid with
states''.  Thus there is a collection~$\Obj(\Cal{C})$ of {\it objects\/}
(states) and for each pair of objects~$C_1,C_2$ there is a set of {\it
morphisms\/}~$\Mor(C_1,C_2) $.  A morphism is represented by an arrow ~$C_1
@>{\varphi }>> C_2$ from the initial state to a final state.  Two morphisms
$C_1 @>{\varphi }>> C_2$ and $C_1' @>{\varphi '}>> C_2'$ compose if and only
if $C_2 = C_1'$, and the composition law is assumed associative.  A monoid is
then a category with a single object.  Notice that we don't assume that
$\Obj(\Cal{C})$ is a set, but we do assume that $\Mor(C_1,C_2)$ is a set for
every pair of morphisms.

Let $X$~be a fixed manifold.  Define the category $\fld X = \fld X^G$ of
$G$~connections as follows.  An object in~$\fld X$ is a connection~$\Theta $
on a principal $G$~bundle $P\to X$.  A morphism $\Theta '@>{\varphi }>>
\Theta $ is a bundle map $\varphi \:P'\to P$ covering the identity map on~$X$
(i.e., a bundle morphism) such that $\Theta '=\varphi ^*\Theta $.  Notice
that the category~$\fld X$ contains identity maps $\Theta @>{\id}>> \Theta $
and that every morphism is invertible.  Such a category is termed a {\it
groupoid\/}.  The category~$\fld X$ carries a topology.\footnote{Here we
ignore whatever technical complications might arise from the fact that the
collection of all principal bundles is not a set.  I hope that readers who
are capable of identifying these difficulties are also capable of coping with
them.} The objects form a union of affine spaces
  $$ \Obj(\fld X) = \bigsqcup _{P} \conn P, \tag{1.30}$$
where $\{P\}$~is the collection of all principal $G$~bundles over~$X$.  The
set of morphisms between two connections is also a topological space in a
natural way.

There is an obvious equivalence relation on~$\fld X$---two connections
$\Theta ,\Theta '$ are equivalent if and only if there exists a morphism
$\Theta @>\varphi >>\Theta '$.  We denote the set of equivalence classes by
$\fldb X$.  Let $\{P_i\}$~be a set of representatives of topological
equivalence classes of $G$~bundles over~$X$.  Then there is a noncanonical
isomorphism
  $$ \fldb X \cong \bigsqcup  _{\{P_i\}} \conn{P_i}\bigm/ \gauge{P_i}
     $$
as a disjoint union.

Heuristically, a groupoid is a ``group with states''.  In gauge theory this
notion provides an illuminating formalism in which to keep track of the
symmetry.  If all $G$~bundles over ~$X$ are isomorphic (which occurs if
$G$~is connected and simply connected and $\dim X\le 3$, for example), then
we might fix $P\to X$ and consider the space of fields to be~$\conn P$ with
the group of gauge transformations~$\gauge P$ acting as symmetries.  This is
the usual picture.  Or more generally we might fix a set of representatives
for the different topological types.  However, this is not adequate.
Consider a situation in which we have two isomorphic bundles $P\to X$ and
$P'\to X'$ we wish to identify, as when we glue connections.  The glued
connection, perhaps even its equivalence class, depends on the particular
identification we choose.  As there is no canonical identification, we need
to keep track of {\it all\/} possible identifications.  This is an additional
symmetry and justifies our consideration of the category of bundles rather
than simply a set of representatives.

A {\it functor\/} $\Cal{F}\:\Cal{C}_1\to\Cal{C}_2$ generalizes the notion of
a homomorphism of monoids.  Namely, $\Cal{F}$~is a map
$\Obj(\Cal{C}_1)\to\Obj(\Cal{C}_2)$, and $\Cal{F}$~assigns to each morphism
$C_1 @>\varphi >> C_2$ a morphism $\Cal{F}(C_1) @>\Cal{F}(\varphi )>>
\Cal{F}(C_2)$ such that composition is preserved.  (If the categories have
identity maps, then these must also be preserved.)  Let $h\:G\hookrightarrow
G'$ be an inclusion of groups.  Then our previous remarks about extensions of
structure group are summarized by the statement that $h$~induces a functor
 $$ h_X\:\fld X^G\longrightarrow \fld X^{G'} \tag{1.31}$$
for any manifold~$X$.

\newpage
\heading
\S{2} The Chern-Simons Action
\endheading
\comment
lasteqno 2@ 30
\endcomment

In this section we integrate the Chern-Simons lagrangian~\thetag{1.26} over
spacetime (a compact, oriented 3-manifold) to construct the Chern-Simons
action.  Since the Chern-Simons form lives on the total space of a bundle,
and not on the base, we choose a section of the bundle to define the
action.\footnote{In this paper we restrict ourselves to trivializable bundles
(cf.~\theprotag{2.1} {Lemma}).  The extension to nontrivializable bundles is
carried out in Part~2~\cite{F2}.} On closed 3-manifolds the integral is
independent of the section, up to an integer, if we make the appropriate
integrality hypothesis on the bilinear form (\theprotag{2.5} {Hypothesis}).
If the 3-manifold has a boundary, then the integral depends only on the
restriction of the section over the boundary, and the dependence is encoded
in a cocycle~\thetag{2.15}.  The Wess-Zumino-Witten action~\thetag{2.13}
enters in the formula for this cocycle; we develop its properties in
Appendix~A.  The cocycle determines a hermitian complex line which only
depends on the restriction of the connection to the boundary.  We term this
line the ``Chern-Simons line''.  These lines vary smoothly, so form a line
bundle over smooth families of connections (\theprotag{2.17} {Proposition}).
This line bundle and cocycle were also considered in~\cite{RSW}
In \theprotag{2.19} {Theorem} we state carefully the properties of the
Chern-Simons action.  This can be viewed as an axiomatization of a local
(topological) classical Lagrangian field theory, in the spirit of Segal's
axioms for conformal field theory~\cite{S} and Atiyah's axioms for
topological quantum field theory~\cite{A}.  The crucial gluing law, which is
an assertion of the locality of the action, is~\thetag{2.27}.

{}From now on we fix a connected, simply connected, compact Lie group~$G$ and
an invariant form~$\form$ on its Lie algebra~$\frak{g}$.  We single out
simply connected groups because of the following topological fact.

     \proclaim{\protag{2.1} {Lemma}}
 If $G$~is simply connected, then any principal $G$~bundle over a
manifold of dimension~$\le 3$ admits a global section, hence is
trivializable.
     \endproclaim

\flushpar
 The proof uses $\pi _0G = \pi _1G = \pi _2G = 0$ and elementary
obstruction theory.  Suppose $\Theta $~is a connection on a principal
$G$~bundle $P\to X$, where $X$~is a closed, oriented 3-manifold.  Let
$p\:X\to P$ be a section.  Recall the Chern-Simons form $\alpha =\alpha
(\Theta )\in \Omega ^3_P$~\thetag{1.26}.  Define
  $$ \ac p \Theta = \int_{X}p^*\alpha (\Theta ). \tag{2.2}$$

     \proclaim{\protag{2.3} {Proposition}}
 Let $\varphi \:P\to P$ be a gauge transformation with associated map
$g_{\varphi }\:P\to G$.  Set $g=g_{\varphi }p\:X\to G$ and let $\phi
_g=g^*\theta $ be the pullback of the Maurer-Cartan form.  Then
  $$ \ac {\varphi p}\Theta = \ac p {\varphi ^*\Theta } = \ac p \Theta -
     \int_{X}\frac16 \langle \phi_g \wedge [\phi_g\wedge
     \phi_g]\rangle . \tag{2.4}$$
     \endproclaim

     \demo{Proof}
 The first equality is immediate from~\thetag{2.2}.  The second equality
follows from ~\thetag{1.28} and Stokes' theorem.
     \enddemo

\flushpar
 We now make the following integrality hypothesis on the bilinear form~$\form$.

     \proclaim{\protag{2.5} {Hypothesis}}
 Assume that the closed form~$-\frac16 \langle \theta \wedge [\theta \wedge
\theta ]\rangle $ represents an {\it integral\/} class in~$H^3(G;\RR)$.
     \endproclaim

\flushpar
 Then the integral in~\thetag{2.4} is an integer.  Since any two sections
of~$P$ are related by a gauge transformation, it follows that
  $$ \acT = \ac p \Theta \pmod1\quad \tag{2.6}$$
is independent of the section~$p$.  This is the Chern-Simons action on closed
manifolds.

We can also deduce the independence of~\thetag{2.6} on the section from the
fact that the Chern-Simons form is closed (cf.~\theprotag{1.27(b)}
{Proposition}).  Thus \thetag{2.2}~doesn't change under homotopies of~$p$.
Then the integrality of the restriction of the cohomology class of the
Chern-Simons form to the fiber gives the rest.

     \proclaim{\protag{2.7} {Proposition}}
 Let $X$ be a closed oriented 3-manifold.  The Chern-Simons action
  $$ S_X\:\fld X\longrightarrow \RR/\ZZ $$
is smooth\footnote{We will not bother with technicalities regarding the
smooth structure of infinite dimensional manifolds here; they are
well-understood.  Rather, we will usually verify smoothness for a family of
connections parametrized by a smooth manifold~$U$, which the reader can take
to be finite dimensional.} and satisfies:\newline
 \rom(a\rom)\ \rom({\it Functoriality\/}\rom)\ If $\varphi \:P'\to P$ is any
bundle map covering an orientation preserving diffeomorphism $\phibar\:X'\to
X$, and $\Theta $~is a connection on~$P$, then
  $$ S_{X'}(\varphi ^*\Theta ) = \acT. $$
 \rom(b\rom)\ \rom({\it Orientation\/}\rom)\ Let $-X$ denote~$X$ with the
opposite orientation.  Then
  $$ S_{-X}(\Theta ) = -\acT. $$
 \rom(c\rom)\ \rom({\it Additivity\/}\rom)\ If $X = X_1\sqcup X_2$ is a
disjoint union, and $\Theta _i$ are connections over~$X_i$, then
  $$ S_{X_1\sqcup X_2}(\Theta _1\sqcup \Theta _2) = S_{X_1}(\Theta
     _1) + S_{X_2}(\Theta _2). $$
     \endproclaim

\flushpar
 It follows from~(a) that there is an induced action
  $$ S_X\:\fldb X\longrightarrow \RZ $$
defined on the fields modulo symmetries.  We will often write the action
as~$\eacT$, in which case (c)~appears as a multiplicative property.

     \demo{Proof}
 The smoothness follows since $\alpha (\Theta )$~is a smooth function
of~$\Theta $~\thetag{1.26}.  For~(a), let $p'\:X'\to P'$ be a section and
$p=\varphi p'\phibar\inv $~the induced section of~$P$.  Then by
\theprotag{1.27(d)} {Proposition}
  $$ \ac p \Theta =\int_{X}(\phibar\inv)^* {p'}^*\varphi ^*\alpha
     (\Theta ) =
     \int_{X'}{p'}^*\alpha (\varphi ^*\Theta ) = S_{X'}(p',\varphi ^*\Theta ).
     $$
Assertions (b) and~(c) follow from standard properties of integration of
differential forms on oriented manifolds.
     \enddemo

The Chern-Simons action of a connection which extends over a 4-manifold can
be computed by integrating the Chern-Weil form.  Let $\tilde{\Theta}$~be a
connection on a principal $G$~bundle $\tilde{P}\to W$ over a compact oriented
4-manifold~$W$, and denote the curvature of~$\tilde{\Theta }$
by~$\tilde{\Omega }$.  Then the action of the restriction~$\partial
\tilde{\Theta }$ to the boundary is
  $$ S_{\partial W}(\partial \tilde{\Theta }) = \int_{W} \langle
     \tilde{\Omega }\wedge \tilde{\Omega }  \rangle \pmod1. \tag{2.8}$$
This follows from \theprotag{1.27(b)} {Proposition} and Stokes' theorem.

The Chern-Simons functional also behaves well under extension of structure
group.

     \proclaim{\protag{2.9} {Proposition}}
 Let $k\:G\hookrightarrow G'$ be an inclusion of connected, simply connected,
compact Lie groups, and $X$~a closed oriented 3-manifold.  Suppose
$\form'$~is an integral form on~$\frak{g}'$ which restricts to~$\form$
on~$\frak{g}$.  Then for any $G$ connection~$\Theta $ over~$X$, its
extension~$\Theta '$ to a $G'$~connection satisfies~$S_X(\Theta ')=\acT$.
     \endproclaim

\flushpar
 In fancier language, \theprotag{2.9} {Proposition} asserts that the diagram
  $$ \CD
     \fldb X(G_1) @>h_X>> \fldb X(G_1')\\
     @VS_XVV @VVS_XV\\
     \RZ @= \RZ\endCD   $$
commutes, where $h_X$~is the functor which extends $G$~connections to
$G'$~connections~\thetag{1.31}.

     \demo{Proof}
 Let $P\to X$ be the $G$~bundle carrying~$\Theta $ and $P'\to X$ its
$G'$~extension.  There is an inclusion $h_P\:P\hookrightarrow P'$.  Recall
from ~\thetag{1.17} that $h_P^*\Theta ' = \dot{h}(\Theta )$.  Since
$\dot{h}^*(\form') = \form$ it follows from~\thetag{1.26} that $h_P^*\alpha
(\Theta ') = \alpha (\Theta )$.  Fix a section $p\:X\to P$; then $h_P\circ
p$~is a section of~$P'$.  Now
  $$ \ac {h_P\circ p}{\Theta '} = \int_{X}p^*h_P^*\alpha (\Theta ') =
     \int_{X}p^*\alpha (\Theta ) = \ac p \Theta . $$
     \enddemo

Next, we consider compact oriented 3-manifolds~$X$ with nonempty boundary.
Suppose $\Theta $ ~is a $G$~connection on $P\to X$.  We retain the
definition~\thetag{2.2}.  But now \theprotag{2.3} {Proposition} is replaced
by

     \proclaim{\protag{2.10} {Proposition}}
 Let $\varphi \:P\to P$ be a gauge transformation with associated map
$g_{\varphi }\:P\to G$.  Set $g=g_{\varphi }p\:X\to G$ and let $\phi
_g=g^*\theta $ be the pullback of the Maurer-Cartan form.  Then
  $$ \ac {\varphi p}\Theta = \ac p {\varphi ^*\Theta } = \ac p \Theta
     + \int_{\partial X}\langle \Adgi p^*\Theta \wedge \phi _g\rangle -
     \int_{X}\frac16 \langle \phi_g \wedge [\phi_g\wedge
     \phi_g]\rangle .  \tag{2.11}$$
     \endproclaim

\flushpar
 It is no longer true that the last two terms in~\thetag{2.11} vanish modulo
integers, but rather

     \proclaim{\protag{2.12} {Lemma}}
 With \theprotag{2.5} {Hypothesis} the functional
  $$ W_{\bX}(g) =  \int_{X} -\frac16 \langle \phi_g \wedge [\phi_g\wedge
     \phi_g]\rangle \pmod1 \tag{2.13}$$
depends only on the restriction of $g\:X\to G$ to~$\partial X$.
     \endproclaim

\flushpar
 $W_{\bX}(g)$~is called the {\it Wess-Zumino-Witten functional\/}; it is the
action of a 1+1~dimensional field theory.  In Appendix~A we sketch the
development of this theory.

     \demo{Proof}
 If $g'\:X'\to G$ and there is an orientation reversing
diffeomorphism~$\partial X\cong \partial X'$ under which the restrictions
of~$g$ and~$g'$ coincide, then these maps patch together into a closed
oriented (singular) 3-{\it chain\/} $\tilde{g}\:X'-X\to G$.  Then
  $$ W_{\bX}(g') - W_{\bX}(g) =  \int_{X'-X}-\frac16 \langle \phi_g \wedge
     [\phi_g\wedge \phi_g]\rangle \, \in \ZZ \tag{2.14}$$
is an integer by~\theprotag{2.5} {Hypothesis}.
     \enddemo

\flushpar
 Therefore, the Chern-Simons action~\thetag{2.2} depends in a controlled
manner on the restriction of~$p$ to~$\partial X$.  We make sense of this by
defining a metrized complex line~$L_{\partial \Theta }$ attached to the
restriction~$\partial \Theta $ of~$\Theta $ to~$\partial X$.  The
Chern-Simons invariant of~$\Theta $ takes values in~$L_{\partial \Theta }$.
(Compare with the discussion at the beginning of this section.)  Notice that
we are only concerned with the elements in~$L_{\partial \Theta }$ of unit
norm, which form a principal homogeneous space for the circle group~$\TT$.
It is easy to go back and forth between the $\TT$-space and the metrized
line, which we do freely (usually without changing the notation).

We remark that we could also treat this dependence on boundary values by
examining homotopies of sections and using the fact that the Chern-Simons
form is closed; we leave this approach to the reader.

First, we abstract the construction of vector spaces (in this case lines) in
situations where one must make a choice (here the choice of a section on the
boundary).  We suppose that the set of possible choices and isomorphisms of
these choices form a groupoid~$\Cal{C}$.  Let $\Cal{L}$~be the category whose
objects are metrized complex lines and whose morphisms are unitary
isomorphisms.\footnote{Alternatively we can think of $\Cal{L}$~as the
category whose objects are principal homogeneous $\TT$-spaces, where $\TT$~is
the circle group.  The elements of unit norm in a metrized line form a
principal homogeneous $\TT$-space.  The tensor product of complex lines is
then a product on the category of $\TT$-spaces.} Suppose we have a functor
$\Cal{F}\:\Cal{C}\to\Cal{L}$.  Define $V_{\Cal{F}}$ to be the inner product
space of invariant sections of the functor~$\Cal{F}$: An element $v\in
V_{\Cal{F}}$ is a collection $\{v(C)\in \Cal{F}(C)\}_{C\in \Obj(\Cal{C})}$
such that if $C_1@>\psi >>C_2$ is a morphism, then $\Cal{F}(\psi )v(C_1) =
v(C_2)$.  Suppose $\Cal{C}$~is {\it connected\/}, that is, there exists a
morphism between any two objects.  Then $\dim V_{\Cal{F}}=0$ or $\dim
V_{\Cal{F}}=1$, the latter occurring if and only if $\Cal{F}$~{\it has no
holonomy\/}, i.e., $\Cal{F}(\psi )=\id$ for every automorphism $C@>\psi >>
C$.  The terminology comes from the example of a metrized line bundle $L\to
M$ with unitary connection over a manifold~$M$.  Form a category~$\Cal{C}(M)$
whose objects are the points of~$M$ and whose morphisms are piecewise smooth
unparametrized paths in~$M$.  The category~$\Cal{C}(M)$ is connected if and
only if $M$~is connected.  The functor~$\Cal{F}$ assigns the fiber~$L_m$
to~$m$ and parallel transport to paths.  The inner product
space~$V_{\Cal{F}}$ is then the space of flat sections.  The functor has no
holonomy if and only if the connection has no holonomy.

Fix a $G$~connection $\eta $ on $Q\to Y$, where $Y$~is an oriented closed
2-manifold.  By \theprotag{2.1} {Lemma} the bundle~$Q$ is trivializable.  Let
$\Cal{C}_Q$~be the category whose objects are sections $q\:Y\to Q$.  For any
two sections~$q,q'$ there is a unique morphism $q @>\psi >> q'$, where $\psi
\:Q\to Q$ is the gauge transformation such that~$q'=\psi q$.  Define the
functor $\Cal{F}_{\eta}\:\Cal{C}_Q\to\Cal{L}$ by $\Cal{F}_{\eta}(q)=\CC$ for
all~$q$, where $\CC$~has its standard metric, and $\Cal{F}_{\eta}(q @>\psi >>
q')$ is multiplication by~$c_Y(q^*\eta ,g_{\psi }q)$, where $g_{\psi }\:Q\to
G$ is the map associated to~$\psi $, and $c_Y$~is the cocycle
y  $$ c_Y(a,g) = \exp\bigl( \tpi\int_{Y}\langle \Adgi a\wedge \phi _g\rangle +
     W_Y(g)\bigr),\qquad
     a\in \Omega ^1_Y(\frak{g}), \quad  g\:Y\to G. \tag{2.15}$$
That $\Cal{F}_{\eta}$~is a functor follows from the cocycle identity
  $$ c_Y(a,g_1g_2) = c_Y(a,g_1) c_Y(a^{g_1},g_2), \tag{2.16}$$
where
  $$ a^g = \Adgi a + \phi _g. $$
We leave the verification of~\thetag{2.16} to the reader, who may wish to
consult~\thetag{A.10} and~\thetag{A.8}.  Since $\Cal{C}_Q$~is connected and
$\Cal{F}_{\eta}$ ~has no holonomy (there are no nontrivial automorphisms), we
obtain the desired metrized line~$L_\eta = L_{Y,\eta} $ of invariant
sections.  From the construction we see that a section $q\:Y\to Q$ induces a
trivialization $L_\eta \cong \CC$.

It is important to observe that these ``Chern-Simons lines'' vary smoothly in
smooth families of connections.

     \proclaim{\protag{2.17} {Proposition}}
 Suppose $\eta _u$~is a smooth family of connections on $Q\to Y$ for
$u$~varying over a smooth manifold~$U$.  Then the lines~$L_{\eta _u}$ form a
smooth hermitian line bundle over~$U$.
     \endproclaim

\flushpar
 The line bundle is trivialized by a section $q\:Y\to Q$.  Transition
functions between different trivializations are smooth since the map
  $$ u\longmapsto c_Y(q^*\eta _u,g_{\psi }q) $$
is smooth, as formula~\thetag{2.15} clearly shows.  Notice that the elements
of unit norm form a circle bundle over~$U$.

Return now to the connection~$\Theta $ over the 3-manifold~$X$.
Equation~\thetag{2.11} shows that
  $$ \partial p\longmapsto e^{ 2\pi i\ac p \Theta } $$
defines an invariant section of unit norm
  $$ e^{ 2\pi i\acT}\in L_{\partial \Theta }, \tag{2.18}$$
where $\partial \Theta $~is the restriction of~$\Theta $ to~$\partial X$ and
$\partial p$~is the restriction of the section~$p$ to~$\partial X$.  This is
the Chern-Simons invariant for manifolds with boundary.  In good Bourbaki
style we allow either~$X$ or~$\partial X$ to be the empty set~$\emptyset $
in~\thetag{2.18}: Set $L_{\emptyset }=\CC$ and $S_{\emptyset }\equiv 0$.  The
following theorem generalizes \theprotag{2.7} {Proposition} and expresses
what we mean by the statement ``$S_X$~is the action of a local Lagrangian
field theory''.

     \proclaim{\protag{2.19} {Theorem}}
 Let $G$~be a connected, simply connected compact Lie group and
$\form\:\frak{g}\otimes \frak{g}\to\RR$~an invariant form on its Lie
algebra~$\frak{g}$ which satisfies the integrality condition \theprotag{2.5}
{Hypothesis}.  Then the assignments
  $$ \alignedat 2
     \eta &\longmapsto L_\eta ,&&\qquad \eta \in \fld Y,\\
     \Theta &\longmapsto \eacT,&&\qquad \Theta \in \fld X,\endalignedat
     \tag{2.20}$$
defined above for closed oriented 2-manifolds~$Y$ and compact oriented
3-manifolds~$X$ are smooth and satisfy:\newline
 \rom(a\rom)\ \rom({\it Functoriality\/}\rom)\ If $\psi \:Q'\to Q$ is a bundle
map covering an orientation preserving diffeomorphism $\overline{\psi}\:Y'\to
Y$, and $\eta $~is a connection on~$Q$, then there is an induced isometry
  $$ \psi ^*\:L_\eta \longrightarrow L_{\psi ^*\eta } \tag{2.21}$$
and these compose properly.  If $\varphi \:P'\to P$ is a bundle map covering
an orientation preserving diffeomorphism $\phibar\:X'\to X$, and
$\Theta $~is a connection on~$P$, then
  $$ (\partial \varphi )^*\(\eacT\) = \eac {X'}{\varphi ^*\Theta },
     \tag{2.22}$$
where $\partial \varphi \:\partial P'\to\partial P$ is the induced map over
the boundary.\newline
 \rom(b\rom)\ \rom({\it Orientation\/}\rom)\ There is a natural isometry
  $$ L_{-Y,\eta } \cong \overline{L_{Y,\eta }}, \tag{2.23}$$
and
  $$ \eac{-X}\Theta = \overline{\eacT}. \tag{2.24}$$
 \rom(c\rom)\ \rom({\it Additivity\/}\rom)\ If $Y=Y_1\sqcup Y_2$ is a disjoint
union, and $\eta _i$~are connections over~$Y_i$, then there is a natural
isometry
  $$ L_{\eta _1\sqcup \eta _2} \cong L_{\eta _1}\otimes L_{\eta _2}.
     \tag{2.25}$$
If $X=X_1\sqcup X_2$ is a disjoint union, and $\Theta _i$~are connections
over~$X_i$, then
  $$ \eac{X_1\sqcup X_2}{\Theta _1\sqcup \Theta _2} = \eac{X_1}{\Theta _1}
     \otimes  \eac{X_2}{\Theta _2}. \tag{2.26}$$
 \rom(d\rom)\ \rom({\it Gluing\/}\rom)\ Suppose $Y\hookrightarrow X$ is a
closed, oriented submanifold and $X\cut$~is the manifold obtained by
cutting~$X$ along~$Y$.  Then $\partial X\cut = \partial X\sqcup Y \sqcup -Y$.
Suppose $\Theta $~is a connection over~$X$, with $\Theta \cut$~the induced
connection over~$X\cut$, and $\eta $~the restriction of~$\Theta $ to~$Y$.
Then
  $$ \eacT = \Tr_\eta \( \eac{X\cut}{\Theta \cut}\), \tag{2.27}$$
where $\Tr_\eta $ is the contraction
  $$ \Tr_\eta \:L_{\partial \Theta \cut} \cong  L_{\partial \Theta }\otimes
     L_\eta  \otimes \overline{L_\eta}\longrightarrow L_{\partial \Theta }
     $$
using the hermitian metric on~$L_\eta $.
     \endproclaim

Several comments are in order.  From a functorial point of view: (a)~implies
that $\eta \mapsto L_\eta $ defines a functor
 $$\fld Y\to\Cal{L} \tag{2.28}$$
and that each~$X$ determines an invariant section~$\eac X\cdot $ of the
composite functor $\fld X\to\fld{\partial X}\to\Cal{L}$, where the first
arrow is restriction to the boundary.  From down here on earth: To each
bundle $Q\to Y$ over a closed, oriented 2-manifold there is associated a
smooth line bundle $L_Q\to\conn Q$ over the space of connections, and the
action of gauge transformations $\gauge Q$ on~$\conn Q$ lifts to~$L_Q$.
Furthermore, a bundle $P\to X$ over a compact, oriented 3-manifold determines
a restriction map $\conn P\to\conn{\partial P}$ and so an induced line bundle
$L_P\to\conn P$; the action of gauge transformations~$\gauge P$ on~$\conn P$
lifts to~$L_P$.  Then $\eac X\cdot $~is an invariant section of~$L_P$.  These
line bundles and sections are smooth, by \theprotag{2.17} {Proposition}.
Although (c)~looks like a multiplicative property, it expresses the
additivity of the classical action~$S_X$.  However, $S_X$~is not defined if
$\partial X\not= \emptyset $, which is why we use the exponential
notation~$e^{2\pi iS_X(\cdot )}$.  \theprotag{2.19} {Theorem} expresses in a
(necessarily) complicated way the fact that $S_X$~is a local functional of
local fields defined as the integral of a local expression (c), ~(d); is
invariant under symmetries of the fields~(a); and is unitary~(b).

     \demo{Proof}
 The smoothness of $\eta \mapsto L_{\eta }$ is \theprotag{2.17} {Proposition}.
Suppose $\Theta _u$~is a smooth family of connections parametrized by a
smooth manifold~$U$.  Fix a section $p\:X\to P$; its restriction to~$\bX$
determines a smooth trivialization of the lines~$L_{\partial \Theta _u}$.
Relative to this trivialization $e^{2\pi iS_X(\Theta _u)}$ is the function
$e^{2\pi iS_X(p,\Theta _u)}$, which is smooth in~$u$ since $u\mapsto \alpha
(\Theta _u)$ is smooth.  For ~(a), fix a section $q'\:Y'\to Q'$ and so an
isometry $L_{\psi ^*\eta }\cong \CC$.  The induced section $q=\psi
q'\bar{\psi} \inv \:Y\to Q$ induces an isometry~$L_\eta \cong \CC$.  Relative
to these trivializations we define $\psi ^*\:L_\eta \to L_{\psi ^*\eta }$ to
be the identity map $\CC\to\CC$.  A routine check shows that this definition
is independent of the choice of section~$q'$.\footnote{For this we need the
functoriality of the Wess-Zumino-Witten action, which is proved
by~\thetag{2.14}.} Equation~\thetag{2.22} is proved similarly using a
trivialization of~$P'$.  For~\thetag{2.23} we observe that the
cocycle~\thetag{2.15} changes sign when the orientation of~$Y$ is reversed.
Similarly, the integral in~\thetag{2.2} changes sign if the orientation
of~$X$ is reversed, from which \thetag{2.24}~follows (after fixing a
trivialization).  The statements in~(c) follow from the fact that the
integral over a disjoint union is the sum of the integrals, and in~(d) from
the fact that $\int_{X\cut}=\int_{X}$.  We leave details to the reader.
     \enddemo

\theprotag{2.9} {Proposition} extends to manifolds with boundary.

     \proclaim{\protag{2.29} {Proposition}}
 Let $k\:G\hookrightarrow G'$ be an inclusion of connected, simply
connected, compact Lie groups.  Suppose $\form'$~is an integral form
on~$\frak{g}'$ which restricts to~$\form$ on~$\frak{g}$.  Then if $\eta
$~is a $G$~connection over a closed, oriented 2-manifold~$Y$, and $\eta
'$~its $G'$~extension, there is a natural isometry
  $$ k_\eta \:L_\eta \longrightarrow L_{\eta '}. \tag{2.30}$$
If $\Theta $~is a $G$~connection over a compact, oriented 3-manifold~$X$,
and $\Theta '$ ~its $G'$~extension, then
  $$ k_{\partial \Theta }\(\eacT \) = \eac X{\Theta '}. $$
     \endproclaim

\flushpar
 In categorical terms, $k$~induces a natural transformation from the functor
$\fld Y^{G}\to\Cal{L}$ to the functor $\fldg Y{G}@>k_Y>> \fldg Y{G'}
@>>> \Cal{L}$ (cf\.~\thetag{2.28}, \thetag{1.31}).  For each~$X$ this induces
a natural transformation from $\fldg X{G}\to \fldg{\partial X}{G} \to
\Cal{L}$ to $\fldg X{G}\to\fldg X{G'}\to\fldg{\partial X}{G'}\to
\Cal{L}$ which preserves the invariant sections $e^{2\pi iS_X(\cdot )}$.
Though the language is complicated, the proof is straightforward, and we omit
it.

\newpage
\heading
\S{3} Classical Solutions and the Hamiltonian Theory
\endheading
\comment
lasteqno 3@ 28
\endcomment

The {\it classical solutions\/} in a Lagrangian field theory are the critical
points of the action, which are computed by a standard calculus of variations
argument.  In the Chern-Simons theory these are the flat connections (up to
equivalence).  A standard result in differential geometry (\theprotag{3.5}
{Proposition}) shows that this space is determined by the fundamental group.
The Hamiltonian theory pertains to spacetimes which are globally products of
a space (here a closed, oriented 2-manifold) with time (the positive reals,
say).  In the Chern-Simons theory the classical solutions are constant in
time, i.e., the hamiltonian function vanishes (\theprotag{3.16}
{Proposition}).  So the space of classical solutions is the moduli space of
flat connections on a bundle over a 2-manifold, which has been the object of
much study recently.  The Chern-Simons lines defined in~\S{2} descend to form
a hermitian line bundle over this moduli space.  Furthermore, the
Chern-Simons action defines a connection on this circle bundle whose
curvature is a symplectic form (\theprotag{3.17} {Proposition}).  This is the
phase space of the Hamiltonian theory.  We also prove some simple facts about
the action of the stabilizer of a connection on the Chern-Simons line
(\theprotag{3.25} {Proposition} and \theprotag{3.26} {Proposition}).

Our first order of business is to calculate the differential of the
action~$S_X$ on a closed oriented 3-manifold.  Recall that the configuration
space of fields is a disjoint union of affine spaces~\thetag{1.30}.

     \proclaim{\protag{3.1} {Proposition}}
 Let $\Theta _t$ be a path in~$\fld X$.  Denote $\Theta =\Theta _0$
and~$\dot{\Theta }=\dtz \Theta _t$.  Then
  $$ \dtz S_X(\Theta _t) = 2\int_{X}\langle \Omega \wedge \dot{\Theta
     }\rangle , \tag{3.2}$$
where $\Omega $ is the curvature of~$\Theta $.
     \endproclaim

\flushpar
 The reader should check that the integrand, which is ostensibly a 3-form
on~$P$, is the lift of a 3-form on~$X$.

     \demo{Proof}
 The connections $\Theta _s$ form a single connection~$\Tb$ on each cylinder
$[0,t]\times X$ with curvature $\Omega _s + ds\wedge \dot{\Theta }_s$, where
$\Omega _s$ is the curvature of~$\Theta _s$.  Then \thetag{2.8}~implies that
for each~$t$,
  $$ S(\Theta _t) - S(\Theta _0) = 2\int_{0}^tdt\int_{X}\langle \Omega
     _s\wedge \dot{\Theta}_s  \rangle. \tag{3.3}$$
The proposition follows by differentiating~\thetag{3.3} at~$t=0$.
     \enddemo

It follows that $dS_X(\Theta )=0$ if and only if~$\Omega =0$, i.e., if and
only if $\Theta $~is flat.  Notice that the {\it Euler-Lagrange equation\/}
  $$ \Omega =0 \tag{3.4}$$
is first order.\footnote{A typical field theory has an action of the form
$S_X(f) = \int_{X}|df|^2$, and then the Euler-Lagrange equation is second
order.}  Since the action is invariant under symmetries, by
\theprotag{2.7(a)} {Proposition}, so is the space of solutions to the
Euler-Lagrange equation.  In this case that fact is also obvious directly.
Let
  $$ \mflat X\subset \fldb X $$
be the space of equivalence classes of solutions to~\thetag{3.4}.

Since \thetag{3.4}~is a local condition and is defined in any dimension, the
space ~$\mflat X$ of equivalence classes of flat connections is defined for
any manifold~$X$.  Nevertheless, we derive these spaces directly from the
Chern-Simons action for certain 2-~and 3-manifolds, as this is the procedure
which generalizes to arbitrary field theories.  First, we note some general
facts about flat connections, which in the context of Chern-Simons theory
reflect the topological nature of the action.

     \proclaim{\protag{3.5} {Proposition}}
 Let $X$~be any smooth manifold.  Choose a basepoint~$x_i$ in each component
of~$X$.  Then the holonomy provides a natural identification
  $$ \mflat X = \prod_i \Hom\bigl( \pi _1(X,x_i),G\bigr)/G \tag{3.6}$$
which is independent of the basepoints.
     \endproclaim

\flushpar
 Here $G$~acts on $\Hom\bigl( \pi _1(X,x_i),G\bigr)/G$ by conjugation.  We
omit the proof, which is standard

The space $\mflat X$ is typically not a manifold.  If $Y$~is a compact
oriented 2-manifold, then $\mflat Y$~is a stratified space whose stratum of
top dimension is a smooth manifold~\cite{G}.  Suppose $\Theta $~is a flat
connection on $P\to X$ (in any dimension), and consider the covariant
derivative~$d_\Theta $ in the adjoint representation.  It defines a complex
  $$ 0 @>>> \Omega ^0_X(\frak{g}_P) @>{d_\Theta}>> \Omega ^1_X(\frak{g}_p)
     @>>> \cdots \tag{3.7}$$
since $d_\Theta ^2=0$.  We denote the cohomology groups of this complex by
$H^\bullet\bigl(X;\frak{g}(\Theta )\bigr)$.  If $\Theta $~represents a smooth
point of~$\mflat X$, then the tangent space at that point is
  $$ T_\Theta \mflat X\cong H^1\bigl( X; \frak{g}(\Theta )\bigr).
     \tag{3.8}$$

If $Y$~is a compact oriented 2-manifold, then for a flat connection~$\eta $
there is a nondegenerate pairing
  $$ H^0\bigl(Y;\frak{g}(\eta )\bigr) \otimes H^2\bigl(Y;\frak{g}(\eta
     )\bigr) \longrightarrow \RR. $$
The zeroth cohomology $H^0\bigl(Y;\frak{g}(\eta )\bigr)$ is the Lie algebra
of the centralizer~$Z_\eta $ of~$\eta $ in the group of gauge
transformations.  For example, it vanishes if $\eta $~is irreducible, since
then $Z_\eta $~is isomorphic to the center of~$G$, which is finite.  In any
case the index theorem gives
  $$ \dim \mflat Y=\dim H^1\bigl(Y;\frak{g}(\eta )\bigr)= - \dim G\cdot \chi
     (Y) + 2\,\dim Z_\eta . \tag{3.9}$$
Of course, this is a formula for the dimension of the moduli space only at
smooth points~$\eta $; at other points this is a formal dimension.  At smooth
points~$\eta $ the right hand side gives a formula for~$\dim \mflat Y$.  Note
in particular that $\mflat Y$~has even dimension.  This is explained by the
theorem of Narasimhan and Seshadri~\cite{NS}, which identifies $\mflat Y$~as
a complex manifold (of {\it stable bundles\/}) when $Y$~is endowed with a
complex structure.

The following lemma also reflects the topological nature of flat connections.

     \proclaim{\protag{3.10} {Lemma}}
 Suppose $\Theta $~is a flat connection on $P\to X$, for some manifold~$X$,
and $\varphi _i\:P'\to P,\; i=0,1$ are bundle maps.  Then if
$\phibar_0\:X'\to X$~is (pseudo)isotopic to~$\phibar_1\:X'\to X$, the
connections~$\varphi _0^*\Theta $ and~$\varphi _1^*\Theta $ are gauge
equivalent.
     \endproclaim

     \demo{Proof}
 This follows directly from \theprotag{3.5} {Proposition}, since
$\phibar_0$~ and $\phibar_1$~ induce the same map on~$\pi _1$.
Alternatively, let $\varphi \:\zo\times P'\to \zo\times P$ be a bundle map
covering a pseudoisotopy $\phibar\:\zo\times X'\to \zo\times X$.  Then by
\theprotag{1.21} {Lemma} there is a unique gauge transformation
$\phitil\:\zo\times P'\to \zo\times P'$ such that $\phitil_0=\id$ and
$\phitil^*\varphi ^*\Theta =\eta _t$ has no $dt$~component, where $t$~is the
coordinate on~$\zo$.  The curvature of~$\phitil^*\varphi ^*\Theta$ is
  $$ \Omega _t - \dot{\eta }_t\wedge dt, \tag{3.11}$$
where $\Omega _t$~is the curvature of~$\eta _t$.  But \thetag{3.11}~
vanishes, since $\Theta $~is flat.  Hence $\dot{\eta }_t=0$, from which~$\eta
_0 = \eta _1$.  In other words $\varphi ^*_0\Theta = \eta _0 = \eta _1 =
\phitil_1^*(\varphi _1^*\Theta )$, which proves the lemma.
     \enddemo

Now consider a compact oriented 3-manifold~$X$, possibly with nonempty
boundary.  For $\eta \in \fld{\partial X}$ we define
  $$ \fld X(\eta ) = \{\,\Theta \in \fld X : \partial \Theta =\eta \,\}
     $$
as the category of fields with boundary value~$\eta $.  Morphisms in~$\fld
X(\eta )$ are required to be the identity over~$\bX$.

     \proclaim{\protag{3.12} {Proposition}}
 Let $\Theta _t$ be a path in~$\fld X(\eta )$.  Then
  $$ \dtz S_X(\Theta _t) = 2\int_{X}\langle \Omega \wedge \dot{\Theta
     }\rangle . $$
     \endproclaim

\flushpar
 We omit the derivation, which is similar to the derivation of ~\thetag{3.2},
except for two points: (i)\ the cylinder $[0,t]\times X$ needs to be smoothed
at the corners; and (ii)\ the action on~$[0,t]\times \partial X$ is zero
since the connection is constant there.  The critical points form the moduli
space of flat connections $\mflat X(\eta )\subset \fldb X(\eta )$ which
restrict to~$\eta $ on the boundary.  This space is nonempty only if $\eta
$~is flat.

We next form the union~$\mflat X$ of the~$\mflat X(\eta )$.\footnote{Of
course, $\mflat X$ is simply the moduli space of flat connections on~$X$.
The algebraic exercise in this paragraph is our understanding of the role
that symmetries play in the classical theory on manifolds with boundary.
Similar considerations also enter in the quantum theory.} For this we need to
divide out the symmetries on the boundary.  Let $\eta '@>\psi >> \eta$ be a
morphism in~$\fld{\bX}$.  We claim that there is an induced functor $\fldb
X(\eta ')\to\fldb X(\eta )$ which maps~$\mflat X(\eta ')$ isomorphically
onto~$\mflat X(\eta )$.  For suppose $\Theta $~is a connection on $P\to X$
with~$\partial \Theta =\eta $.  Choose a bundle $P'\to X$ and a morphism
$\varphi \:P'\to P$ such that $\partial \varphi =\psi $.  Then the induced
functor maps the equivalence class of~$\varphi ^*\Theta $ (in~$\fldb X(\eta '
)$) to the equivalence class of~$\Theta $.  A routine check shows
that this is independent of the choice of~$\varphi $.  The functor is defined
in a similar manner on morphisms in~$\fldb X(\eta ')$.  Therefore, there is a
functor $\eta \mapsto \mflat X(\eta )$ from~$\fld{\bX}$ to the category of
(singular) manifolds.  Then $\mflat X$~is the space of invariant sections of
this functor.  We repeat that $\mflat X$~is simply the moduli space of flat
connections on ~$X$, so is given by~\thetag{3.6}; its topology and manifold
structure should be understood from that point of view.  Finally, the
restriction to~$\bX$ gives a diagram
  $$ \CD
     \mflat X @>>> \fldb X\\
     @VVV @VVV\\
     \mflat{\bX} @>>> \fldb{\bX}\endCD \tag{3.13}$$
If $\eta $~is a flat connection over~$\bX$, then the fiber of~\thetag{3.13}
over the equivalence class of~$\eta $ in~$\mflat{\bX}$ is isomorphic
to~$\mflat X(\eta )$.

If $Y$~is a closed oriented 2-manifold, then $\mflat Y\subset \fld Y$ is the
moduli space of flat connections.  Again we wish to ``derive'' this space
from the Chern-Simons action.  Thus we consider the Hamiltonian formulation
of the theory, i.e., study the action on the cylinder $X=\zi\times Y$.  This
``spacetime''~$X$ is the product of ``space''~$Y$ and ``time''~$\zi$. In the
Hamiltonian formulation we rewrite the space of fields as a space of paths.

     \proclaim{\protag{3.14} {Proposition}}
 Let $\{Q\}$~be a set of representatives for the equivalence classes of $G$~
bundles over~$Y$.  Then there is an identification
  $$ \fldb{\zi\times Y} = \bigsqcup_{\{Q\}} \Map \bigl( \zi,\conn Q\bigr)
     \bigm/ \gauge Q. \tag{3.15}$$
     \endproclaim

     \demo{Proof}
 If $\Theta \in \fld{\zi\times Y}$ is a connection on $P\to \zi\times Y$,
then $\partial P\cong Q$ for a unique~$Q\in \{Q\}$.  Since $\zi$~is
contractible there is an isomorphism $P\cong \zi\times Q$, and by
\theprotag{1.21} {Lemma} we can choose an isomorphism which takes~$\Theta$ to
a path~$\eta _t$ of connections on~$Q$.  As this is unique up to an overall
gauge transformation on~$Q$, the assertion follows.
     \enddemo

The classical solutions on~$\zi\times Y$ are completely determined by their
initial value.\footnote{Contrast with the usual case of a second order
Euler-Lagrange equation, where both an initial value and an initial
derivative must be specified.}

     \proclaim{\protag{3.16} {Proposition}}
 The restriction to the boundary $\mflat {\zi\times Y} \subset \fldb
{\zi\times Y}\to \fldb{\{0\}\times Y}$ is an isomorphism of~$\mflat{\zi\times
Y}$ onto the moduli space~$\mflat Y$ of flat connections over~$Y$.
     \endproclaim

     \demo{Proof}
 Suppose $\Theta$~is a flat connection over~$\zi\times Y$.  After a gauge
transformation this is a path~$\eta _t$ in~$\fld Y$.  But the proof of
\theprotag{3.10} {Lemma} shows that $\eta _t$~is constant.  Thus $\eta
_0$~determines the gauge equivalence class of~$\Theta $.
     \enddemo

One fundamental idea in the Hamiltonian formulation of classical mechanics,
is that the fields over a cylinder are paths in a {\it symplectic
manifold\/}.  \footnote{Many physicists who subscribe to the ``symplectic
philosophy'' emphasize that in a local field theory the symplectic structure
is derived from the lagrangian.  That is our approach.  (See~\cite{GS} for a
survey of symplectic geometry in physics.)} Furthermore, the space of
classical solutions also carries a symplectic structure. For Chern-Simons
theory the identification~\thetag{3.15} indicates that the purported
symplectic structure on each~$\conn Q$ is $\gauge Q$~invariant.  These
assertions form the next proposition.  Recall that the lines~$L_\eta $
attached to~$\eta \in \fld Y$ vary smoothly with~$\eta $.  So for each~$Q$
they form a smooth hermitian line bundle $L_Q\to\conn Q$.

     \proclaim{\protag{3.17} {Proposition}}
 Fix a $G$~bundle $Q\to Y$ over a closed oriented 2-manifold.  Then the
Chern-Simons action defines a unitary connection on the hermitian line bundle
$L_Q\to \conn Q$.  The curvature of this connection times~$i/2\pi $ is
  $$ \omega (\dot{\eta}_1 ,\dot{\eta}_2 )= - 2\int_{Y}\langle \dot{\eta}_1
     \wedge \dot{\eta}_2 \rangle ,\qquad \dot{\eta}_1 ,\dot{\eta}_2 \in
     T\conn Q. \tag{3.18}$$
If $\form$~is nondegenerate, then $\omega $~ is a symplectic form.  The
action of~$\gauge Q$ on~$\conn Q$ lifts to~$L_Q$, and the lifted action
preserves the metric and connection.  The induced moment map is
  $$ \mu _\xi (\eta ) = 2\int_{Y}\langle \Omega (\eta )\wedge \xi \rangle,
     \tag{3.19}$$
where $\xi \in \Omega ^0_Y(\frak{g}_Q)$ is an infinitesimal gauge
transformation.  Over subsets of~$\conn Q$ where $\gauge Q$~acts freely, then
(a subset of) the moduli space~$\mflat Q$ of flat connections is the
symplectic quotient~$\conn Q//\gauge Q$.  There is an induced line bundle
$L_Q\to\mflat Q$ with metric and connection, and $i/2\pi $~times its
curvature is the induced symplectic form on~$\mflat Q$.
     \endproclaim

\flushpar
 That a nondegeneracy hypothesis on the lagrangian is necessary to obtain a
nondegenerate symplectic structure is standard in classical
mechanics~\cite{Ar}.  The symplectic structure of ~$\conn Q$ is
well-known~\cite{AB}, and the symplectic form on~$\mflat Q$ can be defined
using cohomology~\cite{G}.  Here, though, we see both structures arising
directly from the Chern-Simons action in 3~dimensions.

     \demo{Proof}
 Let $\eta _t$ ~be a smooth path in~$\conn Q$.  Then $\et$~determines a
connection~$\eb$ on $\zo\times Q\to \zo\times Y$.  Now $\partial (\eb) = \eta
_1\sqcup \eta _0$ is a connection on $Q\sqcup Q\to Y\sqcup -Y$.  Then
$L_{\partial (\eb)}\cong \overline{L_{\eta_0}}\otimes L_{\eta _1}$ by the
properties in \theprotag{2.19} {Theorem}.  Using the metric on~$L_{\eta _0}$,
we identify $\overline{L_{\eta_0}}\otimes L_{\eta _1}$ as the line of linear
maps $L_{\eta _0}\to L_{\eta _1}$.  Let
  $$ \partrans(\et) = \eac {\zo\times Y}{\eb}\:L_{\eta _0}\longrightarrow
     L_{\eta _1} \tag{3.20}$$
be the Chern-Simons action of~$\eb$.  Note that $\partrans(\et)$~is unitary,
since the corresponding element of $\overline{L_{\eta_0}}\otimes L_{\eta _1}$
has unit norm.  We claim that $\partrans(\et)$~is the parallel transport of a
connection on $L_Q\to\conn Q$.

Fix a section $q\:Y\to Q$.  Then from the construction of lines in~\S{2}
(see~\thetag{2.15}) this induces a (unitary) trivialization $s_q\:\conn Q\to
L_Q$.  Consider the connection~$\eb$ over~$\zo\times Y$.  Using the
formula~\thetag{3.11} for its curvature, we see that the Chern-Simons form is
  $$ q^*\alpha (\eb) = -q^*\langle \et\wedge \dot{\eta }_t\rangle \wedge dt.
     $$
Hence from~\thetag{2.2} the action relative to the section~$q$ is
  $$ S_{\zo\times Y}(q,\eb) = -\int_{0}^1dt\int_{Y}q^*\langle
     \et\wedge \dot{\eta }_t\rangle . \tag{3.21}$$
This motivates us to introduce the (connection) 1-form~$\theta _q$ on~$\conn
Q$ defined by\footnote{Notice that parallel transport is the integral of {\it
minus\/} the connection form relative to some trivialization.}
  $$ (\theta _q)_\eta (\etd) = \tpi\int_{Y}q^*\langle \eta \wedge \dot{\eta
     }\rangle.  \tag{3.22}$$
We must check that this transforms properly under gauge transformations.  If
$\psi \:Q\to Q$ is a gauge transformation, then
  $$ s_{\psi q}(\eta ) = c_Y(q^*\eta ,g_\psi q)\inv s_q(\eta ) $$
according to~\thetag{2.15}.  On the other hand, we easily compute
from~\thetag{1.18} that
  $$ \split
     (\theta _{\psi q})_\eta (\etd) &= \tpi\int_{Y}q^*\langle \psi ^*\eta
\wedge
     \psi ^*\etd\rangle  \\
     &= \tpi\int_{Y}q^*\langle \Ad_{g_\psi \inv }\eta \wedge \Ad_{g_\psi \inv
     }\etd\rangle +\tpi \int_{Y}q^*\langle \phi _{g_\psi \mstrut }\wedge
\Ad_{g_\psi
     \inv }\etd\rangle  \\
     &= (\theta _q)_\eta - \tpi \int_{Y}\langle \Ad_{g\inv }(q^*\etd)\wedge
     \phi _g\rangle ,\endsplit \tag{3.23}$$
where $g=g_\psi q$.  But from~\thetag{2.15} we see that the logarithmic
differential of the cocycle with respect to~$\eta $,
  $$ \frac{dc_Y(q^*\eta ,g_\psi q)(\etd)}{c_Y(q^*\eta ,g_\psi q)} =
     \tpi\int_{Y}\langle \Ad_{g\inv }(q^*\etd)\wedge \phi _g\rangle,
     $$
is minus the last term in~\thetag{3.23}.  It follows that the set of
forms~$\{\theta _q\}$, where $q$~ranges over trivializations of $Q\to Y$,
defines a unitary connection on $L_Q\to \conn Q$.  Furthermore, as the
logarithm of the parallel transport is the integral of {\it minus\/} the
connection form, we see from~\thetag{3.21} that the Chern-Simons
action~\thetag{3.20} is the parallel transport of this connection.  The
curvature times~$i/2\pi $ is
  $$ \omega (\etd_1,\etd_2) = \frac{i}{2\pi }(d\theta _q)(\etd_1,\etd_2) = -
     2\int_{Y}\langle \etd_1\wedge \etd_2\rangle . $$
If $\form$~is nondegenerate, then so is~$\omega $.  Since the curvature of a
line bundle is also closed, it follows that $\omega $~is a symplectic form.

Now \thetag{2.21}~is a lift of the $\gauge Q$~action on~$\conn Q$ to~$L_Q$.
Furthermore, \thetag{2.22}~shows that this lift preserves the parallel
transport~\thetag{3.20}, hence the connection.  (It also preserves the
metric, so acts on the associated circle bundle with connection.)  It follows
that the $\gauge Q$~action on~$\conn Q$ preserves the curvature~$\omega $, so
is symplectic.  We compute a {\it moment map\/} for this action using the
lift to~$L_Q$.\footnote{For a general symplectic group action there may be
many moment maps, if one exists at all.  Here the lift of the action
to~$L_Q$, which is ultimately derived directly from the Chern-Simons action,
singles one out.} In general, if $L\to M$ is a hermitian line bundle (or
circle bundle) over a symplectic manifold, and $\rho \:G\to\Aut(L)$ is a
$G$~action on~$L$ preserving the metric and connection, then the moment map
of the quotient $G$~action on~$M$ is
  $$ \mu _\xi (m) = -\frac{i}{2\pi }\cdot \ver\bigl( \dot{\rho }(\xi )_\ell
     \bigr),\qquad \xi \in \frak{g}, \tag{3.24}$$
where $\ell \in L_m$ is a point of unit norm, $\dot{\rho }(\xi )$~is the
vector field on~$L$ corresponding to~$\xi \in \frak{g}$, and $\ver(\cdot )$
is the vertical part of a vector computed with respect to the connection
on~$L$.  Notice that it is exactly the obstruction to descending the
connection form to the quotient~$L/G$.  We compute this in our situation
using the trivialization $s_q\:\conn Q\to L_Q$ given by a section $q\:Y\to
Q$.  Set~$a=q^*\eta $.  Then a gauge transformation $\psi \:Q\to Q$
corresponds to a map $g=g_\psi q\:Y\to G$.  It acts as multiplication
by~$e^{2\pi ic_Y(a,g)}$ from the (trivialized) fiber at~$\eta $ to the fiber
at~$\psi ^*\eta $.  An infinitesimal gauge transformation corresponds to a
map $\xi \:Y\to \frak{g}$, and it acts as multiplication by the derivative,
which we compute from~\thetag{2.15} to be
  $$ \tpi \int_{Y}\langle a\wedge d\xi \rangle . $$
On the other hand, the connection form in this trivialization
is~\thetag{3.22}, and using this we compute the infinitesimal parallel
transport in the direction~$\xi $ to be multiplication by
  $$ -\tpi \int_{Y}\langle a\wedge d_a\xi \rangle , $$
where $d_a\xi =d\xi +[a\wedge \xi ]$ is the covariant derivative.  The
difference times~$-i/2\pi $ is
  $$ \int_{Y}\langle a\wedge (2d\xi  + [a\wedge \xi ]) = 2\int_{Y}\langle
     \Omega (\eta )\wedge \xi \rangle , $$
where $\Omega (\eta )$~is the curvature of~$\eta $.  Hence by ~\thetag{3.24}
the moment map is~\thetag{3.19}, as desired.  Since the flat connections are
the zeros of~$\mu $, the space of equivalence classes~$\mflat Q$ of flat
connections is the symplectic quotient~$\conn Q//\gauge Q$, if $\gauge
Q$~acts freely.  The line bundle, together with its metric and connection,
pass to a bundle $L_Q\to \mflat Q$ on the quotient.  The symplectic
form~\thetag{3.18} also passes to the quotient.  Its de Rham cohomology class
is an integral\footnote{Thus the symplectic form satisfies the integrality
constraint in the geometric quantization theory.} element of~$H^2(\mflat
Q;\RR)$, since it is the first Chern class of~$L_Q$.
     \enddemo

The connection can also be deduced from the gluing law and diffeomorphism
invariance of the action.  We give this argument in the proof of
\theprotag{5.9} {Proposition}, where we also calculate the curvature and
holonomy directly from the action.  The simplest case of the Chern-Weil
theorem asserts that $i/2\pi $~times the curvature of a line bundle
represents the first Chern class in real cohomology.  This applies here
to~$L_Q$ over the symplectic quotient.  We refine this in Part~2 to calculate
the first Chern class in {\it integral\/} cohomology.

We remark that this connection is consistent with the
isomorphism~\thetag{2.30} associated to a change of gauge group; this follows
from \theprotag{2.9} {Proposition}.

In general $\gauge Q$~does not act freely on~$\conn Q$; the stabilizer
at~$\eta \in \conn Q$ is the subgroup~$Z_\eta $ of parallel gauge
transformations (cf.~\theprotag{1.20} {Proposition}).  The value of such a
gauge transformation at~$y\in Y$, which lies in~$\Aut(Q_y)$, commutes with
the holonomy group at~$y$.  There is an action of~$Z_\eta $ on the
line~$L_\eta $.

     \proclaim{\protag{3.25} {Proposition}}
 The action of~$Z_\eta $ on~$L_\eta $ is constant on components of~$Z_\eta $,
and so factors through an action of the finite group~$\pi _0(Z_\eta )$
on~$L_\eta $.
     \endproclaim

     \demo{Proof}
 If $\psi \in Z_\eta $ then we form the bundle $\Qb_\psi \to\cir\times Y$
with connection~$\eb_\psi $ by gluing the ends of~$\zo\times Q$ using~$\psi
$.  According to~\thetag{3.20} the action of~$\psi $ on~$L_\eta $
is~$\eac{\cir\times Y}{\eb_\psi }$.  But if $\psi $~and $\psi '$~are
connected by a path in~$Z_\eta $, then $\eb_\psi \cong \eb_{\psi '}$.
     \enddemo

Along these lines we also have the following.

     \proclaim{\protag{3.26} {Proposition}}
 The center $Z\subset G$ acts trivially on~$L_\eta $ for all $\eta \in \conn
Q$.
     \endproclaim
\flushpar
 Combining these two propositions we see that the action of~$Z_\eta $
on~$L_\eta $ factors through an action of~$\pi _0(Z_\eta /Z)$.

     \demo{Proof}
 Suppose $g\in Z$, which we identify as a constant function on~$Y$.  Let
$q\:Y\to Q$ be any section and $a=q^*\eta $.  Then the action of~$g$
on~$L_\eta $ is multiplication by the cocycle~$c_Y(a,g)$ in~\thetag{2.15},
which is easily seen to be the identity.
     \enddemo

Next, let $X$ be an arbitrary compact oriented 3-manifold with boundary, and
consider the diagram~\thetag{3.13}.

     \proclaim{\protag{3.27} {Proposition}}
 The image of $r_X\:\mflat X\to\mflat{\bX}$ is a Lagrangian submanifold.
More precisely, the action~$\eac X\cdot $ is a flat section of the pullback
$r_X^*L_{\bX}\to\mflat{\bX}$, and therefore the induced symplectic
form~$r_X^*\omega $ vanishes.
     \endproclaim

\flushpar
 Just as the symplectic form~$\omega $ satisfies an integrality condition---it
is $i/2\pi $~times the curvature of a line bundle---so too does the
Lagrangian submanifold $\mflat X\to\mflat{\bX}$ satisfy an ``integrality
condition''---the pullback line bundle with connection is trivial.  (This
gives a restriction on the holonomy, for example.)  We use the term
``Lagrangian submanifold'' imprecisely.  Rather, we should assert that
$\mflat X\to\mflat{\bX}$ is ``a Lagrangian map''.  Jeffrey and Weitsman also
obtain this result~\cite{JW}; they use the term ``Bohr-Sommerfeld orbits''
for these special Lagrangian submanifolds.

     \demo{Proof}
 If $S^1\to\mflat X$ is a loop of flat connections~$\Theta _t$, there is a
corresponding connection~$\Tb $ over~$S^1\times X$.  Now the log holonomy
of~$r_X^*L_{\bX}$ around~$S^1$ is the action~$S_{S^1\times \partial X}(\Tb
)$, by~\thetag{3.20}.  The curvature of~$\Tb $ is~\thetag{3.11}
  $$ \Omega = - \dot{\Theta }_t\wedge dt. \tag{3.28}$$
Hence from~\thetag{2.8},
  $$ S_{S^1\times \partial X}(\Tb ) = \int_{S^1\times X}\langle \Omega
     \wedge \Omega \rangle =0. $$
So $r_X^*L_{\bX}$~is flat.  The more precise statement that $\eac X\cdot $~is
a flat section is slightly more intricate.  Suppose $\Theta _t$~is a path of
flat connections on~$X$, and consider the induced connection on~$\zo\times
X$.  The action of its restriction to~$\zo\times \bX$ determines the parallel
transport~\thetag{3.20}, whereas the restriction to~$\{t\}\times X$
is~$S_X(\Theta _t)$.  The curvature is still given by~\thetag{3.28}, so that
$\langle \Omega \wedge \Omega \rangle \equiv 0$.  Then equation~\thetag{2.8}
applied to~$\zo\times X$ implies that $\eac X{\Theta _1} $~is obtained
from~$\eac X{\Theta _0}$ by parallel transport.  This proves that the section
is flat.

It remains to show that $2\dim(\image r_X) = \dim\mflat{\bX}$.  Suppose $\Theta
$~is a flat connection on~$X$ representing a smooth point of~$\mflat X$ and
whose restriction~$\partial \Theta $ to~$\bX$ represents a smooth point
of~$\mflat{\bX}$.  Then according to~\thetag{3.8} we must show
  $$ 2 \dim \image \bigl[ (r_X)_*\:H^1\bigl(X;\frak{g}(\Theta )\bigr) \to \dim
     H^1\bigl(\bX;
     \frak{g}(\bold{\Theta )}\bigr)\bigr] = \dim H^1\bigl(\bX;
     \frak{g}(\bold{\Theta )}\bigr). $$
In fact, the image of~$(r_X)_*$ is a Lagrangian subspace.  When $\Theta $~is
trivial the argument is standard.  Consider the commutative diagram
  $$ \CD
     H^1(X) @>(r_X)_* >> H^1(\bX) @>>> H^2(X,\bX)\\
     @VV\cong V @VV\cong V @VV\cong V\\
     H_2(X,\bX) @>>> H_1(\bX) @>i>> H_1(X)\endCD $$
where the coefficients are in a vector space.  The rows are exact and the
vertical isomorphisms are Poincar\'e duality.  By exactness we have $\image\,
(r_X)_* \cong \ker\, i$.  Also, $i$~ and $(r_X)_*$~are adjoint maps, from which
$\image\, (r_X)_*\cong (\ker\, i)^\circ $ is the annihilator of~$\ker\, i$.  It
follows that $\image\,(r_X)_*$~is Lagrangian, as claimed.  The same argument
works if $\Theta $~is nontrivial, provided we prove that the
complex~\thetag{3.7} on~$\bX$ defined by~$d_{\partial \Theta }$ also
satisfies Poincar\'e duality.  This is shown in~\cite{G} using group
cohomology.  There is also a proof in de Rham theory modeled
after~\cite{BT,\S5}.
     \enddemo

In summary, the Hamiltonian theory consists of the assignments
  $$ \aligned
     Y&\longmapsto \mflat Y\\
     X&\longmapsto (\mflat X @>r_X>> \mflat{\bX}) \endaligned
      $$
of a symplectic manifold to each closed oriented 2-manifold and a Lagrangian
map to each compact oriented 3-manifold.  These assignments obey
functoriality, orientation, additivity, and gluing laws analogous to those in
\theprotag{2.19} {Theorem}.  More precisely, one should consider instead the
hermitian line bundle with connection $L_Y\to\mflat Y$ and the section~$\eac
X\cdot \:\mflat X\to r_X^*L_{\bX}$.  Then there is little difference between
the Hamiltonian theory and the Lagrangian theory.\footnote{This somewhat
startling conclusion is only valid for topological theories.  In theories
with local motion there is a nonzero hamiltonian function which must be
incorporated into the formalism.} We leave the precise formulation of these
assertions to the reader (cf\.~\cite{Ax,\S3}).

\newpage
\heading
\S{4} Surfaces with boundary
\endheading
\comment
lasteqno 4@ 46
\endcomment

In the quantum Chern-Simons theory~\cite{W} a key role is played by surfaces
with boundary.  Ultimately, explicit computations depend on their use.  In
this section we develop the classical analogue of this structure.  Based on
\theprotag{2.19} {Theorem} we might guess that there exists a metrized
line~$L_\eta $ depending smoothly on a connection~$\eta $ over a surface with
boundary, and that these lines satisfy a gluing law which constructs the line
of a closed surface by cutting and pasting.  However, there is a topological
obstruction which obstructs such a construction.\footnote{I thank George
Daskalopoulos and Sheldon Chang for (separately) pointing this out to me.}
This will be elucidated in Part~2 of the paper~\cite{F2}.\footnote{The
analogy with \theprotag{2.19} {Theorem} is illuminating.  For a closed
3-manifold the Chern-Simons invariant is a complex number, whereas for a
3-manifold with boundary it is an element in a complex line.  Now for a
closed 2-manifold the Chern-Simons functional constructs a complex line,
whereas for a 2-manifold with boundary it constructs a \dots The
differential-geometric object which replaces the dots in the previous
sentence together with its generalizations form the subject of~\cite{F1}.}
Here we add some ``trivializing data'' on the boundary of the surface which
allows us to construct the desired lines; these lines then obey a gluing law.
There is also a gluing law for the Chern-Simons action where we cut along
manifolds with corners (\theprotag{4.11} {Proposition}). The corresponding
quantum gluing law~\cite{S}, ~\cite{Wal} is quite powerful.  In the middle of
this section we include a long digression about the geometry of connections
over the circle.  This is essentially a universal choice of trivializing data
for each isomorphism class of bundle over the circle.  Then we construct a
smooth line bundle over the space of connections (and basepoints) on the
surface with boundary, provided we suitably restrict the boundary holonomies
(\theprotag{4.26} {Theorem}).  As in~\S{2} these line bundles descend to
various moduli spaces of flat connections.  One of these, by a theorem of
Mehta and Seshadri~\cite{MeS}, can be identified with the moduli space of
{\it stable parabolic bundles\/} when the surface carries a complex
structure.  We compute the dimension of these moduli spaces~\thetag{4.43},
\thetag{4.44} even at reducible connections.  The index calculation in
\theprotag{4.40} {Lemma}, \theprotag{4.41} {Corollary}, and \theprotag{4.42}
{Proposition} may be of independent interest.  In~\S{5} we construct
connections on the line bundles over these moduli spaces.

We begin by mimicking the construction of~$L_\eta $ in \theprotag{2.19}
{Theorem}.  To make sense of the Wess-Zumino-Witten term on a surface with
boundary we rely on the discussion in Appendix~A.  In particular, we use the
lines~$K_\gamma $ which make up a central extension of the loop group.

Suppose $Y$~is a compact oriented 2-manifold and $Q\to Y$ a principal
$G$~bundle.  Let $\sect{\partial Q}$~be the space of sections $r\:\partial
Y\to \partial Q$ of $\partial Q\to \partial Y$.  It is a principal
homogeneous space for the action of the group of gauge
transformations~$\gauge{\partial Q}$.  Then for each connection~$\eta \in
\conn Q$ we construct a metrized line~$L_{\eta ,r}$ which varies smoothly
in~$\eta $ and~$r$.  To do this define the cocycle
  $$ c_Y(a,g) = \exp\bigl(\tpi \int_{Y}\langle \Adgi a\wedge \phi _g\rangle
     \bigr) e^{\tpi W_Y(g)}\in K_{\partial g},\qquad
     a\in \Omega ^1_Y(\frak{g}), \quad  g\:Y\to G \tag{4.1}$$
as in~\thetag{2.15}.  Notice that this cocycle takes values in the
Wess-Zumino-Witten line~$K_{\partial g}$ (cf.~\theprotag{A.1} {Proposition}),
where $\partial g\:\partial Y\to
G$ is the restriction of~$g$ to the boundary.  The cocycle identity
  $$ c_Y(a,g_1g_2) = c_Y(a,g_1)\,c_Y(a^{g_1},g_2) \tag{4.2}$$
holds for these cocycles, as the reader may verify. (Equation~\thetag{A.7}
enters this calculation.)  Let~$\fld Q$ be the category whose objects are
sections $q\:Y\to Q$.  We do {\it not\/} require that these sections restrict
to~$r$ on the boundary; rather the quotient by~$r$ is a map $r\inv \partial
q\:\partial Y\to G$.  In other words, $\partial q=r\cdot (r\inv \partial q)$,
where~ $\cdot $~is the action of~$G$ on the principal bundle~$Q$.  Now for
any two sections~$q,q'$ there is a unique morphism $q@>\psi >> q'$ which is
the gauge transformation $\psi \:Q\to Q$ with~$q'=\psi q$.  Define a functor
$\Cal{F}_{\eta ,r}\:\fld Q\to \Cal{L}$ by
 $$ \Cal{F}_\eta (q)=K_{r\inv\partial q} \tag{4.3}$$
and $\Cal{F}_\eta (q @>\psi >> q')$ equals multiplication by~$c_Y(q^*\eta
,q\inv q')$.  Here $q\inv q'\:Y\to G$ is the unique function so that
$q'=q\cdot (q\inv q')$.  The multiplication is via the isomorphism
(cf\.~\thetag{A.2})
  $$ K_{r\inv \partial q}\otimes K_{(\partial q)\inv (\partial q')}
     \longrightarrow  K_{r\inv \partial q'}. $$
That $\Cal{F}_\eta $~is a functor follows from the cocycle
identity~\thetag{4.2} and the associativity law~\thetag{A.3}.  Define
$L_{\eta ,r}$~to be the line of invariant sections of~$\Cal{F}_\eta $.

     \proclaim{\protag{4.4} {Proposition}}
 Let $G$~be a connected, simply connected compact Lie group and $\form$~an
invariant form on its Lie algebra~$\frak{g}$ which satisfies the integrality
condition \theprotag{2.5} {Hypothesis}.  Suppose $Y$~is a compact oriented
2-manifold and $Q\to Y$ a principal $G$~bundle.  Then the assignment
  $$ \eta,r \longmapsto L_{\eta,r} \, ,\qquad \eta \in \conn Q,\quad r\in
     \sect{\partial Q}, $$
of a metrized line to a connection~$\eta $ on a $G$~bundle $Q\to Y$ and
section $r\:\partial Y\to \partial Q$ is smooth, agrees with the
corresponding assignment~\thetag{2.20} if~$\partial Y=\emptyset $, and
satisfies:\newline
 \rom(a\rom)\ \rom({\it Functoriality\/}\rom)\ If $\psi \:Q'\to Q$ is a bundle
map covering an orientation preserving diffeomorphism
$\overline{\psi}\:Y'\to\nomathbreak Y$, then there is an induced isometry
  $$ \psi ^*\:L_{\eta,r} \longrightarrow L_{\psi ^*\eta ,\psi \inv r}
      \tag{4.5}$$
and these compose properly. \newline
 \rom(b\rom)\ \rom({\it Orientation\/}\rom)\ There is a natural isometry
  $$ L_{-Y;\eta ,r} \cong \overline{L_{Y;\eta ,r}}\,. \tag{4.6}$$
 \rom(c\rom)\ \rom({\it Additivity\/}\rom)\ If $Y=Y_1\sqcup Y_2$ is a disjoint
union, $\eta _i$~are connections over~$Y_i$, and $r_i$~are sections
over~$\partial Y_i$, then there is a natural isometry
  $$ L_{\eta _1\sqcup \eta _2,r_1\sqcup r_2} \cong L_{\eta _1,r_1}\otimes
     L_{\eta _2,r_2}\,.  \tag{4.7}$$
 \rom(d\rom)\ \rom({\it Gluing\/}\rom)\ Suppose $S\hookrightarrow Y$ is a
closed oriented codimension one submanifold and $Y\cut$~the manifold
obtained by cutting along~$S$.  Then $\partial Y\cut = \partial Y\sqcup
S\sqcup -S$.  Furthermore, if $Q\to Y$ is a bundle, and $Q\cut\to Y\cut$ the
cut bundle, then there is an identification of~$Q\cut\res S$
and~$Q\cut\res{-S}$.  Fix trivializations~$r_S,r_{-S}$ over these boundary
pieces which agree under this identification.  Suppose $\eta $~is a
connection on~$Q$ and $\eta \cut$~the induced connection on~$Q\cut$.  Then
for any trivialization~$r_{\partial Y}$ of~$\partial Q$ there is an isometry
  $$ L_{\eta \cut,r_{\partial Y}\sqcup r_S\sqcup r_{-S}}\cong L_{\eta
     ,r_{\partial Y}}\,. \tag{4.8}$$
 \rom(e\rom)\ \rom({\it Change of Trivialization\/}\rom)\ If $r'$~is another
trivialization over~$\partial Y$, then there is an isometry
  $$ L_{\eta ,r'}\cong K_{r^{\prime\inv} r} \otimes L_{\eta ,r}\,. \tag{4.9}$$
These compose in a natural way for three trivializations~$r,r',r''$.
     \endproclaim

     \demo{Proof}
 For smoothness note that a section $q\:Y\to Q$ induces an isomorphism
$L_{\eta ,r}\cong K_{r\inv \partial q}$, and the lines~$K_{r\inv \partial q}$
vary smoothly in~$r$ by \theprotag{A.1} {Proposition}.  Now if $q'$~is any
other section then the patching function is~$c_Y(q^*\eta ,q\inv q')$, which
is independent of~$r$ and varies smoothly in~$\eta $ by the explicit
formula~\thetag{4.1}.  Also, note that if~$\bY=\emptyset $ then the
construction given here for~$L_{\eta ,r}$ reduces to the construction of
lines in~\S{2}.

Next, if $\psi \:Q'\to Q$ is the bundle map in~(a), fix a section $q\:Y\to
Q$.  This induces the section $\psi \inv q\:Y\to Q'$.  Notice that $(\psi
\inv r)\inv \partial (\psi \inv q)=r\inv \partial q$, since $\psi \inv $~is a
gauge transformation.  Then relative to the resulting isomorphisms $L_{\eta
,r}\cong K_{r\inv \partial q}$ and $L_{\psi ^*\eta ,\psi \inv r}\cong
K_{(\psi \inv r)\inv \partial (\psi \inv q)}=K_{r\inv \partial q}$ we
define~\thetag{4.5} to be the identity map.  It is easy to check that this is
independent of the choice of~$q$.

Equations~\thetag{4.6} and~\thetag{4.7} are routine.  For~\thetag{4.8} fix
a section $q\:Y\to Q$, which induces isomorphisms
  $$ \aligned
     L_{\eta \cut,r_{\partial Y}\sqcup r_S\sqcup r_{-S}} &\cong K_{(r_{\partial
     Y})\inv \partial q}\otimes K_{(r_S)\inv \partial q} \otimes
     K_{(r_{-S})\inv \partial q}\\
     L_{\eta ,r_{\partial Y}}&\cong K_{(r_{\partial Y})\inv \partial q}.
     \endaligned \tag{4.10}$$
(In the first equation `$\partial q$' stands for the restriction of~$q$ to
the appropriate piece of the boundary.)  Now the orientation and additivity
properties in \theprotag{A.1} {Proposition} give a trivialization of
$K_{(r_S)\inv \partial q} \otimes K_{(r_{-S})\inv \partial q}$.  Then the
isometries in~\thetag{4.10} combine to give the isometry~\thetag{4.8}.  It
is easy to check that this is independent of the section $q\:Y\to Q$ using
the gluing law for the Wess-Zumino-Witten functional (\theprotag{A.9}
{Proposition}).  Finally, relative to a section $q\:Y\to Q$ the
isometry~\thetag{4.9} becomes the isometry
  $$ K_{r^{\prime\inv}\partial q}\cong K_{r^{\prime\inv}r} \otimes K_{r\inv
     \partial q}\mstrut  $$
of~\thetag{A.2}, which again transforms properly under change of trivializing
section.
     \enddemo

The following gluing law for the Chern-Simons action generalizes
\theprotag{2.19(d)} {Theorem} to ``manifolds with corners''.  Such manifolds
can be smoothed out uniquely up to diffeomorphism by ``straightening the
angle''~\cite{CF}.

     \proclaim{\protag{4.11} {Proposition}}
  Let $X$~be a compact oriented 3-manifold and $Y$~ a compact oriented
2-manifold with boundary.  Suppose $Y\hookrightarrow X$ is an embedding onto
a {\it neat\/} submanifold of~$X$; that is, $Y\cap \bX = \bY$ and $Y$~is
transverse to~$\bX$.  Let $X\cut$~be the manifold with corners obtained by
cutting~$X$ along~$Y$.  Then $\partial X\cut = \partial X\cup Y \cup -Y$
where the union is over~$\partial Y\sqcup -\partial Y$.  Suppose $\Theta $~is
a connection over~$X$.  Let $\Theta \cut$~be the induced connection
over~$X\cut$, let $\eta $~be the restriction of~$\Theta $ to~$Y$, and let
$(\partial \Theta )\cut$ be $\partial \Theta $ cut along~$\bY\hookrightarrow
\bX$.  Then for any trivialization~$r$ over~$\bY$, we have
  $$ \eacT = \Tr_\eta \( \eac{X\cut}{\Theta \cut}\),  $$
where $\Tr_\eta $ is the contraction
  $$ \Tr_\eta \:L_{\partial \Theta \cut} \cong  L_{(\partial \Theta
     )\cut,r}\otimes L_{\eta,r}  \otimes \overline{L_{\eta,r}}
     \longrightarrow L_{(\partial \Theta )\cut,r} \cong L_{\partial \Theta }
     $$
using the hermitian metric on~$L_{\eta ,r}$ and the isomorphism~\thetag{4.8}.
     \endproclaim

\flushpar
 We omit the proof, which contains no new ideas.

Let $\conn Q$~be the space of all connections on a fixed bundle~$Q\to Y$, let
$\gauge Q$~be the space of gauge transformations, and let $\sect Q$~be the
space of sections.  Assume that $\partial Y\not=  \emptyset $.  Then the group
$\gauge Q\times \Map(\partial Y,G)$ acts on~$\conn Q\times \sect Q$ via the
action
  $$ \langle \psi ,\gamma   \rangle\cdot \langle \eta ,r  \rangle
     = \langle \psi ^*\eta ,\psi \inv r\cdot \gamma   \rangle, \qquad \psi
     \in \gauge Q, \quad \gamma \in \Map(\partial Y,G), \quad \eta \in
     \conn Q, \quad r\in \sect Q. \tag{4.12}$$
\theprotag{4.4} {Proposition} asserts that the lines~$L_{\eta ,r}$ form a
smooth line bundle $L\to\conn Q\times \sect Q$, but the action~\thetag{4.12}
only lifts to an action of the {\it central extension\/} $\gauge Q\times
\ce{\partial Y}$ defined in~\thetag{A.5}.  So we do {\it not\/} obtain a line
bundle over the quotient~$\conn Q/\gauge Q$.  From this point of view the
central extension is the obstruction to the existence of that line bundle.
In fact, there are topological obstructions to the existence of a line bundle
over~$\conn Q/\gauge Q$ which obeys the desired gluing law.  So we must
content ourselves with a bundle over a subset of this quotient.  Furthermore,
we introduce additional choices on the boundary to construct the line bundle.

We first digress to discuss connections over the circle.  Without extra
effort we consider connections for arbitrary compact Lie groups~$G$.  Let
$\cir=[0,1]\bigm/0\sim 1$ be the standard circle {\it with basepoint\/}~$0\in
\cir$.  We consider principal bundles $R\to\cir$ with a fixed basepoint
in~$R$ over the basepoint in~$\cir$.  There is a category~$\bfld{\cir}$ of
connections on pointed bundles; morphisms are required to preserve the
basepoints.  Then there are no nontrivial automorphisms of a connection in
this category, since any automorphism which fixes the basepoint is the
identity, according to \theprotag{1.20} {Proposition}.  It follows that if
$\theta _1\cong \theta _2$ are isomorphic connections, there is a unique
(pointed) isomorphism between them.  It is precisely to obtain this rigidity
property that we introduce basepoints.  Now the basepoint determines a
holonomy map $\bfld{\cir}\to G$.  Since isomorphic connections have the same
holonomy, there is an induced map on the equivalence classes $\bfldb{\cir}\to
G$.  It is an easy consequence of \theprotag{1.21} {Lemma} that this latter
map is~$1:1$; that is, a connection with basepoint over~$\cir$ is determined
up to isomorphism by its holonomy.  There is a universal bundle and
connection.  Define
  $$ \Rb = G\times \RR\times G\bigm/ \langle h,s,g\rangle  \sim \langle
     h,s+1,h\inv g\rangle . \tag{4.13}$$
This is a $G$~bundle $\Rb\to G\times\cir$ via projection onto the first two
factors.  Better, it is a family of pointed $G$~bundles $\Rb_{h}\to\cir$
over~$\cir$ parametrized by~${h}\in G$; the basepoint in~$\Rb_{h}$
is~$\{\langle h,0,e\rangle \}$, where $e\in G$~is the identity.  The
Maurer-Cartan form~$\theta $ on~$G$, lifted to~$\{h\}\times \RR\times G$,
drops to a connection~$\theta _{\Rb_{h}}$ on $\Rb_{h}$ with holonomy~${h}$.
There exist connections~$\theta _{\Rb}$ on~$\Rb$ which restricts to~$\theta
_{\Rb_{h}}$ on each slice~$\Rb_{h}$; we construct one explicitly
below~\thetag{4.14}.  Summarizing, if $\theta $~is any (pointed) connection
over~$\cir$, then there is a unique isomorphism $\theta \cong \theta
_{\Rb_{h}}$ for some~${h}$.

We now explicitly construct the connection~$\theta _{\Rb}$.  Let $\theta $ be
the Maurer-Cartan form on~$G$, lifted to $G\times (-0.1,1.1)\times G$ via
projection onto the third factor, and $\theta _h$ the Maurer-Cartan form
lifted via projection onto the first factor.  Fix a smooth cutoff function
$\rho \:(-0.1,1.1)\to[0,1]$ so that $\rho \bigl((-0.1,0.1]\bigr)=0$ and $\rho
\bigl([0.9,1.1)\bigr)=1$.  Set
  $$ \aligned
     (\theta _{\Rb})_{\langle h,s,g  \rangle} &= \theta  + \rho
     (s)\Ad_{g\inv h}\theta _h \\
     &= g\inv dg + \rho (s)g\inv dhh\inv g,\endaligned \tag{4.14}$$
where the second expression makes sense for matrix groups.  Then an easy
calculation with the gluing function in~\thetag{4.13} shows that
\thetag{4.14}~determines a connection form on~$\Rb$.  The restriction to a
slice~$\Rb_{h}$ is~$\theta _{\Rb_h}$.  From~\thetag{1.13} we compute the
curvature 2-form on~$G\times\cir$:
  $$ \Omega _{\Rb} = \rho '(s)ds\wedge \Ad_{g\inv h}\theta _h + \frac 12
     \bigl( \rho (s)^2-\rho (s)\bigr) \Ad_{g\inv h}[\theta _h\wedge \theta
     _h]. \tag{4.15}$$
For an integral form~$\form$ on the Lie algebra~$\frak{g}$, we compute the
Chern-Weil 4-form on~$G\times \cir$ as
  $$ \langle \Omega _{\Rb} \wedge \Omega _{\Rb} \rangle = \rho '(s)\bigl(
     \rho (s)^2-\rho (s)\bigr) ds\wedge \langle \theta _h\wedge [\theta
     _h\wedge \theta _h]  \rangle, $$
in view of~\thetag{1.29}.  Finally, note that
  $$ \aligned
     \int_{\cir}\langle \Omega _{\Rb} \wedge \Omega _{\Rb}  \rangle &=
     \int_{0}^1 ds\,\rho '(s)\bigl( \rho (s)^2-\rho (s)\bigr ) \wedge \langle
     \theta _h\wedge [\theta _h\wedge \theta _h]  \rangle\\
     &= -\frac 16 \langle \theta _h\wedge [\theta _h\wedge \theta _h]
     \rangle.\endaligned \tag{4.16}$$
By \theprotag{2.5} {Hypothesis} this 3-form on~$G$ represents an integral
cohomology class in~$H^3(G;\RR)$.

     \proclaim{\protag{4.17} {Proposition}}
 The bundle $\Rb\to G\times\cir$ is nontrivial if $G\not\cong 1$.
     \endproclaim

     \demo{Proof}
 According to~\thetag{1.25} the Lie algebra~$\frak{g}$ decomposes as a sum of
an abelian algebra and simple algebras.  If there is a simple summand, then
we choose~$\form\:\frak{g}\otimes \frak{g}\to\RR$ to be its Killing form.  It
follows from~\thetag{4.16} that $\langle \Omega _{\Rb}\wedge \Omega _{\Rb}
\rangle$~represents a nonzero cohomology class in~$H^4(G\times\cir)$.  By
Chern-Weil theory this is a characteristic class of $\Rb\to G\times\cir$, and
so the bundle is nontrivial.  If $\frak{g}$~is abelian, then a similar
argument can be made using a 2-form~$\langle \Omega _{\Rb} \rangle$ built
from a trace~$\langle \cdot \rangle\:\frak{g}\to\RR$ on~$\frak{g}$.  If
$G$~is a finite group, then the slice $\Rb_h\to\cir$ is nontrivial if~$h\not=
1$.
     \enddemo

We also record the following lemma about holonomy.  Let $R\to S$ be a
principal $G$~bundle over an oriented, connected 1-manifold.  Though $S$~is
diffeomorphic to a circle, we do not need a fixed parametrization.  Fix a
basepoint~$s\in S$.  Then the space of connections and basepoints in the
fiber is $\conn R\times R_s$, and the holonomy is a smooth map
  $$ \hol\:\conn R\times R_s\longrightarrow G. $$

     \proclaim{\protag{4.18} {Lemma}}
 \newline
 \rom(i\rom)\ $\hol$~is a submersion, i.e., $d\hol_{\langle \eta ,b
\rangle}$ is surjective for all $\eta \in \conn R$, \ $b\in R_s$.\newline
 \rom(ii\rom)\ If $\dot{b}\in T_b(R_s)\cong \frak{g}$, and $\hol(\eta ,b)=h$,
then
  $$ h\inv \, d\hol_{\langle \eta ,b  \rangle}(0,\dot{b}) = (1 -
     \Ad_{h\inv })(\dot{b}). \tag{4.19}$$
 \rom(iii\rom)\ Let $\frak{g}_R$~be the adjoint bundle of~$R$, and suppose
$\zeta \in \Omega ^0_S(\frak{g}_R)$.  With respect to the
basepoint~$b\in R_s$ we identify~$\zeta (s)$ as an element of~$\frak{g}$.
Then if $\hol(\eta ,b)=h$,
  $$ h\inv \, d\hol_{\langle \eta ,b  \rangle}(d_\eta \zeta ,0) = (1 -
     \Ad_{h\inv })\bigl(-\zeta (s)\bigr). \tag{4.20}$$
 \rom(iv\rom)\ Suppose $\hol(\eta ,b)=h$ and $\zeta \in \Omega
^0_S(\frak{g}_R)$ with $d_\eta \zeta =0$.  Then
  $$ \langle\, \zeta (s)\otimes h\inv \, d\hol_{\langle \eta ,b
     \rangle}(\etdr,\dot{b}) \, \rangle = -\int_{S}\langle \zeta \wedge \etd
     \rangle. \tag{4.21}$$
     \endproclaim

\flushpar
 Of course, (i)~implies that $\hol\inv (h)$ is a manifold for each~$h\in G$.
In~(iii) note that the covariant derivative is a map
  $$ \Omega ^0_S(\frak{g}_R) @>d_\eta >> \Omega
     ^1_S(\frak{g}_R). \tag{4.22}$$
The domain~$\Omega ^0_S(\frak{g}_R)$ is the Lie algebra of the group of gauge
transformations~$\gauge R$, the codomain~$\Omega ^1_S(\frak{g}_R)$ is the
tangent space to the affine space~$\conn R$ of connections, and $d_\eta $~is
the infinitesimal action of gauge transformations on connections.  Here
$d_\eta $~is Fredholm; its kernel is the space of covariant constant
sections, which can be identified with the invariants of~$\Ad_h$ by
evaluation at the basepoint.  Assertions~(i) and~(iii) imply that $h\inv
d\hol_{\langle \eta ,r \rangle}$ induces an isomorphism of the cokernel
of~\thetag{4.22} with the coinvariants of~$\Ad_h$, i.e., an isomorphism
$\coker d_\eta \cong \coker(1-\Ad_h)$.

     \demo{Proof}
 Cut the circle~$S$ at the basepoint~$s$ to obtain an interval~$S\cut$ and a
bundle $R\cut\to S\cut$.  Then a connection on~$R$ determines and is
determined by a (horizontal) section of~$R\cut$.  Since the fibers over the
endpoints of~$S\cut$ are identified, we can compare the section at these two
endpoints by an element of~$G$; this is the holonomy.  Relative to a local
trivialization near the terminal endpoint, (i)~translates to the obvious
statement that evaluation at the endpoint is a submersion from the space of
paths in a connected manifold~$A$ to~$A$.  Notice that we can fix the
basepoint, which is the value of the section at the initial endpoint, for
this argument.

For~(ii) suppose $b_t$~is a curve of basepoints with tangent~$\dot{b}$.
Write $b_t=bg_t$ for a curve~$g_t\in G$.  Then the holonomy of~$\eta $ with
respect to~$b_t$ is~$g\inv _thg_t\mstrut $.  Differentiate to
deduce~\thetag{4.19}.

For~(iii) let $\varphi _t$~be a curve in~$\gauge R$ with tangent~$\zeta $.
Then $\varphi ^*_t\eta $~is a path of connections with tangent~$d_\eta \zeta
$.  But
  $$ \hol(\varphi _t^*\eta ,b)=\hol(\eta ,\varphi \inv _tb), $$
and the tangent to the path~$\varphi _t\inv (b)$ is~$-\zeta (s)$.  So~(iii)
follows from~(ii).

To prove~(iv) we first note that \thetag{4.19}~implies
  $$ \langle\, \zeta (s)\otimes h\inv \, d\hol_{\langle \eta ,b
     \rangle}(0,\dot{b}) \, \rangle = \langle \,\zeta (s)\otimes (1 -
     \Ad_{h\inv })(\dot{b})\,  \rangle = \langle \,(1 -
     \Ad_{h\inv })\bigl(\zeta (s)\bigr)\otimes \dot{b}\,  \rangle = 0.
     $$
In the last step we use the fact that $\zeta $~is parallel, so commutes with
the holonomy.  Hence we may set~$\dot{b}=0$.  Let $\eta _t$~be a path of
connections on~$R$ with initial velocity~$\etd$, and $b_t$~a path in~$R_s$
with initial velocity~$\dot{b}$.  We cut at the basepoint~$s\in S$ to obtain
a path of connections on $R\cut\to S\cut$.  Let $r$~be a flat section
of~$R\cut$ relative to~$\eta $ with initial value~$b$, and define
$g_t\:S\cut\to G$ so that $rg_t$~is a flat section of~$R\cut$ relative
to~$\eta _t$ with initial value~$b$.  Then from~\thetag{1.18} and the fact
that $(rg_t)^*\eta _t=0$ we deduce $r^*\etd = - d_{r^*\eta }\dot{g}$.
Therefore, by Stokes' theorem we have
  $$ \int_{S}\langle \zeta \wedge \etd  \rangle = \int_{S\cut}\langle
     r^*\zeta \wedge r^*\etd  \rangle = -\int_{S\cut}\langle r^*\zeta \wedge
     d_{r^*\eta }\dot{g}  \rangle = -\langle r^*\zeta \otimes \dot{g}
     \rangle\res{\partial S\cut}. \tag{4.23}$$
At the initial point of~$S\cut$ we have $\dot{g}=0$, whereas at the terminal
point $\dot{g}=h\inv \,d\hol_{\eta ,b}(\etdr,\dot{b})$.  Thus~\thetag{4.23}
implies~\thetag{4.21}.
     \enddemo

Return now to the case of a connected and simply connected group~$G$.
Since the bundle $\Rb\to G\times\cir$ is not trivial, we cannot
continuously trivialize the bundles $\Rb_{h}\to\cir$ simultaneously for
all~$h\in G$.  Hence restrict to a subset~$V\subset G$ where this is
possible, and thus choose a smooth trivialization
 $$ r\:V\times \cir\to\Rb, \tag{4.24}$$
i.e., a smooth family of trivializations $r_h\:S^1\to\Rb_{h}$.  Our
constructions in the next paragraphs and in~\S{5} depend on the
trivializations~$r_{h}$ through~\thetag{4.9}.

To use these standard trivializations on 2-manifolds we must identify
boundary components with the standard circle.  Let $Y$~be a compact oriented
2-manifold, and fix a diffeomorphism $\cir\to(\partial Y)_i$ for each
component of~$\partial Y$.  We do not require that these boundary
parametrizations preserve orientation, and so we distinguish `$+$'~ boundary
components and `$-$'~boundary components according to whether the
parametrization preserves or reverses the orientation.  The images of the
basepoint in~$\cir$ give a basepoint on each component of~$\partial Y$.  We
consider connections~$\eta $ on bundles $Q\to Y$ together with
basepoints~$b_i\in Q\res{(\partial Y)_i}$ over the basepoints on
the~$(\partial Y)_i$.  We denote the collection of basepoints as $b=\langle
b_1,\dots ,b_k \rangle$.  Bundle morphisms are required to preserve
the basepoints.  These connections on bundles with basepoints together with
these morphisms form a category~$\bfld Y$ of {\it pointed\/} connections.

Now suppose $\langle \eta,b \rangle \in \bfld Y$ is a pointed connection on
$Q\to Y$.  Then the restriction of~$\eta $ to any boundary
component~$(\partial Y)_i$ is isomorphic to a standard connection~$\theta
_{\Rb_{h_i}}$, where $h_i\in G$ is the holonomy of~$\eta $ around~$(\partial
Y)_i$, computed with respect to the basepoint.  Hence there is a unique
pointed isomorphism $Q\res{(\partial Y)_i}\cong \Rb_{h_i}$ which
pulls~$\theta _{\Rb_{h_i}}$ back to the restriction of~$\eta $.  Assume that
all of the boundary holonomies fall within the distinguished subset~$V\subset
G$.  Then the trivializations~$r_{h_i}$ chosen above induce a trivialization
$r(\eta ,b)\:\partial Y\to\partial Q$ of~$Q$ over the boundary.  Now let
  $$L_{\eta, b}=L_{\eta ,r(\eta ,b)}  \tag{4.25}$$
be the line defined above relative to this distinguished trivialization.

     \proclaim{\protag{4.26} {Theorem}}
 Let $G$~be a connected, simply connected compact Lie group and $\form$~an
invariant bilinear form on its Lie algebra~$\frak{g}$ which satisfies the
integrality \theprotag{2.5} {Hypothesis}.  Fix a subset~$V\subset G$ over
which we choose a smooth trivialization~\thetag{4.24}.  Suppose $Y$~is a
compact oriented 2-manifold with parametrized boundary.  Let $\bfld
Y(V)$~denote the subset of pointed connections whose boundary holonomies all
lie in~$V$.  Then the assignment
  $$ \langle \eta,b \rangle \longrightarrow L_{\eta,b} ,\qquad \langle \eta,b
     \rangle \in \bfld Y(V), \tag{4.27}$$
of a metrized line to such pointed connections is smooth and agrees with the
corresponding assignment~\thetag{2.20} if~$\bY=\emptyset $.  It satisfies the
functoriality property~\thetag{2.21} for bundle maps which preserve the
basepoints and the boundary parametrizations.  Furthermore, it satisfies the
orientation property~\thetag{2.23} and the additivity property~\thetag{2.25}.
Finally, it satisfies:\newline
 \rom(d\rom)\ \rom({\it Gluing\/}\rom)\ Suppose $S\hookrightarrow Y$ is a
closed, oriented codimension one submanifold and $Y\cut$~the manifold
obtained by cutting along~$S$.  Then $\partial Y\cut = \partial Y\sqcup
S\sqcup -S$ and we use parametrizations which agree on~$S$ and~$-S$.  Suppose
$\langle \eta ,b \rangle $~is a pointed connection over~$Y$ and $\langle \eta
\cut,b\cut \rangle$~the induced pointed connection over~$Y\cut$.  \rom(We
choose basepoints which agree over~$S$ and~$-S$ to form~$b\cut$
from~$b$.\rom) Then there is a natural isometry
  $$ L_{\eta,b} \cong L_{\eta \cut,b\cut}. $$
     \endproclaim

     \demo{Proof}
 For smoothness, it suffices to remark that the holonomy is a smooth function
of a connection.  For the functoriality suppose $\psi \:Q'\to Q$ is a bundle
map preserving the basepoints and the boundary parametrizations.  Let
$R',R$~be the restrictions of~$Q',Q$ to some boundary components which
correspond under~$\overline{\psi }$, and let $\psi _h\:R\to \Rb_h$ be the
unique pointed map with $\psi ^*_h(\theta _{\Rb_h})=\eta \res R$.  Then $\psi
_h\psi \:R'\to \Rb_h$ satisfies $(\psi _h\psi )^*(\theta _{\Rb_h})=\psi
^*\eta \res{R'}$.  Now the functoriality follows from~\thetag{4.5}.  The
other properties follow directly from the corresponding properties in
\theprotag{4.4} {Proposition}.
     \enddemo

Fix a bundle $Q\to Y$ together with parametrizations of the components of
$\bY = \sqcup _{i=1}^k(\bY)_i$.  Let $y_i$~be the basepoint of~$(\bY)_i$.
The holonomy is a map
  $$ \hol_Q\:\conn Q\times Q_{y_1}\times \dots \times Q_{y_k}\longrightarrow
     G\times \dots \times G. \tag{4.28}$$
Here $Q_{y_i}$ is the fiber of~$Q$ over~$y_i$, the space of basepoints
at~$y_i$.  Let
 $$ \conn{Q;V,\dots ,V}=\hol\inv _Q(V\times \dots \times V). $$
\theprotag{4.26} {Theorem} asserts the existence of a smooth line bundle
 $$ L_{Q;V,\dots ,V}\to\conn{Q;V,\dots ,V}. \tag{4.29}$$
Furthermore, the map~\thetag{4.28} is invariant under the $\gauge Q$~action,
and \theprotag{4.26} {Theorem} asserts that this action lifts to the line
bundle~$L$.  Notice that this $\gauge Q$ action is free, since we include
basepoints.

There is a larger symmetry group which acts when the boundary holonomies are
fixed.  Fix $h_1,\dots ,h_k\in V\subset G$ and let
 $$ \conn{Q;h_1,\dots,h_k}=\hol_Q\inv (h_1,\dots ,h_k); $$
by \theprotag{4.18(i)} {Lemma} this is a smooth manifold.  Consider the
restriction
  $$ L_{Q;h_1,\dots ,h_k}\longrightarrow \conn{Q;h_1,\dots ,h_k} \tag{4.30}$$
of~$L_{Q;V,\dots ,V}$ to this space.  Let $Z_{h_i}\subset G$ be the
centralizer of~$h_i$.  A result of Bott and
Taubes~\cite{BTau,Prop.~10.2} asserts that $Z_{h_i}$~is
connected.\footnote{This depends on the fact that $G$~is connected and simply
connected.} If $h_i$~is a regular element of~$G$, then $Z_{h_i}$~is a maximal
torus in~$G$.  Let $Z_{h_i}$~act on the fiber~$Q_{y_i}$ via the right
principal bundle action of~$G$.  Then the group
  $$ \gauge{Q;h_1,\dots ,h_k} = \gauge Q\times Z_{h_1}\times \dots \times
     Z_{h_k} \tag{4.31}$$
acts on~$\hol_Q\inv (h_1,\dots ,h_k)$ via the formula
  $$ \langle \eta ;b_1,\dots ,b_k  \rangle\cdot \langle \psi
     ;g_1,\dots ,g_k
     \rangle = \langle \psi ^*\eta ;\psi \inv (b_1)\cdot g_1,\dots ,\psi
     \inv (b_k)\cdot g_k  \rangle, \tag{4.32}$$
where $\eta \in \conn Q$, \ $\psi \in \gauge Q$, \ $b_i\in Q_{y_i}$, and
$g_i\in Z_{h_i}$.  In general this action is neither free nor effective.  We
determine the stabilizer at~$\langle \eta ;b_1,\dots ,b_k \rangle$ in case
$Y$~has no closed components.  Any gauge transformation that preserves~$\eta
$ is parallel, by \theprotag{1.20} {Proposition}, and its value
in~$\Aut(Q_y)$ at~$y\in Y$ commutes with the holonomy group at~$y$.  Let
$Z_{\eta }\subset \gauge Q$ be the subgroup of all such~$\psi $.  Evaluation
at the basepoint~$b_i$ gives an embedding $Z_\eta \hookrightarrow G$, and it
is easy to see that the image is contained in~$Z_{h_i}$.  This gives an
embedding of~$Z_\eta $ in~$\gauge{Q;h_1,\dots ,h_k}$, and the image is the
stabilizer of~$\langle \eta ;b_1,\dots ,b_k \rangle$.  Notice that the kernel
of the action, which is the intersection of these stabilizers, contains the
center $Z\subset G$, which sits in~$\gauge Q$ as the subgroup of {\it
global\/} gauge transformations.

Let $\fconn{Q;h_1,\dots ,h_k}$ denote the space of flat connections
in~$\conn{Q;h_1,\dots ,h_k}$.  This is not necessarily a manifold, but the
subset of {\it irreducible\/} flat connections~$\ficonn\Qfix$ does form a
submanifold.  In any case the group of gauge transformations~$\gauge Q$ acts
freely, if $Y$~has no closed components, and we denote the quotient by
  $$ \Cal{M}'_{Y;h_1,\dots ,h_k} = \fconn{Q;h_1,\dots ,h_k}\bigm/\gauge Q.
     \tag{4.33}$$
The prime reminds us of the basepoints.  This quotient is independent of the
choice of bundle $Q\to Y$ up to {\it canonical\/} diffeomorphism, since all
$G$~bundles over~$Y$ are isomorphic (they are all trivial).  It is easy to
see that up to {\it noncanonical\/} diffeomorphism this moduli space only
depends on the conjugacy classes of the~$h_i$.  Thus we can restrict~$h_i$ to
be in a maximal torus of~$G$.  If we divide out by the full symmetry
group~\thetag{4.31}, we obtain a smaller moduli space
  $$ \Cal{M}_{Y;\hb_1,\dots ,\hb_k} = \fconn{Q;h_1,\dots ,h_k}\bigm/\gauge
     {Q;h_1,\dots ,h_k}.  \tag{4.34}$$
This moduli space depends only on the conjugacy classes~$\hb_i$ of the~$h_i$.
It is the space of equivalence classes of flat connections on~$Q$ whose
boundary holonomies are conjugate to the~$h_i$.  Notice that no basepoints
are needed to describe this moduli space.  There is a map
  $$ \Cal{M}'_{Y;h_1,\dots ,h_k}\longrightarrow \Cal{M}_{Y;\hb_1,\dots ,\hb_k}.
     \tag{4.35}$$
The fiber over a connection~$\eta $ is the set of basepoints with respect to
which the boundary holonomies equal the~$h_i$.  The group
   $$ Z_h=Z_{h_1}\times\dots \times Z_{h_k}  \tag{4.36}$$
acts on the fiber with kernel~$Z_\eta $.  If $\eta $~is irreducible, then
$Z_\eta \cong Z$~is the group of global gauge transformations, which
is isomorphic to the center of~$G$; it is a normal subgroup of~$Z_h$.  Thus
over the subspace
  $$ \Cal{M}^*_{Y;\hb_1,\dots ,\hb_k}\subset \Cal{M}_{Y;\hb_1,\dots ,\hb_k}
     \tag{4.37} $$
of irreducible connections, \thetag{4.35}~is a principal bundle with
structure group~$Z_h/Z$.

To determine the tangent space to ~$\Cal{M}'_{Y;h_1,\dots ,h_k}$ at~$\langle
\eta ;b_1,\dots ,b_k  \rangle$ we examine the differential of the equations
which assert that $\eta $~is flat and that the boundary holonomies
are~$h_1,\dots ,h_k$.  This leads to the complex
  $$ 0 \;@>>>\; \Omega ^0_Y(\frak{g}_Q) \;@>{d_\eta \oplus \ev_b}>>\; \Omega
     ^1_Y(\frak{g}_Q) \oplus \frak{g}^{\oplus k} \;@>{d_\eta \oplus h\inv
     d\hol_{\langle \eta ,b  \rangle}}>>\; \Omega ^2_Y(\frak{g}_Q)\oplus
     \frak{g}^{\oplus k} \;@>>>\; 0. \tag{4.38}$$
Here $\ev_b$~is evaluation at the basepoints.  The fact that \thetag{4.38}~is
a complex follows from~\thetag{4.19} and~\thetag{4.20}.  We identify the
cohomology of~\thetag{4.38} with the {\it compactly supported\/} cohomology
of the interior~$\Yo$ with coefficients in~$\gQ$, i.e., with the cohomology
of the complex
  $$ 0 @>>> \cfrm0 @> d_\eta >> \cfrm1 @> d_\eta >> \cfrm2 @>>> 0 \tag{4.39}$$
of compactly supported differential forms on~$\Yo$ with coefficients
in~$\gQ$.  Let $\tilde{H}^\bullet=\tilde{H}^\bullet(Y;\eta ,b)$ denote the
cohomology of~\thetag{4.38} and
$H^\bullet_c=H^\bullet_c\bigl(\Yo;\frak{g}(\eta )\bigr)$ the cohomology
of~\thetag{4.39}.

     \proclaim{\protag{4.40} {Lemma}}
 The inclusion of~\thetag{4.39} into~\thetag{4.38} induces an isomorphism
$\tilde{H }^\bullet\cong H^\bullet_c$ on cohomology.
     \endproclaim

\flushpar
  A more direct approach might be to show that $\tilde{H}^\bullet$~is a de Rham
model for the twisted relative cohomology $H^\bullet\bigl(Y,\bY;\frak{g}(\eta
)\bigr)$.  The following would be a de Rham proof of excision in that
context.

     \demo{Proof}
 A simple check shows that the inclusion commutes with the differentials.  We
first show that the induced map on cohomology is injective.  At degree~0
there is nothing to check.  For degree~1 suppose $\etd\in \cfrm1$ is closed
and maps to zero in~$\tilde{H}^1$.  Then there exists $\zeta \in \frm0$ such
that $\etd=d_\eta \zeta $ and $\zeta _b=0$.  Since $\etd=0$ near~$\bY$ we see
that $\zeta $~is parallel near~$\bY$, and then the condition~$\zeta _b=0$
implies~$\zeta =0$ near~$\bY$.  Thus $\zeta $~has compact support, and so
$\etd$~vanishes in~$H^1_c$.  For degree~2 suppose $\tau \in \cfrm2$ is closed
and maps to zero in~$\tilde{H}^2$.  Then there exists $\langle \etdr,\bd
\rangle\in \frm1\oplus \frak{g}^{\oplus k}$ such that $d_\eta \etd=\tau $ and
$h\inv d\hol_{\langle \eta ,b \rangle}(\etdr,\bd)=0$.  Now
\thetag{4.19}~implies that $h\inv d\hol_{\langle \eta ,b \rangle}(\etd,0)$ is
in the image of~$1-\Ad_{h\inv }$, and then the remarks following
\theprotag{4.18} {Lemma} imply that $\partial \etd$ ~is in the image
of~$d_\eta $ on~$\bY$.  Extending away from the boundary we construct~$\zeta
\in \frm0$ such that $\etd=d_\eta \zeta $ near~$\bY$.  So $\tau =d_\eta
(\etd-d_\eta \zeta )$ and $\etd-d_\eta \zeta \in \cfrm1$, whence $\tau
$~vanishes in~$H^2_c$.

It remains to show that the inclusion of~\thetag{4.39} into~\thetag{4.38}
induces a surjection on cohomology.  The only possible cohomology
of~\thetag{4.38} in degree~0 occurs on closed components of~$Y$, since a flat
section of~$\gQ$ which vanishes at a basepoint vanishes identically on the
component of~$Y$ containing the basepoint.  The compact cohomology and
ordinary cohomology agree on closed components, so we are done.  For degree~1
suppose $\langle \etdr,\bd \rangle\in \frm1\oplus \frak{g}^{\oplus k}$
satisfies
  $$ \aligned
     d_\eta \etd &=0\\
     h\inv d\hol_{\langle \eta ,b \rangle}(\etdr,\bd)&=0.\endaligned
     $$
As above we can find $\zeta \in \frm0$ so that $\etd=d_\eta \zeta $
near~$\bY$.  We can alter any choice of~$\zeta $ by a parallel section near
the boundary, and in this way we arrange that $\zeta _b=\bd$.  Then $\langle
\etd-d_\eta \zeta ,0 \rangle$ is cohomologous to~$\langle \etdr,\bd \rangle$,
and $\etd-d_\eta \zeta \in \cfrm1$ has compact support.  This completes the
proof for degree~1.  There is nothing to prove in degree~2.
     \enddemo

Since the compact cohomology and the ordinary cohomology are in (Poincar\'e)
duality, we deduce the following.

     \proclaim{\protag{4.41} {Corollary}}
 If $Y$~has no closed components, then
  $$ \aligned
     \tilde{H}^0(Y;\eta ,b) &= 0\\
     \dim \tilde{H}^2(Y;\eta ,b) &= \dim H^0\bigl(Y;\frak{g}(\eta )\bigr).
     \endaligned $$
     \endproclaim

\flushpar
 Note that $H^0\bigl(Y;\frak{g}(\eta )\bigr)$ is the Lie algebra of~$Z_\eta $,
the subgroup of gauge transformations which fix~$\eta $.

\theprotag{4.40} {Lemma} implies that the index of~\thetag{4.38} is the index
of~\thetag{4.39}, which by Poincar\'e duality is the Euler characteristic
of~$Y$ with coefficients in~$\gQ$.  (Since \thetag{4.38}~differs from the
twisted de Rham complex by a finite dimensional operator of index~0, this
also follows directly.)

     \proclaim{\protag{4.42} {Proposition}}
 The index of~\thetag{4.38} is
  $$ \dim \tilde{H}^0(Y;\eta ,b) - \dim \tilde{H}^1(Y;\eta ,b) + \dim
     \tilde{H}^2(Y;\eta ,b) =\dim G\cdot \chi (Y). $$
     \endproclaim

     \demo{Proof}\!\!\footnote{I thank Frank Quinn for pointing out this
proof.  My original proof, based on Atiyah-Patodi-Singer, was much less
elementary.}
 By the de Rham theorem this twisted Euler characteristic can be computed
using other models of cohomology, e.g.~cellular theory.  It suffices to
consider the case where $Y$~is connected.  Fix a basepoint~$y\in Y$ and a
trivialization of the fiber of~$\gQ$ at~$y$.  Choose a finite CW~structure
on~$Y$, and let $\tilde{C}_\bullet(Y)$~be the cellular chain complex of the
universal cover of~$Y$; it is a finite dimensional module over the group
algebra~$\RR[\pi _1(Y,y)]$.  Since the flat connection~$\eta $ determines a
representation of~$\pi _1(Y,y)$ on~$\frak{g}$, we can form a cochain complex
$\Hom_{\pi _1(Y,y)}\bigl(\tilde{C}_\bullet(Y),\frak{g}\bigr)$.  Its
cohomology is the cohomology of~$Y$ with coefficients in~$\gQ$ (twisted by
the flat connection~$\eta $).  By counting dimensions we see that its Euler
characteristic, which equals the Euler characteristic of its cohomology, is
$\dim G\cdot \chi (Y)$.
     \enddemo

Combining \theprotag{4.42} {Proposition} with \theprotag{4.41} {Corollary} we
obtain a formula for the dimension of the moduli space $\Cal{M}'_{Y;h_1,\dots
,h_k}$~\thetag{4.33} at a smooth point, assuming that $Y$~has no closed
components:
  $$ \dim \Cal{M}'_{Y;h_1,\dots ,h_k} = \dim \tilde{H}^1(Y;\eta ,b) = -\dim
     G\cdot \chi (Y) + \dim Z_\eta . \tag{4.43}$$
{}From the discussion following~\thetag{4.35} we see that the dimension
of $\Cal{M}_{Y;\hb_1,\dots ,\hb_k}$~\thetag{4.34} is
  $$ \aligned
     \dim \Cal{M}_{Y;\hb_1,\dots ,\hb_k} &= -\dim G\cdot \chi (Y) -
     \sum\limits_{i} \dim Z_{h_i} + 2\dim Z_\eta \\
     &= -\dim G\cdot \chi (\overline{Y}) \, + \,
     \sum\limits_{i} (\dim G - \dim Z_{h_i}) \,+\, 2\dim
     Z_\eta  \endaligned
     \tag{4.44}$$
at the connection~$\eta $.  Here $\overline{Y}$ is the compactification
of~$Y$ obtained by gluing in standard disks.  The second equation
in~\thetag{4.44} makes clear that $\dim\Cal{M}_{Y;\hb_1,\dots ,\hb_k}$ is an
even number.  This is explained by the theorem of Mehta and
Seshadri~\cite{MeS}, which identifies~$\dim\Cal{M}_{Y;\hb_1,\dots ,\hb_k}$ as
a complex manifold (of {\it stable parabolic bundles\/}) when $Y$~is endowed
with a complex structure.\footnote{There is a recent gauge theoretic proof of
the Mehta-Seshadri theorem by Jonathan Poritz~\cite{Po}.}

The passage from~\thetag{4.43} to~\thetag{4.44} and the map~\thetag{4.35} are
illuminated by the exact cohomology sequence of the pair~$(Y,\bY)$, modified
using excision:
  $$ \CD
     0 @>>> H^0\bigl(Y;\frak{g}(\eta )\bigr) @>j>> H^0\bigl(\bY;\frak{g}(\eta
     )\bigr) @>\delta >> H^1_c\bigl(\Yo;\frak{g}(\eta )\bigr) @>i>>
     H^1\bigl(Y;\frak{g}(\eta )\bigr)\\
     @. @VV\cong V @VV\cong V @VV\cong V\\
     0 @>>> \Lie Z_\eta @>j>> \oplus \Lie Z_{h_i} @>\delta >> T_\eta
     \Cal{M}'_{Y;h_1,\dots ,h_k} \endCD \tag{4.45}$$
We still assume that $Y$~has no closed components.  The formal tangent space
$T_\eta \Cal{M}_{Y;\hb_1,\dots ,\hb_k}$ is the quotient of the formal tangent
space~$T_\eta \Cal{M}'_{Y;h_1,\dots ,h_k}$ by the image of~$\delta $.

\theprotag{4.26} {Theorem} states that the bundle~\thetag{4.30} descends to
the quotient~$\Cal{M}'_{Y;h_1,\dots ,h_k}$.  In fact, this bundle also
descends further to the moduli space~$\Cal{M}_{Y;\hb_1,\dots ,\hb_k}$.

     \proclaim{\protag{4.46} {Proposition}}
 The action of~$\gauge\Qfix$ on~$\fconn\Qfix$ lifts to the line bundle
$L_{\Qfix}\to\fconn\Qfix$.  The center $Z\hookrightarrow \gauge\Qfix$ acts
trivially.
     \endproclaim

     \demo{Proof}
 If $g\in Z_{h_i}$, then the change of basepoint $g_i\to b_ig$ changes the
fiber~$L_\eta $ over a connection~$\eta $ according to~\thetag{4.9}.  The
Wess-Zumino-Witten line~$K_g$ which enters is canonically
trivial~\thetag{A.4} since $g$~is constant, and this trivialization is
compatible with multiplication in~$Z_{h_i}$.  This gives a lift of the
$Z_{h_i}$~action to~$L_{Q;h_1,\dots,h_k}$, as desired.

An element of the center~$Z$ is represented as a constant map $g\:Y\to G$.
Fix a section $q\:Y\to Q$ and let $a=q^*\eta $.  Then the action of~$g$
on~$L_\eta $ is multiplication by $c_Y(a,g)$~\thetag{4.1}, which is trivial.
(We use the trivialization~\thetag{A.4} here.)
     \enddemo

\newpage
\heading
\S{5} Hamiltonian Theory on Surfaces with Boundary
\endheading
\comment
lasteqno 5@ 31
\endcomment

As in~\S{3} we use the Chern-Simons functional on paths of connections over a
surface, now possibly with boundary, to define a connection on a line
bundle~\thetag{4.30} over the space of connections (\theprotag{5.9}
{Proposition}).  Rather than proceed by direct calculation, we use the
abstract construction of a connection from its holonomy explained in
Appendix~B.  Just as the construction of the bundle depends on the choice of
trivializing section~\thetag{4.24}, so too does the connection.\footnote{We
can also construct a connection on~\thetag{4.29} where the boundary
holonomies are allowed to vary over a subset~$V\subset G$ on which the
universal bundle is trivialized.  Then the connection also depends on a
choice of universal connection~\thetag{4.14}.  We omit that here, but return
to it in Part~2 where we can make better sense of the dependence on these
choices.} This line bundle with connection descends to the moduli spaces
$\Cal{M}'_{Y;h_1,\dots ,h_k}$~\thetag{4.33} and $\Cal{M}^*_{Y;\hb_1,\dots
,\hb_k}$~\thetag{4.37}.  Because our constructions are local, these line
bundles obey a gluing law (\theprotag{5.29} {Proposition}).  The analogue of
this gluing law in the quantum theory leads to a modular functor~\cite{S},
from which one obtains the Verlinde algebra~\cite{V}.  For details in the
case of a finite gauge group, see~\cite{FQ}.  One can twist the line bundle
over~$\Cal{M}^*_{Y;\hb_1,\dots ,\hb_k}$ by representations of the
centralizers of the boundary holonomies.  These twisted line
bundles~\thetag{5.31} are considered by Daskalopoulos and
Wentworth~\cite{DasW}.  They only construct them for certain conjugacy
classes of the fixed boundary holonomies, but we find no such restriction on
the holonomy.\footnote{If we did not include the 1-form~\thetag{5.8} in our
connection, then we would find their restriction.}

Let $Y$~be a compact oriented 2-manifold with parametrized boundary.  Fix a
$G$~bundle $Q\to Y$.  Now a path of connections and basepoints~$\langle \eta
_t,b_t  \rangle$ on~$Q$ determines a connection~$\eb$ and a path~$\bold{b}$
of basepoints on~$\zo\times Q$.  The connection~$\eb$ has no $dt$~component.
Restricting to the boundary we obtain a unique map to the universal bundle
  $$ \CD
     \zo\times \partial Q @>\varphi >> \Rb\\
     @VVV @VVV \\
     \zo\times \bY @>\bar{\varphi }>> G\times\cir\endCD \tag{5.1}$$
which restricted to a boundary component~$(\bQ)_i$ and a fixed ~$t$ satisfies
$\varphi ^*(\theta _{\Rb_{h_i(t)}}) = \partial \eta _i(t)$ and preserves the
basepoints.  Here $h_i(t)$~is the holonomy around~$(\bQ)_i$.  The pullback of
the universal connection is
  $$ \varphi ^*(\theta _{\Rb}) = \partial \eb + \xi dt \tag{5.2}$$
for some $\xi \:\zo\times \bQ\to\frak{g}$.  An easy argument shows that for
fixed~$t$ the infinitesimal gauge transformation $\xi _t\in \Omega
^0_{\bY}(\frak{g}_Q)$ depends only on the first derivative~$\langle
\etdr_t,\dot{b}_t  \rangle$ of~$\langle \eta _t,b_t  \rangle$.  So we have a
function
  $$ \xi \:T(\conn Q\times Q_{y_1}\times \dots \times Q_{y_k})\longrightarrow
     \Omega ^0_{\bY}(\frak{g}_Q). \tag{5.3}$$
Now from~\thetag{5.2} we compute the curvature of~$\varphi ^*(\theta
_{\Rb})$ to be
  $$ dt\wedge (\partial \etd - d_\eta \xi ), $$
where $d_\eta $~is the covariant derivative.  On the other hand this
curvature is the pullback of~\thetag{4.15} under~\thetag{5.1}.  In
particular, if the boundary holonomies are constant in~$t$, then this
curvature vanishes and
  $$ \partial \etd = d_\eta \xi . \tag{5.4}$$

Fix a subset~$V\subset G$ and a trivialization~\thetag{4.24} of the universal
bundle over~$V$.  (In a moment we will fix the boundary holonomies.)  Then
for a path $\langle \eta _t,b_t \rangle\in \conn{Q;V,\dots ,V}$ with boundary
holonomies in~$V$ we obtain from~\thetag{5.1} a section
  $$ \bold{r}(\eb,\bold{b})\:\zo\times \bY\longrightarrow \zo\times \bQ.
     \tag{5.5}$$
These sections glue properly when we glue together paths.  According to the
construction~\thetag{4.25} and~\thetag{4.3} of the lines~\thetag{4.27}, and
the trivialization~\thetag{A.4}, the section~\thetag{5.5} induces an
isometry
  $$ L_{\zo\times \bY;\partial \eb,b_0,b_1}\cong \CC  \tag{5.6}$$
These isometries are compatible with gluing and gauge transformations.

Now fix the boundary holonomies to be $h_1,\dots ,h_k\in V$.  Then the
pullback
  $$ a = r(\eta ,b)^*\partial \eta\in \Omega ^1_{\bY}(\frak{g}),\qquad
     \langle \eta ,b \rangle \in \conn\Qfix, \tag{5.7}$$
is independent of~$\langle \eta ,b  \rangle$.  Define the 1-form $\alpha \in
\Omega ^1_{\conn\Qfix}$ by
  $$ \alpha (\etdr,\dot{b}) = \tpi\int_{\bY}\langle \xi (\etdr,\bd)\wedge a
     \rangle. \tag{5.8}$$
This 1-form enters in the following extension of \theprotag{3.17}
{Proposition}.

     \proclaim{\protag{5.9} {Proposition}}
  Fix a $G$~bundle $Q\to Y$ over a compact oriented 2-manifold with
parametrized boundary $\bY = \sqcup _{i=1}^k(\bY)_i$.  Choose also a
universal connection~\thetag{4.14} and a trivialization~\thetag{4.24} of the
universal bundle over points $h_1,\dots ,h_k\in G$.  Then the Chern-Simons
action, modified by the 1-form~\thetag{5.8}, defines a unitary connection on
the hermitian line bundle~\thetag{4.30}.  The curvature of this connection
times~$i/2\pi $ is
  $$ \omega _{\langle \eta ,b \rangle}(\etdr_1 ,\bd_1;\etdr_2,\bd_2)= -
     2\int_{Y}\langle \etd_1 \wedge \etd_2 \rangle \,+\, 2\int_{\bY}\langle
     \xi _1\wedge d_\eta \xi _2  \rangle, \tag{5.10}$$
where $\xi _i=\xi _{\langle \eta ,b \rangle}(\etdr_i,\bd_i)$ is defined
by~\thetag{5.3}.  The action of~$\gauge Q$ on~$\conn \Qfix$ lifts
to~$L_{Q;h_1,\dots,h_k}$, and the lifted action preserves the metric and
connection.  The induced ``moment map'' is
  $$ \mu _\xi (\eta ) = 2\int_{Y}\langle \Omega (\eta )\wedge \xi \rangle,
      \tag{5.11}$$
where $\xi \in \Omega ^0_Y(\frak{g}_Q)$ is an infinitesimal gauge
transformation.
     \endproclaim

\flushpar
 We do not have a good geometric explanation of the ``correction''
1-form~\thetag{5.8}.  Even if $\form$~is nondegenerate, the form~$\omega $ is
degenerate; the kernel consists of variations where $\etd=0$ and $\bd$~is
arbitrary.  The ``moment map''~\thetag{5.11} is still well-defined as the
obstruction to descending the connection to the quotient by~$\gauge Q$.

We remark that Chang~\cite{Ch} constructs a connection on a determinant line
bundle with similar formulas for the curvature and moment map.

One can give a proof of \theprotag{5.9} {Proposition} by calculations
similar to those in the proof of~\theprotag{3.17} {Proposition}.  For this
one needs detailed geometry of the Wess-Zumino-Witten line bundle, explained
in Appendix~A.  We opt for an easier route, based on the gluing law in
\theprotag{4.11} {Proposition} and the theorem in Appendix~B which constructs
a connection from the parallel transport.

     \demo{Proof}
 Let $\langle \eta _t,b_t \rangle$ be a smooth path in~$\conn\Qfix$.  As above
we obtain a connection~$\eb$ on~$\zo\times Q\to\zo\times Y$.  The boundary
$\partial (\eb)=\eta _1\sqcup \eta _0\sqcup \partial \eta _t$ is a connection
on $Q\sqcup Q\sqcup \zo\times \bQ\to Y\sqcup -Y \sqcup -\zo\times \bY$.
Using the inverse of the trivialization~\thetag{5.6} we identify the
Chern-Simons action on~$\eb$ as a map
  $$ \partrans(\et) = \eac {\zo\times Y}{\eb}\:L_{\eta _0,b_0}\longrightarrow
     L_{\eta _1,b_1}.  \tag{5.12}$$
$\partrans(\cdot )$~is a smooth function of the path, since the Chern-Simons
action is smooth.  For constant paths \thetag{5.12}~is the identity map,
since then the connection~$\eb$ bounds a connection on~$D^2\times Y$ which is
constant in the $D^2$~factor.  Furthermore, $\partrans(\cdot )$~is
reparametrization invariant by the functoriality property~\thetag{2.22}.  The
generalized gluing law \theprotag{4.11} {Proposition} and the gluing law for
the trivializations~\thetag{5.6} imply that \thetag{5.12}~composes properly
when we glue paths; in the terminology of Appendix~B, it is an additive
function.  Hence the hypotheses of {\theprotag{B.6} {Corollary}} are
satisfied, and this theorem asserts the existence of a connection on
$L_{Q;h_1,\dots,h_k}\to\conn\Qfix$ with parallel
transport~\thetag{5.12}.\footnote{Clearly this argument generalizes to an
axiomatic framework, where one deduces the Hamiltonian theory (symplectic
structure) from the general properties of the action.  We will not attempt
this here (but see~\cite{Ax,\S3}).} Notice that \thetag{B.3}~is an explicit
formula for the connection form.

However, this is not the connection we want.  The connection
on~$L_{Q;h_1,\dots,h_k}$ we consider is this connection plus the
1-form~$\alpha $ defined in~\thetag{5.8}.  We will now show that the
curvature and moment map of this modified connection are~\thetag{5.10}
and~\thetag{5.11}.

To verify the curvature formula~\thetag{5.10}, fix $x,y$~small and consider
the connection
  $$ \eta  + sx\etd_1 + ty\etd_2  \tag{5.13}$$
on the bundle~$\zo_s\times \zo_t\times Q$.  Choose basepoints which agree
with~$b$ at $s=t=0$, have derivative~$\bd_1$ and~$\bd_2$ along the $s$~and
$t$~directions and have constant boundary holonomies.  Let
  $$ \gamma _{x,y}\:\partial (\zo\times \zo)\longrightarrow \conn\Qfix $$
be the resulting map on the boundary of the $s,t$~rectangle.  Then by the
calculus of differential forms we see that the curvature of the connection
on~$L_{Q;h_1,\dots,h_k}$ evaluated in the directions~$\langle \etdr_1,\bd_1
\rangle$, $\langle \etdr_2,\bd_2 \rangle$ is
  $$ -\frac{d}{dx}\res{x=0}\frac{d}{dy}\res{y=0} \log \hol(\gamma _{x,y})
     \, + \, d\alpha (\etdr_1,\bd_1;\etdr_2,\bd_2), \tag{5.14}$$
where $\hol(\gamma _{x,y})$ is the holonomy around the loop.
Equation~\thetag{5.12} defines this holonomy as the exponential of
$\tpi$~times the Chern-Simons action, which we compute using~\thetag{2.8}
applied to $W=\zo\times \zo\times Y$.  Note
  $$ \partial W = \partial (\zo\times \zo)\times Y\;\cup\; \zo\times
     \zo\times \bY. \tag{5.15}$$
Now the curvature of~\thetag{5.13} is
  $$ \Omega =\Omega _{s,t} + x\,ds\wedge \etd_1 + y\,dt\wedge \etd_2,
     \tag{5.16}$$
where $\Omega _{s,t}$~is the curvature of~\thetag{5.13} restricted
to~$Y\times \{s\}\times \{t\}$.  A straightforward computation shows
  $$ \int_{\zo\times \zo\times Y}\langle \Omega \wedge \Omega   \rangle =
     -2xy \int_{Y}\langle \etd_1\wedge \etd_2\rangle . \tag{5.17}$$

It remains to calculate the action on $\zo\times \zo\times \bY$
(c.f.~\thetag{5.15}) and the contribution from~$d\alpha $ in~\thetag{5.14}.
The definition~\thetag{5.12} instructs us to use the
trivialization~\thetag{5.6}, and so to calculate the action on $X=\zo\times
\zo\times \bY$ relative to the canonical section~$r$ on~$\bX$.  In fact, we
may as well use~$r$ over all of~$X$.  Consider the connection~\thetag{5.13}
on $\zo\times \zo\times \bQ$.  (We now omit the `$\partial $' from the
notation for the connection.)  As in~\thetag{5.2} we have
  $$ r^*(\eta  + sx\etd_1 + ty\etd_2 ) = a -x\xi _1ds - y\xi _2dt,
     $$
where~$a$, defined in~\thetag{5.7}, is independent of~$s$ and~$t$, and $\xi
_1,\xi _2\:X\to\frak{g}$ are defined by~\thetag{5.3}.  From~\thetag{5.16} we
see that
  $$ r^*(\Omega ) = xds\wedge \etd_1 + ydt\wedge \etd_2. \tag{5.18}$$ 
The action relative to the trivialization~$r$ is computed from~\thetag{1.26}
and~\thetag{2.2}:
  $$ S_X(r,\eta  + sx\etd_1 + ty\etd_2 ) = xy\int_{\zo\times \zo}ds\wedge dt
     \int_{\bY} -\langle \xi _1\wedge \etd_2  \rangle + \langle \xi _2\wedge
     \etd_1  \rangle + \langle [\xi _1,\xi _2]\wedge a  \rangle. \tag{5.19}$$
Combining~\thetag{5.17} and~\thetag{5.19}, substituting~\thetag{5.4}, and
using Stokes' theorem we see that
  $$ -\frac{i}{2\pi } \frac{d}{dx}\res{x=0}\frac{d}{dy}\res{y=0} \log
     \hol(\gamma _{x,y}) = -2\int_{Y}\langle \etd_1\wedge \etd_2  \rangle
     + \int_{\bY} 2\langle \xi _1\wedge d_\eta \xi _2  \rangle -
     \int_{\bY}\langle [\xi _1,\xi _2]\wedge a  \rangle. $$
It is easy to see that $\frac{i}{2\pi }d\alpha $ in~\thetag{5.14} cancels the
last term, thereby proving~\thetag{5.10}.

Next, we verify~\thetag{5.11}.  Suppose $\xi \in \Omega ^0_Y(\frak{g}_Q)$ and
$\psi _t\in \gauge Q$ is a path of gauge transformations with $\psi _0=\id$
and $\dot\psi _0=\xi $.  Consider the path $\langle \psi _t^*\eta ,\psi
_t\inv b \rangle\in \conn\Qfix$.  Now the path $\psi _t^*\eta $ forms a
connection~$\eb$ on~$[0,x]\times Y$.  Fix a section $q\:Y\to Q$ which has
$\partial q=r(\eta ,b)$, and let $\qb\:[0,x]\times Y\to[0,x]\times Q$ be the
section~$\psi _t\inv q$.  According to~\thetag{3.24} and~\thetag{5.12} the
moment map is
  $$ \mu _\xi (\eta ) = -\frac{d}{dx}\res{x=0} S_{[0,x]\times Y}(\qb,\eb)
     \,-\, \frac{i}{2\pi }\alpha (d_\eta \xi ,\xi \res{\text{basepoints}} ).
     \tag{5.20}$$
Now
  $$ \qb^*\eb = q^*\eta -\xi dt, $$
from which the pullback of the curvature~$\Ob$ of~$\eb$ is
  $$ \qb^*\Ob = \Omega +dt\wedge d_\eta \xi , $$
where $\Omega =\Omega (\eta )$ is the curvature of~$\eta $.  A short
computation shows
  $$ S_{[0,x]\times Y} (\qb,\eb) = \int_{0}^x dt\Bigl[
     -2\int_{Y}\langle \Omega \wedge \xi (t)  \rangle + \int_{\bY}\langle \xi
     (t)\wedge a  \rangle \Bigr]. \tag{5.21}$$
We deduce~\thetag{5.11} easily from~\thetag{5.20} and~\thetag{5.21}.  This
completes the proof of  \theprotag{5.9} {Proposition}.
     \enddemo

     \proclaim{\protag{5.22} {Corollary}}
 The line bundle with connection in \theprotag{5.9} {Proposition} descends
to a line bundle with connection
  $$ L_{Y;h_1,\dots,h_k}\longrightarrow \mflat{Y;h_1,\dots ,h_k}'
     =\fconn\Qfix\bigm/\gauge Q\tag{5.23}$$
over the moduli space~\thetag{4.33}.  The curvature times~$i/2\pi $ is given
by~\thetag{5.10}.
     \endproclaim

\flushpar
 This follows directly from the fact that the ``moment map''~\thetag{5.11},
which is the obstruction to descending the connection, vanishes on the space
$\fconn\Qfix$ of flat connections.  The quotient line bundle with connection
is independent of~$Q$ up to canonical diffeomorphism.

We next show that this connection descends to~$\mflat{Y;\hb_1,\dots ,\hb_k}$,
the moduli space of flat connections whose boundary holonomies are conjugate
to the~$h_i$ (cf.~\thetag{4.34}), at least over the irreducible connections.

     \proclaim{\protag{5.24} {Proposition}}
 The lift of $\gauge\Qfix$ to $L_{\Qfix}\to\fconn\Qfix$, defined in
\theprotag{4.46} {Proposition}, is parallel with respect to the connection
on~$L_{\Qfix}$ defined in \theprotag{5.9} {Proposition} and preserves that
connection as well.
     \endproclaim

 \demo{Proof}
 By \theprotag{5.9} {Proposition} we need only consider the action of the
subgroup $Z_h$~\thetag{4.36}.  An easy argument shows that it preserves the
holonomy along paths, hence preserves the connection, so we need only check
that the action is parallel.

Suppose $\langle \eta ,b \rangle\in \conn\Qfix$.  Fix a section $q\:Y\to Q$
such that $\partial q=r(\eta ,b)$.  Now let $\lambda \in \Lie(Z_{h_i})$ for
some fixed~$i$, and consider the path $g_t=e^{t\lambda }\in Z_{h_i}$.  Recall
that $Z_{h_i}$~is connected, so every element can be written~$e^{\lambda }$
for some~$\lambda $.  Acting on~$\langle \eta ,b \rangle$ we obtain the path
  $$ t\longmapsto \langle \eta ;b_1,\dots ,b_ig_t,\dots ,b_k \rangle
     \tag{5.25}$$
in~$\conn\Qfix$.  The connection on~$\zo\times Q$ it is independent of the
first factor, but the basepoint on the $i^{\text{th} }$ boundary component
varies.  Now the constant section $q\:\zo\times Y\to\zo\times Q$ does not
induce the (inverse) trivialization~\thetag{5.6} on $\zo\times (\bY)_i$, as
required by~\thetag{5.12}.  Instead, we need to use the section~$qg_t$.  Note
that $e^{\tpi W_{\zo\times (\bY)_i}(g_t)}$ induces the
trivialization~\thetag{A.4} of~$K_{g_t}$ on the boundary.  Hence the parallel
transport~\thetag{5.12} of the lift of~\thetag{5.25}
to~$L_{Q;h_1,\dots,h_k}$, computed relative to the trivialization
of~$L_{Q;h_1,\dots,h_k}$ induced by~$q$, is (cf.~\thetag{4.1})
  $$ \split
     c_{-\zo\times (\bY)_i}(r^*(\partial \eta ),g_t)
     &= c_{\zo\times (\bY)_i}\bigl( (rg_t)^*(\partial \eta ),g_t\inv \bigr)\\
     &=\exp\Bigl( -\tpi\int_{0}^1 dt\int_{(\bY)_i}\langle \Ad_{g_t}a\wedge
     g_t\inv \dot{g}\mstrut _t  \rangle \Bigr)\\
     &= \exp\Bigl( -\tpi \int_{0}^1\int_{(\bY)_i}\langle a\wedge \lambda
     \rangle\Bigr).\endsplit $$
This is canceled by the correction 1-form~\thetag{5.8}.
     \enddemo

The space~$\ficonn\Qfix$ of {\it irreducible\/} flat connections is a
manifold, and $\gauge\Qfix$~acts with constant stabilizer~$Z$ (cf.~the
discussion following~\thetag{4.32}).  Since $Z$~acts trivially on the line
bundle $L_{Q;h_1,\dots,h_k}\to\ficonn\Qfix$ by \theprotag{4.46}
{Proposition}, it follows from \theprotag{5.24} {Proposition} that this line
bundle with connection descends to a line bundle with connection
  $$ L_{Y;\hb_1,\dots ,\hb_k}\longrightarrow \mflat{Y;\hb_1,\dots
     ,\hb_k}^* \tag{5.26}$$
over the moduli space of irreducible flat connections with boundary
holonomies conjugate to the~$h_i$.

     \proclaim{\protag{5.27} {Proposition}}
 If $\form$~is a nondegenerate form on~$\frak{g}$, then the curvature
of~\thetag{5.26} is a symplectic form on~$\mflat{Y;\hb_1,\dots ,\hb_k}^*$.
     \endproclaim

     \demo{Proof}
 By \theprotag{4.40} {Lemma} we can realize tangent vectors at~$\eta $
to~$\mflat{Y;\hb_1,\dots ,\hb_k}^*$ by compactly supported forms~$\etd$, and
evaluated on two such forms the curvature~\thetag{5.10} is
  $$ \omega _\eta (\etd_1,\etd_2) = -2\int_{Y}\langle \etd_1\wedge \etd_2
     \rangle. \tag{5.28}$$
If $\form$~is nondegenerate, then Poincar\'e duality (for twisted
coefficients) implies that $\omega _\eta $~induces a nondegenerate pairing
  $$ H^1_c\bigl(\Yo;\frak{g}(\eta )\bigr)\otimes H^1\bigl(Y;\frak{g}(\eta
     )\bigr) \longrightarrow \RR; $$
here we do not require that $\etd_2$~in \thetag{5.28}~have compact support.
The exact sequence~\thetag{4.45} implies that $\omega _\eta $~is also
nondegenerate as a bilinear pairing on $H^1_c\bigl(\Yo;\frak{g}(\eta
)\bigr)/\im\delta \cong T_\eta \Cal{M}_{Y;\hb_1,\dots ,\hb_k}^*$.
     \enddemo

As with all of our constructions there is a gluing law for the connection
constructed in \theprotag{5.9} {Proposition}.  One consequence of the
basepoints, which rigidify the connection over the boundary, is that gluing
is well-defined for the pointed moduli space~\thetag{5.23}.  It is not,
however, well-defined once we remove the basepoint~\thetag{5.26}.

     \proclaim{\protag{5.29} {Proposition}}
 Suppose $Q\to Y$ is a $G$~bundle over a compact oriented 2-manifold with
parametrized boundary.  Fix boundary holonomies, which we collectively
denote~$h_{\bY}$.  Now suppose $S\hookrightarrow Y$ is a closed oriented
codimension one submanifold, and $Q\cut\to Y\cut$ the cut bundle.  Fix
holonomies~$h_S$ for the components of~$S$, and denote by~$h_{-S}$ their
inverses.  Then \thetag{4.8}~leads to a diagram of maps
  $$ \CD
     L_{Q\cut;h_{\bY},h_S,h_{-S}} @>>> L\mstrut _{Q;h_{\bY}}\\
     @VVV @VVV\\
     \mflat{Y\cut;h_{\bY},h_S,h_{-S}}' @>>> \mflat{Y;h_{\bY}}' \endCD
     \tag{5.30}$$
which is compatible with the connection constructed in \theprotag{5.22}
{Corollary}.
     \endproclaim

\flushpar
 There is a corresponding gluing law for the connection in \theprotag{5.9}
{Proposition}; it is proved using the gluing law \theprotag{4.11}
{Proposition} for the action and the definition~\thetag{5.12}, \thetag{5.8}
of the connection.  Then \thetag{5.30}~ is the quotient by~$\gauge Q$.

We also remark that this connection is compatible with change of gauge group,
as in \theprotag{2.9} {Proposition} and \theprotag{2.29} {Proposition}.
However, we must be careful to use compatible trivializations of the
universal bundles of connections over the circle.

Finally, we construct twistings of the line bundle~\thetag{5.26}.  Recall
that the map~\thetag{4.35}, restricted to the subspace of irreducible
connections, is a principal bundle with structure group~$Z_h/Z$, where
$Z_h$~is defined in~\thetag{4.36}.  Suppose
  $$ \lambda \:Z_h/Z\longmapsto \TT $$
is a unitary representation.  Then there is an induced hermitian line
bundle\footnote{We did not succeed in finding a unitary connection
on~$L_\lambda $.  Perhaps one should be induced from a connection
on~\thetag{4.35}, but we did not find a connection there either.}
  $$ L\mstrut _\lambda \longrightarrow \mflat{Y;\hb_1,\dots ,\hb_k}^*.
     $$
In the quantum Chern-Simons theory one considers the tensor product bundles
  $$ L_{Y;\hb_1,\dots ,\hb_k}\otimes L\mstrut _\lambda \longrightarrow
     \mflat{Y;\hb_1,\dots ,\hb_k}^* \tag{5.31}$$
for various~$\lambda $.  We do not find a relationship between the
holonomies~$h_i$ and the representation~$\lambda $, as was found
in~\cite{DasW,\S5}.

\newpage
\heading
\S{6} Special Cases
\endheading
\comment
lasteqno 6@  3
\endcomment

We first note some special features of the Hamiltonian theory~(\S{3}) on
surfaces of genus~$0$ and genus~$1$.  We also state the usual
formula~\thetag{6.3} for the Chern-Simons functional when $G=SU(2)$.

The 2-sphere $Y=S^2$ is simply connected, so according to \theprotag{3.5}
{Proposition} the moduli space~$\mflat Y$ consists of a single point.  Of
course, the dimension formula~\thetag{3.9} predicts that $\dim\mflat Y=0$.
The line corresponding to a trivial connection is canonically trivial (over
any surface).

If $Y=\cir\times \cir$~is a torus, then $\pi _1(\cir\times \cir,*)\cong
\ZZ\times \ZZ$ for the standard basepoint~$*\in \cir\times \cir$.  Hence
$\Hom\bigl(\pi _1(\cir\times \cir,*),G\bigr)$ is given by the set of
commuting pairs of elements $g_1,g_2\in G$.  We can simultaneously conjugate
both elements into a fixed maximal torus $T\subset G$, since they commute.
Then the normalizer~$N(T)$ acts by conjugation on pairs of elements in~$T$,
with $T\subset N(T)$ acting trivially.  Thus the quotient Weyl group
$W=N(T)/T$ acts, and
  $$ \mflat {\cir\times \cir} \approx T\times T\bigm/ W. \tag{6.1}$$
Note that for a generic connection~$\eta $ on the torus the
centralizer~$Z_\eta $ is isomorphic to the maximal torus~$T$.  If the
holonomies are both singular elements of~$G$, then the centralizer is bigger.
Thus the $W$~action degenerates at pairs of singular points of~$G$ (fixed by
the same element of~$W$).  Away from these singular points
\thetag{3.9}~predicts the proper formula for the dimension of~\thetag{6.1}.

We now specialize to $G=SU_2$.  Fix the standard maximal torus~$T$ of
diagonal matrices.  The nontrivial element of the Weyl group~$W\cong \zmod2$
acts on~$T$ by permuting the elements in a diagonal matrix.  The fixed points
are $1,-1\in T$, so there are four singular points in~\thetag{6.1}.

The integral forms on~$\frak{g}=\frak{s}\frak{u}_2$ are parametrized by an
integer~$k$:
  $$ \langle a\otimes b  \rangle = -\frac{k}{8\pi ^2}\Tr(ab),\qquad a,b\in
     \frak{s}\frak{u}_2. \tag{6.2}$$
Then the Chern-Simons functional~\thetag{2.2} takes the form
  $$ S_X(A) = -\frac{k}{8\pi ^2}\int_{X}\Tr(A\wedge dA + \frac 23 A\wedge
     A\wedge A), \tag{6.3}$$
where $A\in \Omega ^3_X(\frak{s}\frak{u}_2)$ is the connection form relative
to some trivialization.

\newpage
\heading
\S{A} Appendix:  Classical Wess-Zumino-Witten Theory\footnotemark
\endheading
\comment
lasteqno A@ 19
\endcomment

\footnotetext{I thank Jacek Brodzki for discussions about material in this
section.}

The Wess-Zumino-Witten functional~\thetag{2.13} is the action of a
$1+1$~dimensional topological classical field theory.  It arises in the
Chern-Simons theory as a term in the cocycle which measures the dependence of
the Chern-Simons action on the boundary trivialization.  In this appendix we
briefly study it as a field theory in its own right.  In fact, it is simpler
than the Chern-Simons theory and is a nice first example of the geometry of
classical field theory.  Our main interest, however, is in the metrized
lines~$K_\gamma $ attached to loops $\gamma \:\cir\to G$, which together form
a central extension of the loop group.  We restrict ourselves to connected,
simply connected, compact Lie groups~$G$ endowed with an integral bilinear
form~$\form\:\frak{g}\otimes \frak{g}\to\RR$ on the Lie algebra.  There is a
theory for arbitrary compact~$G$ endowed with a class in~$H^3(G)$; this can
be constructed using the techniques of Part~2 (cf.~\cite{Ga}).  Our
constructions of the lines here is quite simple, based on the fact that every
field on a ``space'' is the boundary of a field on some ``spacetime''.  This
procedure is elaborated in~\cite{Ax,\S4} in a more general context.

For $Y$~a closed oriented 2-manifold, equation~\thetag{2.13} defines a
function~$W_{Y}(g)$ on the space of (piecewise) smooth maps $g\:Y\to G$;
this is the content of \theprotag{2.12} {Lemma}.  To generalize to
2-manifolds with boundary, we must construct lines attached to the boundary
fields.

     \proclaim{\protag{A.1} {Proposition}}
 Let $S$~be a closed oriented 1-manifold \rom(union of circles\rom).  Then for
each smooth map $\gamma \:S\to G$ there is attached a metrized complex
line~$K_{\gamma }=K_{S,\gamma }$.  The assignment $\gamma \mapsto K_\gamma $
is smooth; satisfies functoriality, orientation, and additivity conditions
analogous to~\thetag{2.21}, \thetag{2.23}, and~\thetag{2.25}; and for $\gamma
_1,\gamma _2\:S\to G$ with pointwise product~$\gamma _1\gamma _2$ there is an
isometry
  $$ K_{\gamma _1}\otimes K_{\gamma _2}\longrightarrow K_{\gamma _1\gamma _2}.
     \tag{A.2}$$
Furthermore, for three loops~$\gamma _1,\gamma _2,\gamma _3$ we have the
commutative diagram
  $$ \CD
     K_{\gamma _1}\otimes K_{\gamma _2}\otimes K_{\gamma _3}  @>>> K_{\gamma
     _1}\otimes K_{\gamma  _2\gamma _3} \\
     @VVV @VVV\\
     K_{\gamma _1\gamma _2}\otimes K_{\gamma _3} @>>> K_{\gamma _1\gamma
     _2\gamma _3}
     \endCD \tag{A.3}$$
Finally, if $\gamma $~is a constant loop then there is a trivialization
 $$ K_\gamma \cong \CC  \tag{A.4}$$
which respects~\thetag{A.2}.
     \endproclaim

\flushpar
 It follows from~\thetag{A.2} and~\thetag{A.3} that the set of elements of
unit norm in the lines~$K_{\gamma }$ form a group~$\ce S$ which is a
central extension of the loop group:
  $$ 1 @>>> \TT @>>> \ce S @>>> \Map(S,G) @>>> 1. \tag{A.5}$$
This construction of the central extension first appeared in
Mickelsson~\cite{M}.  The last assertion means that this central extension is
split over the constant loops.  The functoriality means that the group of
orientation preserving diffeomorphisms~$\Diff^+(S)$ acts on~$\ce S$.  For
more on this central extension, see~\cite{PS,\S4}.

     \demo{Proof}
 We only sketch the constructions.  Fix an oriented 2-manifold~$D$
with~$\partial D=S$.  Let $\Cal{C}_\gamma $~be the category of extensions
$\Gamma \:D\to G$ of $\gamma \:S\to G$; it is nonempty by our assumptions on
the topology of~$G$.  Endow~$\Cal{C}_\gamma $ with a unique morphism
between any two objects.  Define a functor $\Cal{F}_\gamma \:\Cal{C}_\gamma
\to\Cal{L}$ by $\Cal{F}_\gamma (\Gamma )=\CC$ and $\Cal{F}_\gamma
(\Gamma _0\to\Gamma _1)$ is multiplication by
  $$ \exp\bigl( \tpi W_{D^{\text{double}} }(\Gamma _0\cup -\Gamma _1)\bigr).
     \tag{A.6}$$
Here $D^{\text{double} }$ is the double of~$D$ and $\Gamma _0\cup-\Gamma
_1$~is the (piecewise smooth) map which is~$\Gamma _0$ on one copy of~$D$ and
$\Gamma _1$~on the other, with the indicated orientations.  It is easy to
check that $\Cal{F}_\gamma $~is a functor; its invariant sections form the
line~$K_\gamma $.

To construct the isometry~\thetag{A.2} fix extensions~$\Gamma _1,\Gamma _2$
of~$\gamma _1,\gamma _2$.  Then $\Gamma _1\Gamma _2$~is an extension
of~$\gamma _1\gamma _2$.  These extensions trivialize the lines
in~\thetag{A.2}, and relative to these trivializations the
isometry~\thetag{A.2} is multiplication by
  $$ \exp\bigl( \tpi \int_{D}\langle \Gamma _1^*\theta  \wedge
     \Ad_{\Gamma _2}\Gamma _2^*\theta \rangle\bigr)
     =\exp\bigl( \tpi \int_{D}\langle \Gamma _1\inv d\Gamma \mstrut _1 \wedge
     d\Gamma \mstrut _2\,\Gamma _2\inv   \rangle\bigr), \tag{A.7}$$
where $\theta $~is the Maurer-Cartan form on~$G$.  This calculation proceeds
from the following formula.  For any 3-manifold~$X$ and $g\:X\to G$ a smooth
map, we set
  $$ \omega  (g) = -\frac 16 \langle g^*\theta \wedge [g^*\theta \wedge
     g^*\theta ]  \rangle = -\frac 16\langle g\inv dg\wedge [g\inv dg\wedge
     g\inv dg]  \rangle, $$
the Wess-Zumino-Witten lagrangian.  Then for $g_1,g_2\:X\to G$ we have
  $$ \omega (g_1g_2) = \omega (g_1)+ \omega (g_2) + d\sigma (g_1,g_2),
     \tag{A.8}$$
where
  $$ \sigma (g_1,g_2)= d\langle g_1^*\theta \wedge \Ad_{g_2}g_2^*\theta
     \rangle = d\langle g_1\inv dg_1\wedge dg_2g_2\inv   \rangle.
     $$

For the functoriality, suppose $f\:S'\to S$ is a diffeomorphism, $\gamma
\:S\to G$ a map, and $\gamma '=\gamma \circ f$.  Then an extension $\Gamma
\:D\to G$ gives a trivialization $K_\gamma \cong \CC$.  Let $D'$~be the
2-manifold $D'=S'\cup_f D$.  Then $\partial D'=S'$ and $\gamma '$~extends to
a map $\Gamma '\:D'\to G$ which agrees with~$\Gamma $ on~$D$.  Now $\Gamma
'$~gives a trivialization $K_{\gamma '}\cong \CC$.  Relative to these
trivializations we define $f^*\:K_\gamma \to K_{\gamma '}$ to be the
identity.

The trivialization for constant~$\gamma $ comes from the constant
extension~$\Gamma $.
     \enddemo

     \proclaim{\protag{A.9} {Proposition}}
 Let $Y$~be a compact oriented 2-manifold with boundary and $g\:Y\to G$ a
\rom(piecewise\rom) smooth map.  Then the action
  $$ e^{\tpi W_Y(g)}\in K_{\partial Y,\partial g} $$
is defined and satisfies functoriality, orientation, additivity, and gluing
conditions analogous to~\thetag{2.22}, \thetag{2.24}, \thetag{2.26},
and~\thetag{2.27}.  In addition, if $g_1,g_2\:Y\to G$ with pointwise
product~$g_1g_2$, then
  $$ e^{\tpi W_Y(g_1g_2)} = \exp(\tpi\int_{Y} \langle g_1^*\theta \wedge
     \Ad_{g_2}g_2^*\theta   \rangle) \, e^{\tpi W_Y(g_1)}\otimes e^{\tpi
     W_Y(g_2)}.  \tag{A.10}$$
     \endproclaim

\flushpar
 To compute the action we choose an oriented 2-manifold~$D$ with $\partial
D=\partial Y$ and extend~$\gamma =\partial g$ to $\Gamma \:D\to G$.  Then we
apply formula~\thetag{2.13} to $g\cup-\Gamma \:Y\cup -D\to G$.  The
dependence on the extension transforms by the cocycle~\thetag{A.6}.

One consequence of~\thetag{A.10} is
  $$ e^{\tpi W_Y(g\inv )} = e^{\tpi W_{-Y}(g)}. $$

We record here the differential of the Wess-Zumino-Witten functional.  For
any manifold~$M$ we identify the tangent space to~$\Map(M,G)$ at any point
with $\Map(M,\frak{g}) = \Omega ^0_M(\frak{g})$ via left translation.

     \proclaim{\protag{A.11} {Proposition}}
 Let $Y$~be a closed oriented 2-manifold.  Suppose $g\:Y\to G$ and $\Xi \in
\Omega ^0_Y(\frak{g})$.  Set $\Phi =g^*\theta \in \Omega ^1_Y(\frak{g})$,
where $\theta $~is the Maurer-Cartan form on~$G$.  Then
  $$ d(W_Y)_g(\Xi ) = \int_{Y}\langle \Xi \wedge d\Phi   \rangle. \tag{A.12}$$
     \endproclaim

\flushpar
 If $Y$~has nonempty boundary, then the same formula serves to compute the
covariant derivative of the action (relative to some trivialization) with
respect to the connection defined below.

     \demo{Proof}
 Let $X$~be an oriented 3-manifold with~$\partial X=Y$.  We extend $g,\Phi
,\Xi $ over~$X$ without introducing new notation.  Let $\Gamma _t$~be a path
of maps $X\to G$ with~$\Gamma \inv \dot{\Gamma }=\Xi $, where we use the
Leibnitz notation for the derivative at~$t=0$.  Set $\Phi _t=\Gamma _t^*\theta
$.  Then we calculate
  $$ \aligned
     d\Phi &= -\frac 12 [\Phi \wedge \Phi ]\\
     d\Xi &= \dot{\Phi} + [\Xi \wedge \Phi ].\endaligned \tag{A.13}$$
Now the Wess-Zumino-Witten lagrangian is
  $$ -\frac 16\langle \Phi \wedge [\Phi \wedge \Phi ]  \rangle = \frac
     13\langle \Phi \wedge d\Phi   \rangle. $$
Some calculation with~\thetag{A.13} shows
  $$ \frac{d}{dt}\res{t=0} \frac 13\langle \Phi \wedge d\Phi   \rangle =
     d\langle \Xi  \wedge d\Phi   \rangle, $$
and then \thetag{A.12}~follows by Stokes' theorem.
     \enddemo

The analogue of \theprotag{3.17} {Proposition} is the following.

     \proclaim{\protag{A.14} {Proposition}}
 Fix a closed oriented 1-manifold~$S$.  Then the Wess-Zumino-Witten action
defines a unitary connection~$\alpha $ on the hermitian line bundle
$K\to\Map(S,G)$.  The curvature of~$\alpha $ times~$i/2\pi $ is
  $$ \tau _\gamma (\xi _1,\xi _2) = - \int_{S}\langle [\xi _1,\xi _2]\wedge
     \phi \rangle, \qquad \xi _1,\xi _2\in \Omega ^0_S(\frak{g}), \tag{A.15}$$
where $\phi =\gamma ^*\theta $ is the pullback of the Maurer-Cartan form by
$\gamma \:S\to G$.  The central extension $\ce S$ acts on~$K$ by both left
and right multiplication.  The action of the center preserves the
connection~$\alpha $.  Left multiplication by $\gamma _0\:S\to G$
changes~$\alpha $ by the 1-form
  $$ \lambda _\gamma (\xi ) = \tpi\int_{S}\langle \gamma _0^*\theta \wedge
     \Ad_\gamma \xi   \rangle. \tag{A.16}$$
Right multiplication by $\gamma _0\:S\to G$ changes~$\alpha $ by the 1-form
  $$ \rho _\gamma (\xi ) = \tpi\int_{S}\langle\xi \wedge \Ad_{\gamma _0}
     \gamma _0^*\theta  \rangle. \tag{A.17}$$
     \endproclaim

\flushpar
 Note that since the center of~$\ce S$ preserves~$\alpha $, the action of an
element in $\ce S$ depends only on its image in~$\Map(S,G)$.
Formulas~\thetag{A.16} and~\thetag{A.17} give the difference between the
pullback of~$\alpha $ and~$\alpha $.  Also, the curvature~\thetag{A.15} is
the transgression of~$-\frac 16\langle \theta \wedge [\theta \wedge \theta ]
\rangle\in \Omega ^3_G$ to~$\Map(S,G)$.

     \demo{Proof}
 We use `$\alpha $' to denote both the connection (a 1-form on the circle
bundle of unit vectors in~$K$) and its local expression on~$\Map(S,G)$
relative to a trivialization.  We suppose that for each $\gamma \:S\to G$ in
some open set in the loop group we are given a smoothly varying extension
$\Gamma \:D\to G$, where $\partial D=S$, and so a trivialization of~$K$ over
that subset.  Note that these extensions determine extensions of tangent
vectors as well.  Now if $\gamma _t$~is a path in this subset, and
$\boldsymbol{\gamma }\:[0,1]\times S\to G$ the resulting map,
then\footnote{The sign in~\thetag{A.18} caused the author great confusion.
It is simply explained by the observation
  $$ \partial (-\zo\times D) = -\{1\}\times D \,\cup\, \{0\}\times D \,\cup\,
     \zo\times S. $$}
  $$ W_{[0,1]\times S}(\boldsymbol{\gamma }) = - \int_{0}^1dt\int_{D}\langle
     \Gamma \inv \dot{\Gamma }\wedge d\Phi   \rangle, \tag{A.18}$$
where $\Phi =\Gamma _t^*\theta $ is the pullback of the Maurer-Cartan form by
the extension~$\Gamma _t$ of~$\gamma _t$.  Consider the (connection)
1-form\footnote{Recall that parallel transport is the integral of {\it
minus\/} the connection form relative to a trivialization.}
  $$ \alpha _\gamma (\xi ) = \tpi \int_{D}\langle \Xi \wedge d\Phi
\rangle,\qquad
     \gamma \:S\to G,\quad \xi \in \Omega ^0_S(\frak{g}), \tag{A.19}$$
where $\Gamma \:D\to G$ and $\Xi \in \Omega ^0_D(\frak{g})$ are the given
extensions of~$\gamma $ and~$\xi $, and $\Phi =\Gamma ^*\theta $.  If
$\tilde{\Gamma }$~is a different extension of~$\gamma $, and it induces the
extension~$\tilde{\Xi }$ of~$\xi $, then the trivialization of~$K$ changes by
the {\it inverse\/} of the cocycle~\thetag{A.6}.  The logarithmic
differential of this cocycle is computed by~\thetag{A.12} as
  $$ - \tpi\int_{D}\langle \tilde{\Xi }\wedge d\tilde{\Phi }  \rangle \,+\,\tpi
     \int_{D}\langle \Xi \wedge d\Phi   \rangle, $$
which is precisely~$-(\tilde{\alpha }-\alpha )$, {\it minus\/} the difference
of the connection forms relative to the two trivializations.  This
consistency shows that $\alpha $ defines a unitary connection on
$K\to\Map(S,G)$.

Now if $\xo,\xt\in \Omega ^0_S(\frak{g})$ with extensions $\Xo,\Xt\in \Omega
^0_D(\frak{g})$, then the derivative of $\langle \Xt\wedge d\Phi   \rangle$ in
the
direction~$\Xo$ is
  $$ \split
     \langle \Xt\wedge d\dot{\Phi }  \rangle &= \langle \Xt\wedge
     -d[\Xo,\Phi ] \rangle \\
     &= \langle [d\Xo,\Xt]\wedge \Phi   \rangle + \langle [\Xo,\Xt]\wedge d\Phi
     \rangle\endsplit $$
by ~\thetag{A.13}.  Hence the curvature times~$i/2\pi $ is
  $$ \split
     \frac{i}{2\pi } d\alpha _\gamma (\xo,\xt) &= -\int_{D}\bigl(\,\langle
     [d\Xo,\Xt]\wedge  \Phi \rangle + \langle [\Xo,\Xt]\wedge d\Phi
     \rangle\,\bigr) - \bigl(\,\langle [d\Xt,\Xo]\wedge  \Phi \rangle +
     \langle [\Xt,\Xo]\wedge d\Phi \rangle\,\bigr)  - \langle [\Xo,\Xt]\wedge
     d\Phi  \rangle\\
     &= -\int_{D}\langle [d\Xo,\Xt]\wedge \Phi   \rangle + \langle
     [\Xo,d\Xt]\wedge \Phi   \rangle + \langle [\Xo,\Xt]\wedge d\Phi
     \rangle \\
     &= -\int_{S}\langle [\xo,\xt]\wedge \phi   \rangle.\endsplit $$

The assertion about the action of the center is one of the defining
properties of a connection form on a circle bundle---it is invariant under
the $\TT$~action.  Formulas~\thetag{A.16} and~\thetag{A.17} follow most
easily by computing the parallel transport~\thetag{A.18} along~$\gamma
_0\boldsymbol{\gamma }$ and~$\boldsymbol{\gamma }\gamma _0$,
using~\thetag{A.10}. Alternatively, compute directly with~\thetag{A.19}, but
include a factor~\thetag{A.7} for the identification of tensor products.
     \enddemo

\newpage
\heading
\S{B} Appendix:  Functions of paths and 1-forms
\endheading
\comment
lasteqno B@  8
\endcomment

Let $N$~be a smooth manifold, possibly infinite dimensional, and~$\pN$ the
space of smooth parametrized paths~$\ell \:\zo\to N$.  Then $\pN$~is also a
smooth manifold; a tangent to a path~$\ell $ is a vector field~$v(t)\in
T_{\ell (t)}N$ along~$\ell $.  A function $g\:\pN\to\RR$ is invariant under
reparametrization if $g(\tilde{\ell })=g(\ell )$, where $\tilde{\ell }(t) =
\ell \bigl(s(t)\bigr)$ for some diffeomorphism $s\:\zo\to\zo$ which fixes the
endpoints.  Suppose $g$~is such a function, $\ell $~is any path, and
$t_i$~are chosen with $0=t_0\le t_1\le \cdots\le t_M=1$.  Let $\ell _i = \ell
\res{[t_{i-1},t_i]}$.  Then we call~$g$ {\it additive\/} if
  $$ g(\ell ) = \sum\limits_{i=1}^M g(\ell _i), $$
where $g(\ell _i)$~makes sense as $g$~is invariant under reparametrization.
A {\it point path\/} is a constant function~$\ell \:\zo\to N$.

     \proclaim{\protag{B.1} {Proposition}}
 Suppose $g\:\pN\to\RR$ is a smooth function such that: (i)\ $g$~vanishes on
point paths; (ii)\ $g$~is invariant under reparametrization; and (iii)\
$g$~is additive.  Then there is a smooth 1-form~$\theta $ on~$N$ such that
  $$ g(\ell ) = \int_{\ell }\theta . \tag{B.2}$$
     \endproclaim

     \demo{Proof}
 Let $\pi \:TN \to N$ be the projection, and define the linear inclusion
  $$ \aligned
     i\:TN &\longrightarrow T\pN\\
     v &\longmapsto (t\mapsto tv),\endaligned $$
where $t\mapsto tv$ is a tangent vector to the constant path at~$\pi (v)$.
Now set
  $$ \theta (v) = dg\bigl( i(v) \bigr). \tag{B.3}$$
Since $g$~is smooth, so is~$\theta $; and $\theta $~is linear since $i$
and~$dg$ are.  Hence $\theta $~is a smooth 1-form on~$N$.  Fix a path~$\ell $
and a large integer~$M$.  Set $\ell _i = \ell \res{[\frac{i-1}{M},\frac
iM]}$.  Then by additivity,
  $$ g(\ell ) = \sum\limits_{i=1}^M g(\ell _i). \tag{B.4}$$
Reparametrize~$\ell _i$ by
  $$ \ell _i(t) = \ell (\frac{t+i-1}{M}),\qquad 0\le t\le 1, $$
and set
  $$ \ell ^\epsilon _i(t) = \ell (\frac{\epsilon t+i-1}{M}),\qquad 0\le t\le
     1. $$
Then $\epsilon \mapsto \ell ^\epsilon ,\; 0\le \epsilon \le 1$, is a smooth
path in~$\pN$ from the point path at~$\ell (\frac{i-1}{M})$ to~$\ell _i$.
The tangent to this path at~$\epsilon =0$ is $t\mapsto \frac tM \,\dot{\ell
}(\frac{i-1}{M})$.  By Taylor's theorem and the definition~\thetag{B.3}
of~$\theta $,
  $$ g(\ell _i) = \frac 1M\, \theta \bigl( \dot{\ell }(\frac{i-1}{M})\bigr) +
     O(\frac{1}{M^2}). \tag{B.5}$$
Finally, combine~\thetag{B.4} and~\thetag{B.5} and take~$M\to\infty $ to
obtain~\thetag{B.2}.
     \enddemo

     \proclaim{\protag{B.6} {Corollary}}
 Let $L\to M$ be a smooth hermitian line bundle.  Suppose that for each
path~$\lb\in \pM$ there is given an isometry
  $$ \partrans(\ell )\:L_{\lb(0)} \longrightarrow L_{\lb(1)} \tag{B.7}$$
depending smoothly on~$\ell $ such that: \rom{(i)}\ $\partrans$~is the
identity map on point paths; \rom{(ii)}\ $\partrans$~is invariant under
reparametrization; and \rom{(iii)}\ $\partrans$~is additive.  Then there is
a unitary connection on~$L$ for which $\partrans$~is the parallel transport.
     \endproclaim

     \demo{Proof}
 Let $N\subset L$ be the vectors of unit norm.  Then $N\to M$ is a circle
bundle.  If $\ell \in \path(N)$ is a smooth path in~$N$, then its
projection~$\lb$ is a smooth path in~$ M$.  Using the parallel
transport~\thetag{B.7} we find an element~$\partrans(\lb)\ell (0)\in
L_{\lb(1)}$ of unit norm.  Hence it differs from~$\ell (1)$ by an element
$h(\ell )\in \CC$ of unit norm:
 $$ \ell (1) = \partrans(\lb)\ell (0)\cdot h(\ell ), \tag{B.8}$$
where the multiplication~$\cdot $ occurs in the fiber~$L_{\lb(1)}$.  In other
words, $h$~determines a smooth function
  $$ h\:\path(N)\longrightarrow \TT  $$
into the unit complex numbers.  The hypotheses imply: (i)\ $h\equiv 1$~on
point paths; (ii)\ $h$~is invariant under reparametrization; and (iii)\
$h$~is additive.  In addition, we claim that $h$~has a logarithm; that is,
there exists a function $g\:\pN\to i\RR$ with $h(\ell ) = e^{g(\ell )}$.  This
is because $\path(N)$~retracts onto the space of constant paths and $h\equiv
1$ on constant paths.  Therefore, by \theprotag{B.1} {Proposition} there
exists a smooth (imaginary) 1-form~$\theta $ on~$N$ such that
  $$ h(\ell ) = e^{\int_{\ell }\theta }. $$
{}From ~\thetag{B.8} it is obvious that the restriction of~$\theta $ to a
fiber of the circle bundle $N\to M$ is the standard Maurer-Cartan form on the
circle.  Similarly, from~\thetag{B.8} we see that $h$~is invariant under the
circle action on~$N$, whence $\theta $~is also.  Therefore, $\theta $~is a
connection on~$N$ and \thetag{B.7}~is the associated parallel transport.
     \enddemo

\enddocument